%% file: MAIN.tex
\documentclass[astrosymb, twocolumn]{aastex631}
\usepackage{subfigure}
\usepackage{rotating}
\usepackage{mathtools}
\usepackage{xspace}

\received{}
\revised{}
\accepted{}
\submitjournal{\apj}
\shorttitle{UV Variable AGN}
\shortauthors{Wasleske et al.}
\graphicspath{{./}{./}}

\begin{document}

\title{Variable AGN in the GALEX Time Domain Survey}

\author[0000-0003-3986-9427]{Erik J. Wasleske}
\affiliation{Department of Physics and Astronomy, Washington State University, Pullman, WA 99163, USA}

\author[0000-0003-4703-7276]{Vivienne F. Baldassare}
\affiliation{Department of Physics and Astronomy, Washington State University, Pullman, WA 99163, USA}

\author[0000-0003-3574-2963]{Christopher M. Carroll}
\affiliation{Department of Physics and Astronomy, Washington State University, Pullman, WA 99163, USA}

%%%%%%%%%%%%%%%%%%%%%%%%%%%%%%%%%%%%%%%%%%%%%%%%%%%%%%%%%%%%%%%%%%%%%%%%%%%%%%%%
\begin{abstract} 
We searched the Northern Hemisphere Fields of the GALEX Time-Domain Survey (TDS) for galaxies with UV variability indicative of active galactic nuclei (AGNs). We identified 48 high-probability candidate AGNs from a parent sample of 1819 galaxies in the NASA Sloan Atlas (NSA) catalog. We further characterized these systems using optical spectroscopic diagnostics, WISE IR color selection criteria, and spectral energy distribution (SED) modeling. Of the 48 candidates, eight were identified as AGNs  from optical emission lines, two were identified by their IR colors, and 28 were identified through spectral energy decomposition. Observational biases of each selection method are discussed in connecting these AGNs subsamples to another. By selecting AGNs based on UV variability, we also identified six low-mass AGNs candidates, all of which would have been missed by spectroscopic selection. 

\end{abstract}

%\keywords{Active galactic nuclei, Intermediate-mass black holes }

%%%%%%%%%%%%%%%%%%%%%%%%%%%%%%%%%%%%%%%%%%%%%%%%%%%%%%%%%%%%%%%%%%%%%%%%%%%%%%%%

\section{Introduction} \label{sec:intro}

Supermassive black holes (SMBHs) have been found to reside in the centers of all relatively massive galaxies \citep{1998AJ....115.2285M}. SMBH masses also correlate with properties of their host galaxy such as stellar mass and stellar velocity dispersion \citep{2000ApJ...539L...9F, 2000ApJ...539L..13G, Kormendy:2013ve}. However, there are many open questions surrounding black hole (BH) formation and growth, as well as the abundance of SMBHs in low-mass galaxies ($\text{M}_{\ast}<10^{10}\;\text{M}_{\Sun}$). Answering these questions relies on fully characterizing the population of SMBHs. 

Identifying BHs in low-mass galaxies  is of key importance for constraining models of BH seed formation \citep{IMBH_Handbook_Greene_2020}. The low-mass end of the BH occupation fraction and BH mass function can help illuminate whether BH seeds were common or rare, as well as their likely masses \citep{Volonteri2008MNRAS.383.1079V, VolonteriARAA, natarajan2014GReGr..46.1702N}. Unfortunately, detecting BHs in these systems is difficult. Based on scaling relations at higher galaxy masses, we expect BHs in dwarf galaxies to be in the mass range of $\sim10^{4-5}\;\text{M}_{\Sun}$. Thus, dynamical detection is only currently possible for systems within $\sim10$ Mpc. It is therefore necessary to search for signs of accretion (i.e., active galactic nuclei) to build statistical samples of dwarf galaxies with BHs. Much work has been dedicated to this effort, including searches with optical spectroscopic signatures \citep{Reines:2013pia, Moran2014AJ....148..136M, Molina2021ApJ...922..155M}, X-rays  \citep{Pardo2016ApJ...831..203P, Mezcua2018MNRAS.478.2576M}, and radio emission \citep{Reines2020ApJ...888...36R}. Each method has particular advantages and limitations. In this work, we focus on searches using UV variability. 

AGNs are known to vary over an array of time scales across the entire electromagnetic spectrum. Variability is a prolific tool to identify AGNs (e.g., \citealt{Ulrich_et_al_1997, Geha2003AJ....125....1G, Macleod2011ApJ...728...26M, Caplar2017ApJ...834..111C}). Importantly, it can do so in systems that lack optical spectroscopic signatures due to star formation dilution \citep{Trump_2015}, metallicity \citep{groves2006MNRAS.371.1559G}, or obscuration. Indeed, variability has been shown to select AGNs in dwarf galaxies that are missed by spectroscopic selection \citep{Baldassare2018ApJ..868..152, Baldassare2020ApJ...896...10B}. The timescale of variability depends on the origin of the emission. X-ray variability tends to occur on relatively shorter timescales than optical variability. This is because X-rays are generated closer to the BH in the hot corona, while optical/UV emission comes from the accretion disk. 

Much work has been done studying the optical variability of AGNs, thanks to ground-based time domain surveys such as the Sloan Digital Sky Survey (SDSS; \cite{york2000}) Stripe 82 Legacy field \citep{abzajian2009}, Palomar Transient Factory \citep{Law2009PASP..121.1395L}, Zwicky Transient Facility \citep{Graham2019PASP..131g8001G}, and Pan-STARRS \citep{Chambers2016arXiv161205560C}. Comparatively fewer studies have been carried out using UV observations (GALEX; \cite{inital_GALEX_paper} despite the fact that the UV should probe emission from hotter, interior parts of the accretion disk. The Galaxy Evolution Explorer (GALEX) Time Domain Survey (TDS; \citealt{Gezari_2013}) offers the opportunity to study the UV variability of thousands of nearby galaxies. GALEX was a NASA Explorer mission satellite telescope, launched in early 2003, surveying galaxies in the near and far ultraviolet (1350--2800 \AA \xspace) \citep{inital_GALEX_paper}. In line with its goals of studying star formation and galaxy evolution, the GALEX TDS collected repeat observations over three years of approximately $40\;\rm{deg}^{2}$ of sky. The long baseline and sensitivity of GALEX presents an optimal opportunity to search for variable AGN \citep{Gezari_2013}. 

\cite{Gezari_2013} searched the GALEX TDS for strongly variable sources. They selected sources as variable if they vary at the $5\sigma$ level in at least one epoch. They identified over a thousand variable sources, including 358 quasars and 305 AGNs. \cite{Hung_2016} explored the optical spectra and wavelength-dependent variability of 23 active galaxies from the GALEX TDS with optical light curves from Pan-STARRS. They find that almost all of their sources have optical spectra consistent with an AGN and that this emission becomes bluer when the AGN is more luminous.

We aim to build on previous studies using the GALEX TDS by searching for UV variable AGN candidates in nearby galaxies. Our primary goal was to identify elusive AGNs, such as AGNs in low-mass galaxies and those lacking optical spectroscopic AGN signatures. 

This paper is organized as follows: Section \ref{sec:Data_Set} will discuss the selection of reference catalogs and the GALEX survey; Section \ref{sec:Methods} will outline the methodology used to identify variable objects through construction of light curves and fitting their emission spectrum to measure BH mass; Section \ref{sec:Results} will present the results of this study and Section \ref{sec:Discussion} will consider these results and the framework of further studies. Throughout this paper, we assume a $\lambda$CDM cosmology with parameters $h = 0.7$, $\Omega_m = 0.3$, and $\Omega_\lambda = 0.7$ \citep{spergel2007}.

%%%%%%%%%%%%%%%%%%%%%%%%%%%%%%%%%%%%%%%%%%%%%
\section{Data}
\label{sec:Data_Set}

%%%%
\subsection{GALEX Time Domain Survey}
\label{subsec: GALEX TDS}
\begin{figure}
    \centering
    \subfigure{\includegraphics[scale=0.30]{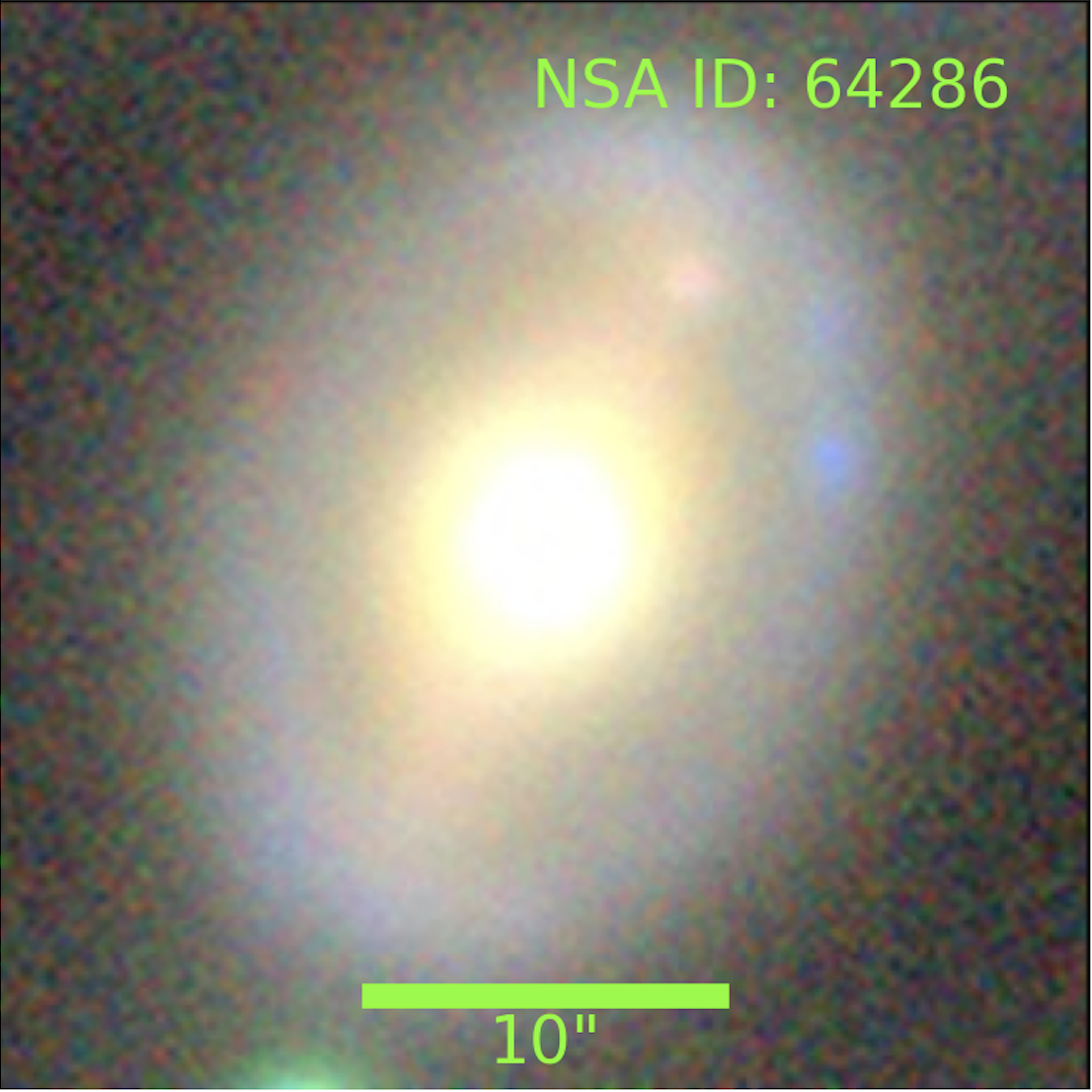}}
    \hspace{1em}% Space between image A and B
    \subfigure{\includegraphics[scale=0.30]{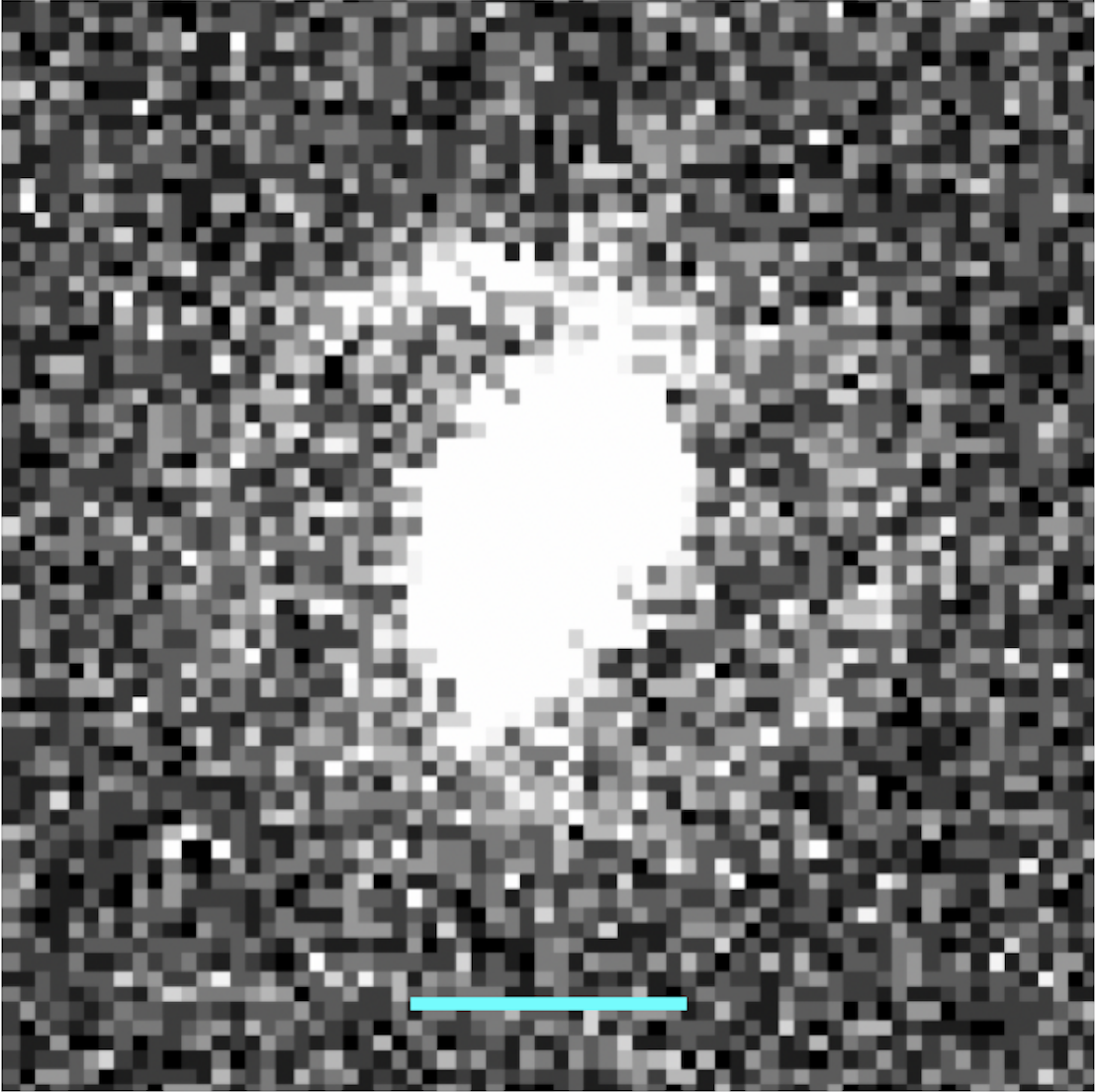}}%\quad
    \caption{\emph{Top:} Legacy Survey Image of galaxy J100207.04+030327.6, NSA ID:64286, with a scale bar of 10". \emph{Bottom:} 
    GALEX NUV z-scaled image of galaxy 64286 observed on February $14^{th}$, 2008. Scale bar shows 25" scale.}
    \label{fig:J100207.04+030327.6_images}
\end{figure}

The GALEX TDS is a dataset collected in the near-UV (NUV) band of the GALEX satellite. This NUV channel covered a range of 1750\AA \xspace to 2800\AA \xspace. GALEX TDS consists of seven $1.1^{\circ}$ diameter circular field pointings within six of the PS1 MDS fields, being the XMMLSS, CDFS, COSMOS, GROTH, ELAISN1, and VVDS22H fields \citep{Gezari_2013, Hung_2016}. Taken over approximately three years with a cadence of roughly two days, the GALEX TDS generated a catalog particularly useful for searches for variable AGNs. Figure \ref{fig:J100207.04+030327.6_images} displays an image of the galaxy with IAU designation J100207.04+030327.6 for the GALEX images as well as the Legacy Survey. 

%%%%
\subsection{Additional Catalogs}
\label{subsec: Data_catalogs}

For aperture photometry, a reference catalog was needed to identify galaxies within the imaging data to build the parent sample. In addition, we selected a star catalog in order to characterize the field around each galaxy.

We used the NASA Sloan Atlas (NSA) as our parent sample. The NSA combines SDSS and GALEX images to construct a list of roughly $641,500$ galaxies  \citep{2009ApJ...691..394M, 2009ARA&A..47..159B, 2010ApJ...722..491Z}. The NSA covers galaxies observed up to SDSS DR13 and provides various photometric and spectroscopically derived quantities, including stellar masses.

The NSA catalog's most recent iteration extends to a redshift of $z=0.15$. With a declination angle range in $-24$ deg to $85$ deg, the NSA does not provide ample coverage for the southern fields (XMMLSS and CDFS) of the GALEX TDS, which will be discussed in a future publication using a separate catalog to generate the parent sample of galaxies. 

We used the LAMOST Catalog, which consists of roughly 5.7 million sources, to develop a list of stars surrounding our galaxy targets \citep{Luo_2015}. \cite{Bai_2018} to crossmatch stars observed by GALEX and acquire UV aperture photometry. About $2.2$ million stars of the LAMOST Catalog were detected by GALEX, thus LAMOST provides a check to the stars amongst the GALEX galaxies. This catalog is our stellar sample for the star selection process described in Section \ref{subsubsec:Aperture_Photometry_Application}.  

Finally, we obtained multiwavelength photometry for modeling the spectral energy distributions (SEDs) of our sources. We use our GALEX photometry (Section \ref{subsubsec:Aperture_Photometry_Application}) with data from SDSS DR16 \citep{SDSSDR16paper}, The United Kingdom Infrared Telescope (UKIRT; \cite{UKIRT_Lawrence2007MNRAS.379.1599L}), Wide-field Infrared Survey Explorer (WISE; \cite{WISE_Wright2010AJ....140.1868W}), unblurred and unofficial co-adds of the WISE imaging (unWISE; \cite{unWISE_Lang2014AJ....147..108L}), and The Two Micron All-Sky Survey (2MASS; \cite{2MASS_2006AJ....131.1163S}).
%%%%%%%%%%%%%%%%%%%%%%%%%%%%%%%%%%%%%%%%%%%%%
\section{Methods}
\label{sec:Methods}
We constructed light curves for every galaxy and then searched for sources with significant variability,  which we labeled as AGN candidates. We further analyzed these sources and studied their optical spectroscopy, IR colors, and spectral energy distributions. The methodology used to execute these steps is described below. 
%%%%
\subsection{Light Curve Construction } \label{subsec:Light_Curve_Const}
Here we describe the construction of light curves of stars and galaxies within the GALEX TDS using aperture photometry.

\cite{Gezari_2013} stated that as most galaxies are unresolved by the GALEX point-spread function (PSF) and space-based observations are not seeing-dependent, aperture photometry was a sufficiently accurate method for determining variability. Two apertures were applied to each object to measure the signal. A circular aperture with 9$''$ radius---the full-width half-maximum (FWHM) of the GALEX PSF---was centered over the object's position to determine the total signal. A secondary annulus, with inner radius 27$''$ (3x the FWHM) and outer radius 54$''$ (6x the FWHM) was used to measure the local background surrounding each object. This was repeated for each object and each epoch that the pointing was observed.

Typical photometric uncertainties were calculated using Poissonian errors. Poisson error has been shown to underestimate the uncertainties in GALEX photometry by a factor of $\sim$2 \citep{2007AJ....133.1780T, Gezari_2013}. To compensate for this, we estimated uncertainties empirically following the technique described in \cite{Gezari_2013}. This is discussed in further detail below and demonstrated in Figure \ref{fig:star_std_template}. 

\subsubsection{Aperture Photometry} \label{subsubsec:Aperture_Photometry_Application}
We use aperture photometry to generate a light curve for each galaxy in the parent sample. The aperture photometry routine was then repeated for stars from the LAMOST catalog within a $0.1$ deg. radius of the galaxy. This produced a short list of $\sim$ten stars near each galaxy that provided a representation of how the local field varied. These stellar light curves were also used to establish the empirical uncertainties.

\begin{figure}
    \centering
    \includegraphics[width=0.5\textwidth]{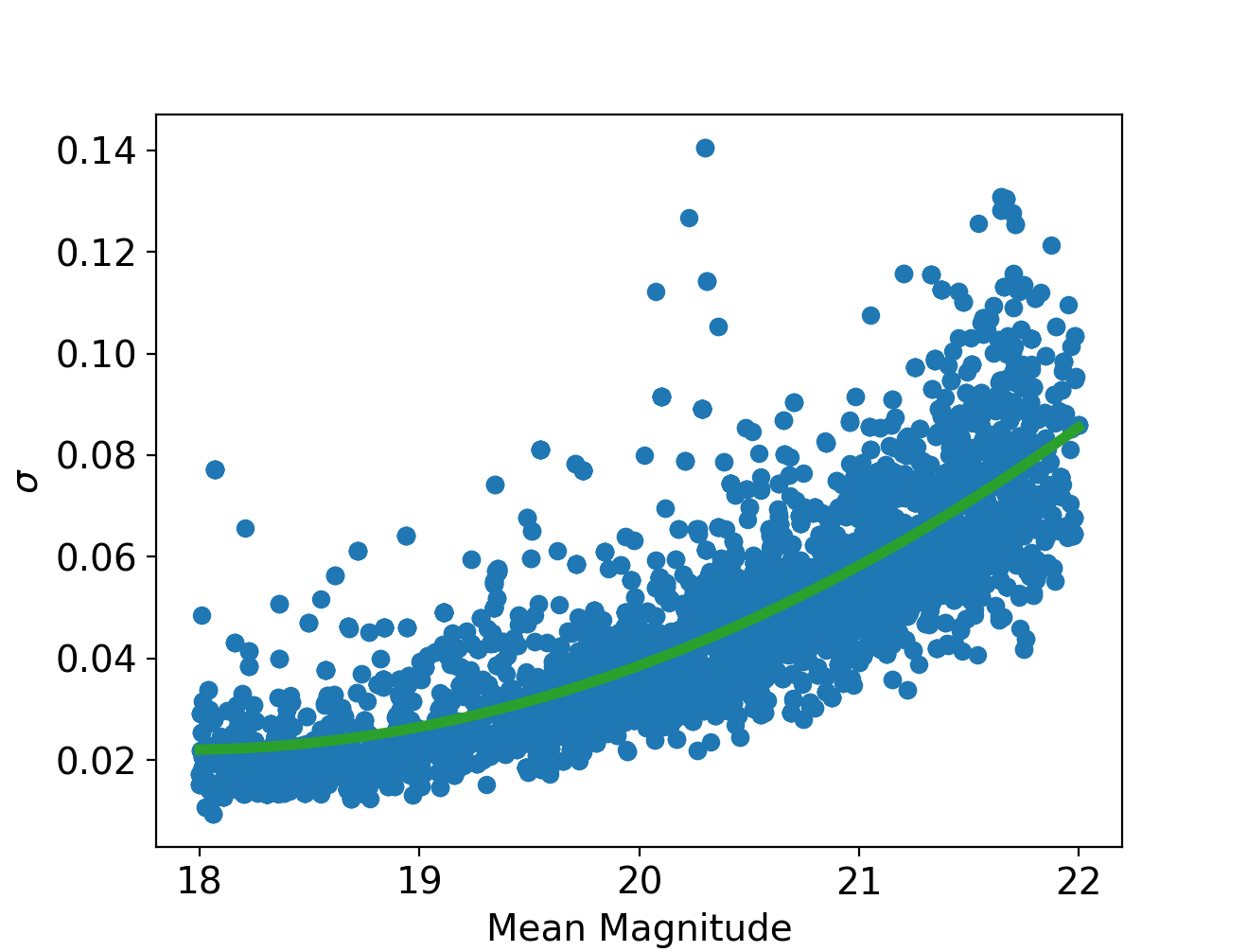}
    \caption{Standard deviation per mean magnitude for sigma-clipped selected stars within $\protect 0.1$ deg of each galaxy located in the GALEX TDS. This population contains 9583 stars. The green fitting line is a polynomial of degree 2.}
    \label{fig:star_std_template}
\end{figure}

Stars from this population were used to build a relation between the standard deviation and mean magnitude. We remove outliers using a sigma-clipping of 3$\sigma$ \citep{1960mspa.book.....C}. Our results are shown in Figure \ref{fig:star_std_template}, where a quadratic polynomial fit was applied to this clipped set of stars. We then adopted the quadratic fit of Figure \ref{fig:star_std_template} to estimate magnitude-dependent uncertainties in each galaxy light curve.

\begin{figure*}
    \centering
    \includegraphics[width=\textwidth]{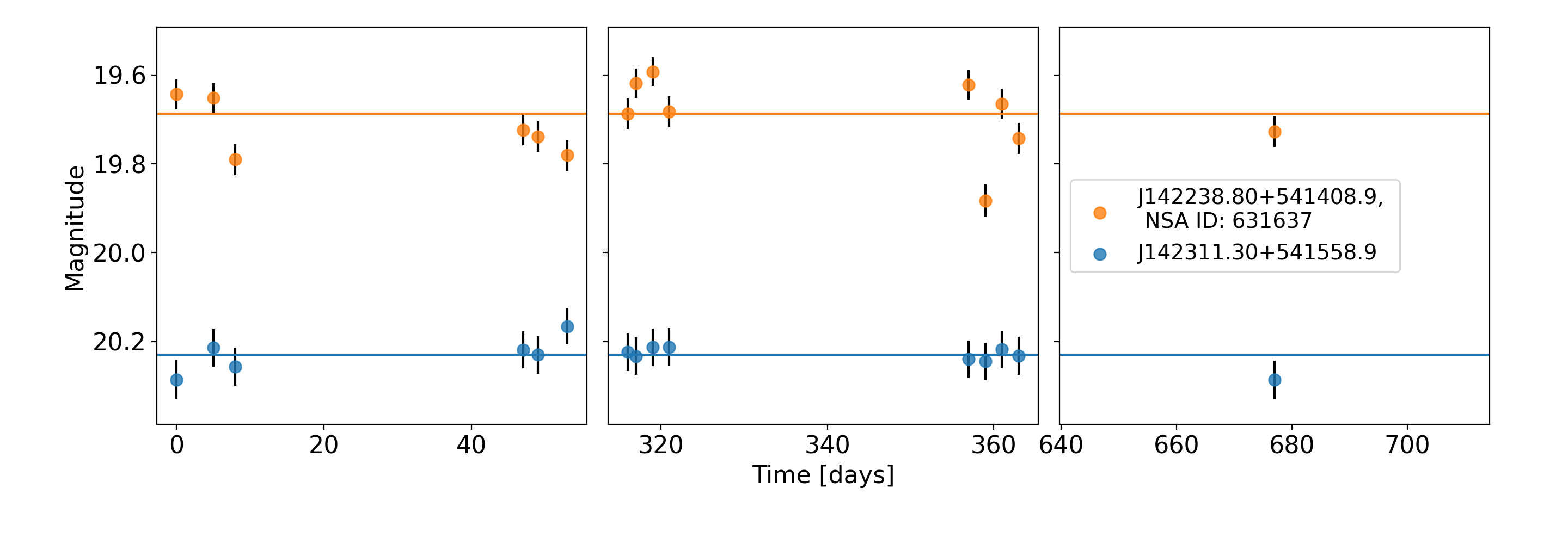}
    \caption{Light curve examples for a variable galaxy (orange) and a nearby star(blue). Median magnitude values for the galaxy and star are displayed by the horizontal lines. The error bars are set from the relation demonstrated in Figure \ref{fig:star_std_template}. The galaxy's magnitude range spans over $0.3$ magnitude while the star varies in values within $0.2$ magnitude. }
    \label{fig:Light_Curve_Ex}
\end{figure*}

\subsubsection{Variable Galaxy Selection}
\label{subsubsec: galaxy_selection}
Galaxies were selected as highly variable from calculations of the standard deviation of the objects light curve. We confirmed their variability by calculating the objects fractional variance, such that
\begin{equation}
    F_{\text{var}} = \frac{\sqrt{\sigma_{\text{var}}^2 - \delta^2}}{\left<f\right>} ,
    \label{eq:fractional_var}
\end{equation}
where $\sigma_{\text{var}}^2$ is the variance of the flux signal, $\delta^2$ is the mean square of the uncertainty of the fluxes, and $\left<f\right>$ is the mean flux for all observations \citep{1990ApJ...359...86E, Rodriguez_Pascual_1997, Characterize_Variable_Vaughan_et_al_03}. Figure \ref{fig:Light_Curve_Ex} provides and example light curve of a highly variable galaxy and a nearby star. In this figure, the magnitude of the galaxy varies in value by $0.30$ magnitude, more than twice that of the star. This comparison shows how non-varying stars set the reference for varying galaxies.

After calculating the standard deviation of each object's light curve, the objects were binned by their mean magnitude values, in bins of $0.5$ mag. In each bin, galaxies were selected as highly variable if their standard deviation value was greater than twice the deviation above the bin's mean value. The highly variable galaxies can be seen compared to the less variable galaxies in Figure 4. 

This process was repeated for the values of fractional variation from the object's light curves. This resulted in recovering 23 of the variable galaxies identified through high standard deviation values with no new objects. The 25 galaxies not selected still demonstrated comparable $F_{\text{var}}$ to the rest of the population, as shown in Table \ref{tab:results_table}. Objects that were selected based on their standard deviation but not fractional variation were lost in the more populated higher magnitude bins. Of these objects, 16 had NUV magnitudes greater than 21 mag. As $F_{\text{var}}$ has dependence on the uncertainty of the photometry, dimmer objects would be more likely have higher fractional variance values. For those bins, our threshold value is higher, so we collect a less complete sample.  

\begin{figure}
    \centering
    \includegraphics[width=0.5\textwidth]{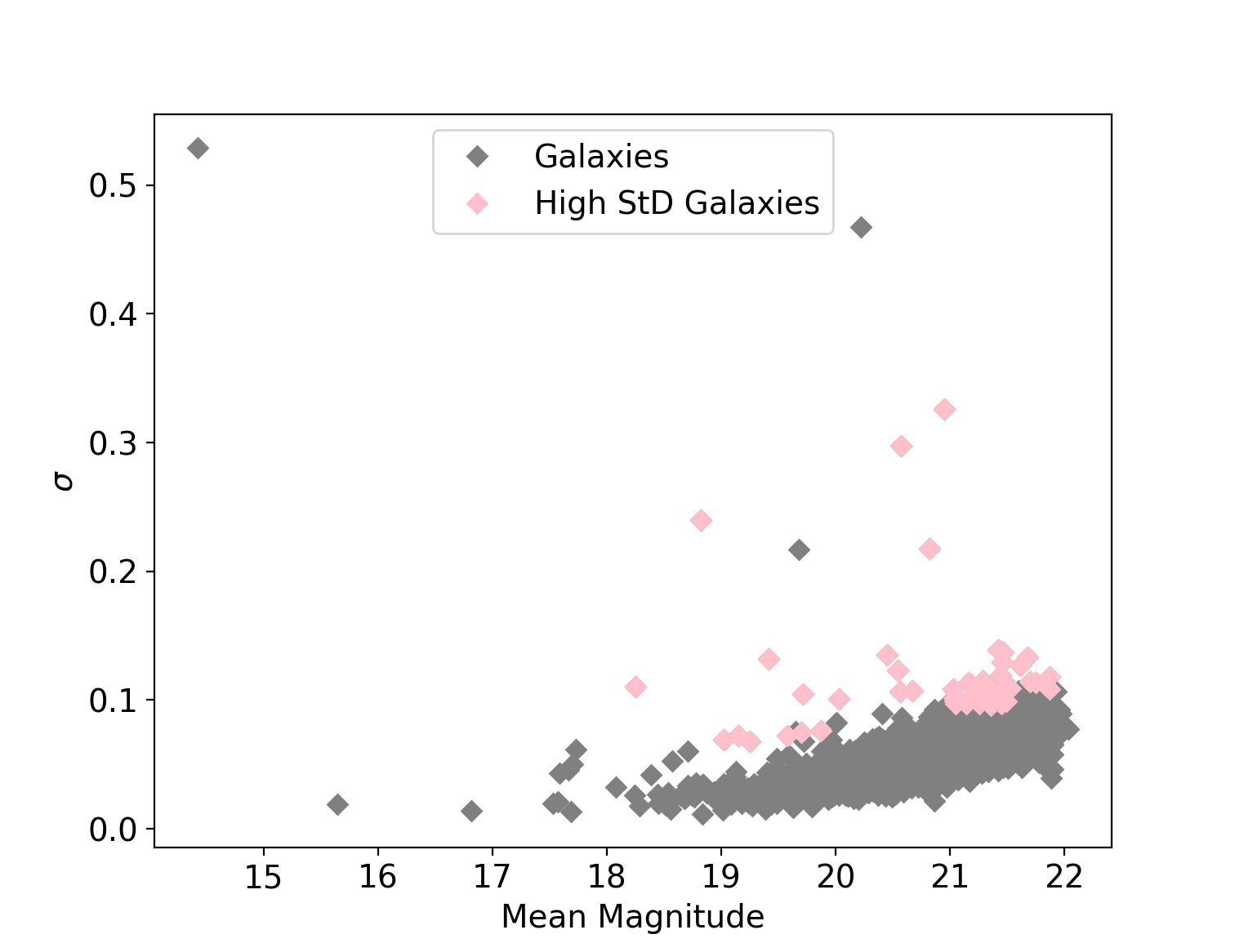}
    \caption{Galaxies standard deviation $\protect\sigma$ value plotted against the objects mean magnitude value. For every $0.5$ mag step along the mean magnitude axis, the deviation of that bin was calculated. High standard deviation galaxies were selected if they exhibited a $\protect\sigma$ value greater than twice the deviation over the mean $\protect\sigma$ value of its magnitude bin. Two galaxies met this threshold with their light curves consisted of one flare event against otherwise constant magnitude. These were thus removed from the High StD population}
    \label{fig:gals_std_select}
\end{figure}

\begin{figure*}[htp]
  \centering
    \includegraphics[width=\textwidth]{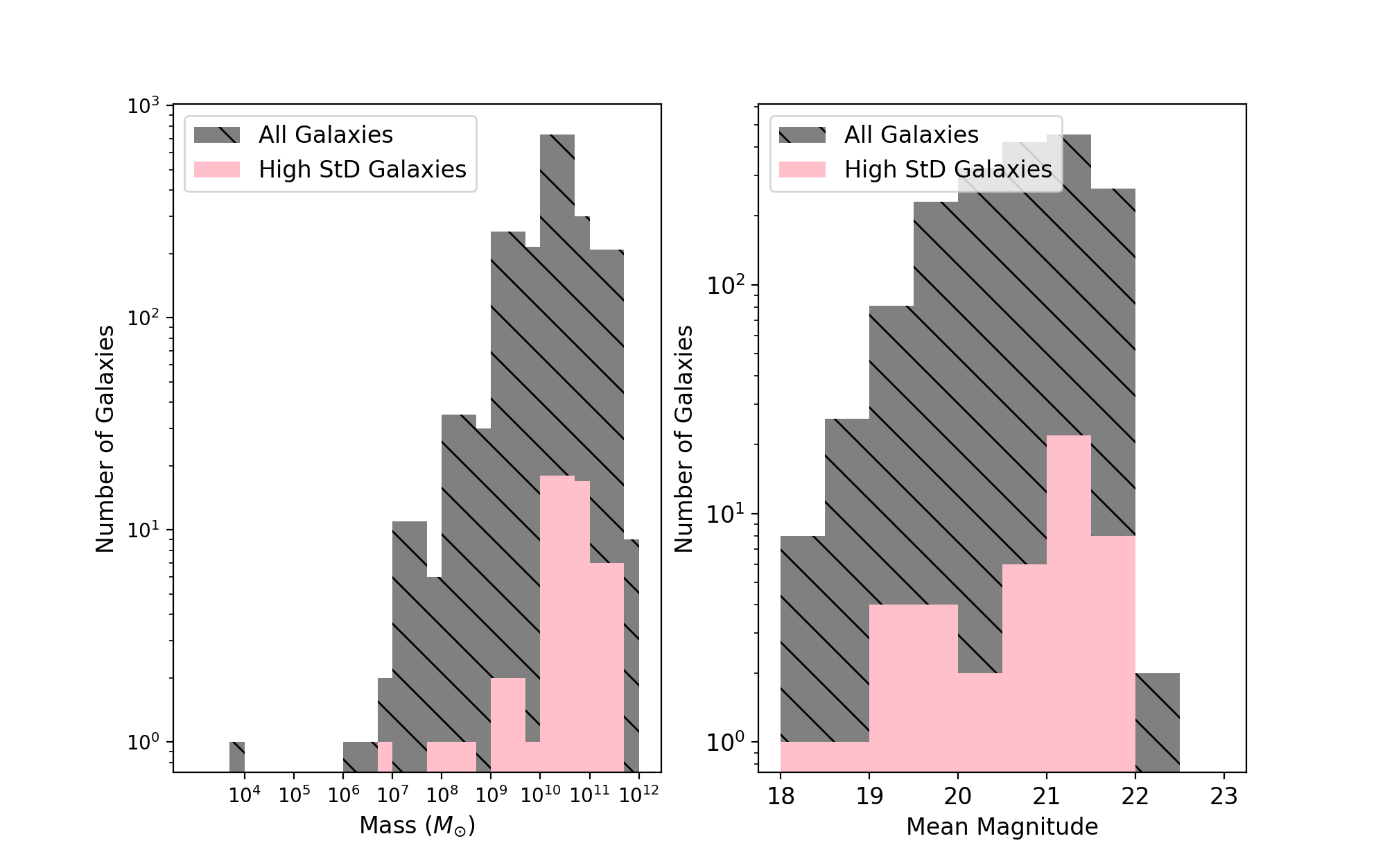}
  \caption{\emph{Left:} Histogram of the the stellar mass listed in the NSA given in solar masses for the parent and high standard deviation populations of galaxies. \emph{Right:} Histogram of the calculated mean magnitude for the parent and high standard deviation populations of galaxies.}
  \label{fig:gals_histograms}
\end{figure*}

The high standard deviation population of galaxies well represented the distribution of mass and mean magnitude within the parent population, as show in the histograms of Figure \ref{fig:gals_histograms}. These 48 objects comprised our final group of variable AGN candidates.

%%%%

\subsection{Emission Line Fitting} \label{subsec:emission_fitting}
After selecting variable galaxies, we used SDSS DR7 observations to study their spectroscopic properties. All spectral lines were fitted with a sequence of 1D Gaussian models following  \cite{Reines:2013pia}. Prior to fitting, the local continuum around each peak was modeled by a linear line which was subtracted from the spectrum. Flux densities were corrected from reddening using the dust-maps of \cite{1998ApJ...500..525S}. This method is outlined as such:
\begin{enumerate}
    \item First we fit the [\ion{S}{2}] doublet with a single Gaussian for each line.
    
    \item We fit the H$\alpha$ + [\ion{N}{2}] complex following the steps below.
    \begin{enumerate}
        \item The [\ion{S}{2}] fit sets the template for the narrow lines in the H$\alpha$ + [\ion{N}{2}] complex. The narrow lines for this complex were fit by fixing the centers of the [\ion{N}{2}] lines to the values of 6584\AA \xspace and 6548\AA \xspace. In addition, the flux ratio of the 6584\AA \xspace to the 6548\AA \xspace [\ion{N}{2}] line was held to $2.96$. This was achieved by recursively adjusting the amplitudes of the fitting of the [\ion{N}{2}] lines as the width of these peaks were held constant to those of the [\ion{S}{2}] doublet. The H$\alpha$ width was allowed to increase up to $25\%$ of the line width value.
    
        \item The broad component was then included with the model of the narrow lines and the entire complex was fitted once more. This new fitting is used if the broad component's FWHM is greater than $500\;\rm{km\;s^{-1}}$ and the reduced $\chi^2$ of this model is at least $20\%$ better than that of narrow line fitting.
    \end{enumerate}
    
    \item The H$\beta$ line was fit once more using a narrow line fit first, adjusting the width up to an additional $25\%$ of template widths. It was then fitted with the addition of the broad component.
    
    \item The H$\beta$, [\ion{O}{3}] and [\ion{O}{1}] lines were also fit using Gaussian models.
\end{enumerate}

\begin{figure*}
    \centering
    \includegraphics[width=\textwidth]{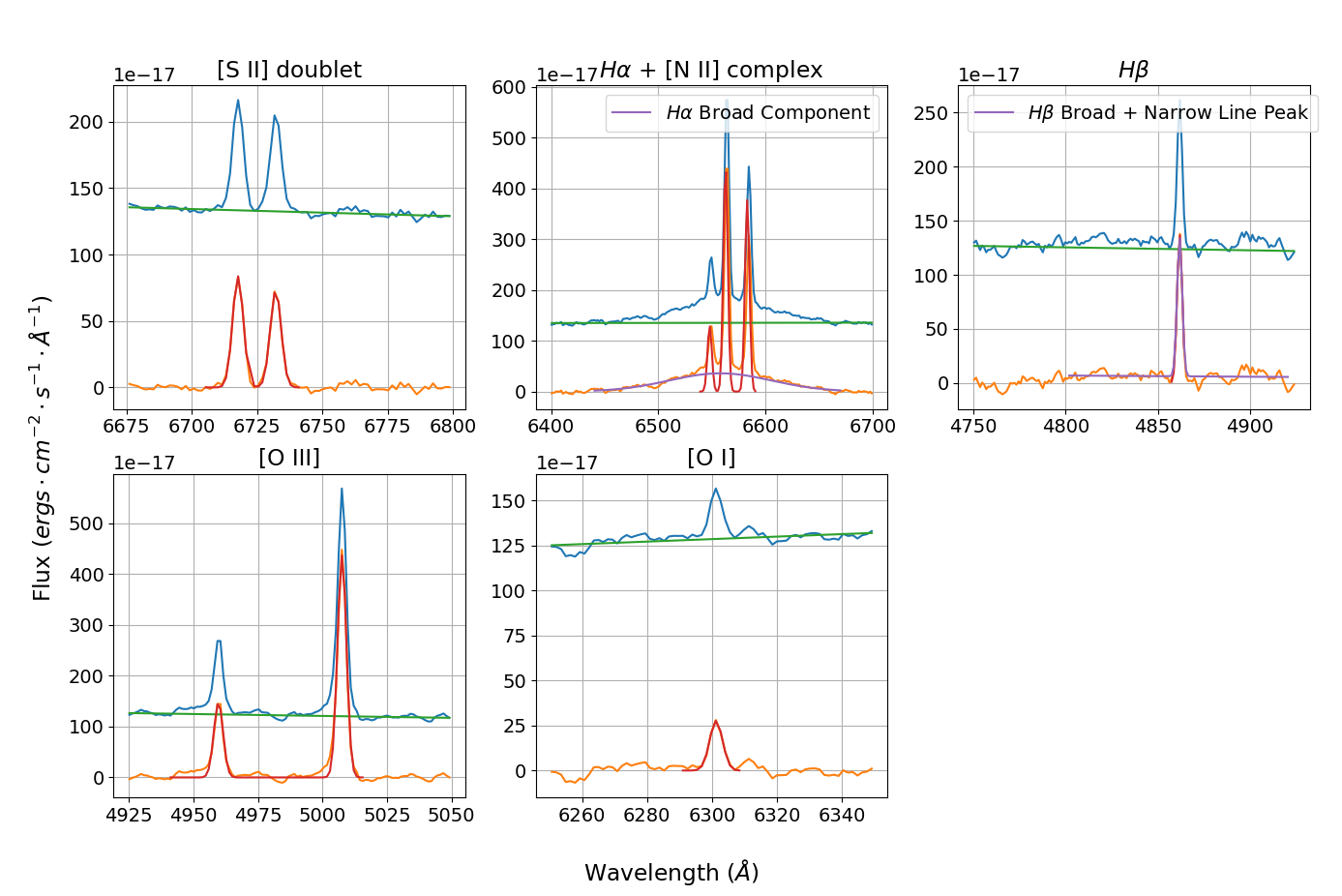}
    \caption{Emission line fits for NSA 64286. In each plot, the blue line represents the initial spectrum, green is the linear continuum fit, orange is the continuum subtracted spectrum, red is the Gaussian models for each line, and purple represents the broad component fits for H$\protect\alpha$ and H$\protect\beta$.}
    \label{fig:Emission_Fit_Ex}
\end{figure*}

An example spectrum is shown in Figure \ref{fig:Emission_Fit_Ex}. This methodology takes special care of the H$\alpha$ + [\ion{N}{2}] complex as the broad H$\alpha$ associated with gas in close proximity to the BH can become harder to detect at lower BH mass. As mentioned in \cite{Reines:2013pia}, the [\ion{S}{2}] doublet has previously been used to closely model the narrow lines of the H$\alpha$ + [\ion{N}{2}] complex \citep{Ho_1997, Greene_2004}. The H$\beta$, [\ion{O}{3}], and [\ion{O}{1}] fits are included to apply optical spectroscopic diagnostics as further discussed in Section \ref{sec:Results}.

We estimated BH masses using the luminosity ($L_{\text{H}\alpha}$) and FWHM of the broad H$\alpha$ component \citep{greene2005ApJ...630..122G, Reines:2013pia, Burke21_MNRAS}. Specifically, we used the results from \cite{Burke21_MNRAS}, where:

\begin{equation}
\label{eq:Mass_BH}
\begin{split}
    \log \left(\frac{\text{M}_{\text{BH}}}{\text{M}_\odot} \right) = 6.57  + 0.47\log \left( \frac{L_{\text{H}\alpha}}{10^{42}\,\text{erg s}^{-1}} \right)  \\
     +  2.06 \log \left( \frac{\text{FWHM}_{\text{H}\alpha}}{10^{3}\,\text{km s}^{-1}}\right) .
\end{split}
\end{equation}
Our BH mass estimates are presented in Table \ref{tab:results_table} with uncertainties collected from bootstrapping over 100 iterations of the H$\alpha$ broad component fitting.

We discuss the origins of the broad H$\alpha$ emission in Section \ref{subsec:BPT_analysis}.
%%%%
\subsection{Spectral Energy Distribution Modeling}
\label{subsec: Spectral_Energy_Distribution_Modeling} 
We then used our multiwavelength photometry to  model SEDs for each object in our sample. Using the AGN and galaxy templates of \cite{Assef_2010} and \cite{Kirkpatrick_2015}, which were constructed empirically based on the NOAO Deep-Wide Field Survey's Bo\"{o}tes field, the iterative method of non-negative linear combinations constructs the SEDs model over a 300\AA \xspace (0.03$\mu$m) to 300,000\AA \xspace (30.0$\mu$m) wavelength range. 

Templates for AGN, elliptical (ELL), star-forming (SFG), and irregular (IRR) galaxies are consider in the decomposition of the SEDs. As 15$\mu$m emission can mainly be linked to AGN contributions \citep{Lambrides_2020}, we used the fractional AGN contribution of the total SED at 15$\mu$m ($f_{\text{AGN}}$) as a threshold for selection. We follow \cite{Carroll_2021} in adopting $f_{AGN} \geq$ 0.7 as a quality cut for AGN identification. For objects with $f_{AGN} \geq$  0.7, we identified them as SED AGNs.  

%%%%%%%%%%%%%%%%%%%%%%%%%%%%%%%%%%%%%%%%%%%%%
\section{Results}
\label{sec:Results}

A record of our sample properties is presented is given in Table \ref{tab:results_table}. Sky position, stellar mass, and redshift are taken from the NSA catalog, with stellar masses converted from units of  $\text{M}_{\odot} h^{-2}$ to $\text{M}_{\odot}$. NUV magnitude, standard deviation and fractional variation were determined for each galaxy by the methods of Section \ref{subsec:Light_Curve_Const}. 
Below we present an analysis of optical spectroscopic properties, IR properties, and SEDs. We then compare the samples selected with each AGN selection technique.

\input{table1}
% % % %

\subsection{Optical Spectroscopic Properties}
\label{subsec:BPT_analysis}
\begin{figure*}
    \centering
    \includegraphics[width=\textwidth]{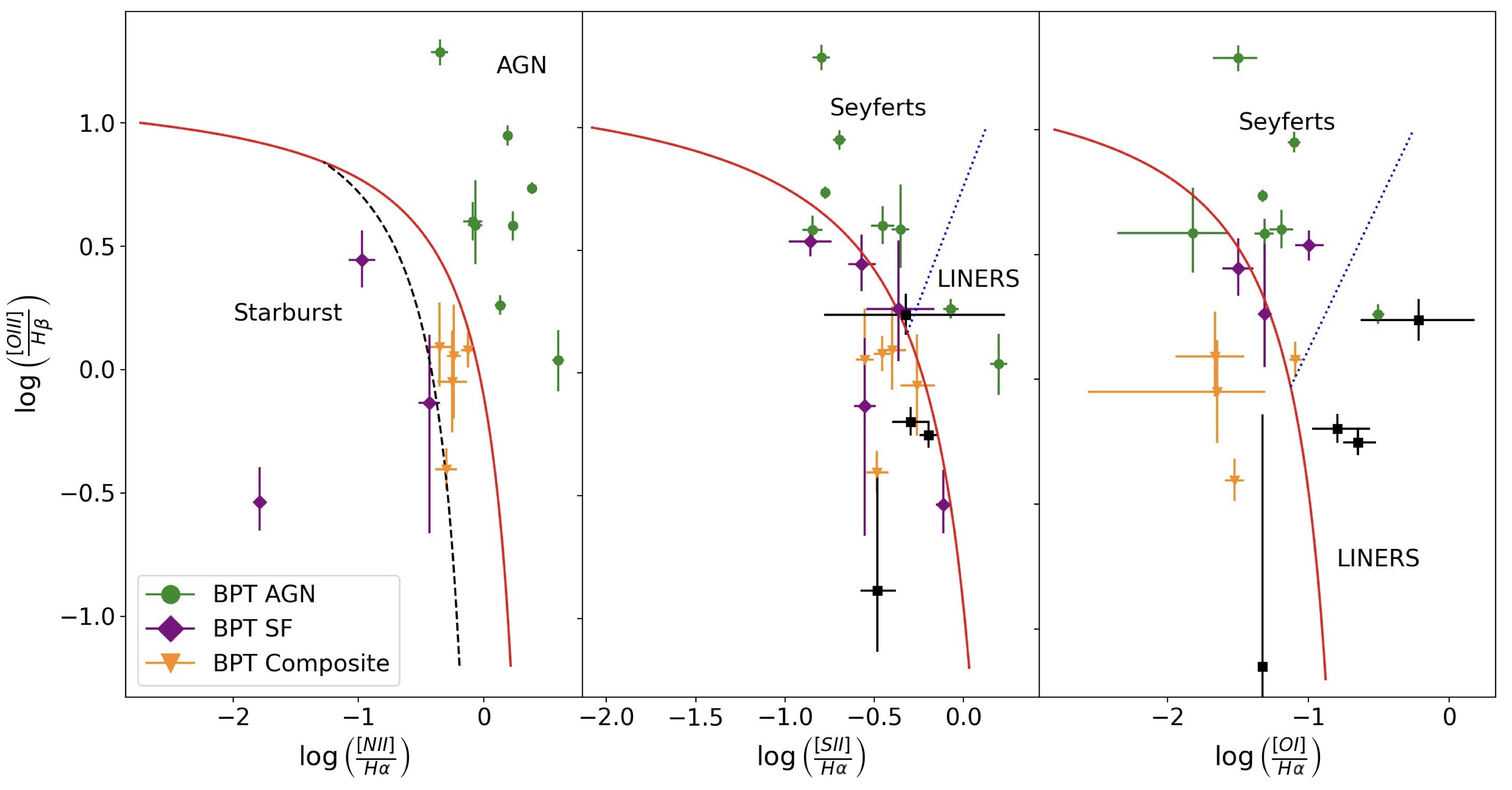}
    \caption{Optical diagnostic diagrams. Markers and colors set by classification of left subplot and carries through to other subplots. Black squares represent objects that could not be plotted on the first diagram due to weak [NII] emission, but could be plotted on other diagnostic diagrams. Two objects of the left panel in the star forming regime are not shown for adequate scaling of the axes. Table \ref{tab:class_table} shows classification of each galaxy.}
    \label{fig:BPT_diagram}
\end{figure*}

We use our emission line fluxes to plot objects on the BPT diagram (\citealt{1981_BPT_Paper}; see Figure \ref{fig:BPT_diagram}). Classification lines were taken from \cite{Kewley_2001_Starburst_Galaxies}, \cite{Kauffmann_03_Host_Gals_of_AGN}, and \cite{Classify_AGN_Kewley_et_al_06}.

The classification lines of \cite{Kewley_2001_Starburst_Galaxies} separate regions of AGNs from starburst galaxies. \cite{Kauffmann_03_Host_Gals_of_AGN} added an additional classification to the standard log([\ion{O}{3}]/\text{H}$\beta$) vs log([\ion{N}{2}]/\text{H}$\alpha$) BPT diagram to distinguish pure star formation objects from those that are composite HII-AGN types. This classification line sets the lower limit of composite galaxies such that objects located between this and the Kewley (Ke01) line of Figure \ref{fig:BPT_diagram} are defined to be composite galaxies. Lastly, \cite{Classify_AGN_Kewley_et_al_06} presented an additional categorization to separate Seyfert \citep{1990ApJ...359...86E, Dong_2012} from LINER (low-ionization nuclear emission-line region) galaxies. In separating these two types of galactic nuclei, \cite{Classify_AGN_Kewley_et_al_06} divides the AGN region of two of the optical diagrams, where Seyfert galaxies lie above the line and LINERs below. Table \ref{tab:class_table} provides our classifications for each spectroscopic diagnostic diagram. 

% Figure: BPT Diagram

In Figure \ref{fig:BPT_diagram}, it is shown that some points positioned near classification lines in the original BPT cross into new classifications with alternate optical spectroscopic diagnostics. This is especially apparent in a fraction of AGNs and starburst galaxies near the Ke01 line in comparing the [\ion{N}{2}] to [\ion{S}{2}] and [\ion{O}{1}] diagnostics. These cases will be discussed in more detail below. From the original BPT diagram (left panel of Figure~\ref{fig:BPT_diagram}), 8 galaxies were selected as AGNs and 5 as composite galaxies. 

All but one of the BPT AGNs are also classified as AGNs on the [\ion{S}{2}] and [\ion{O}{1}] diagrams. These sources also tend to have high AGN contributions to their SED models. The 5 BPT composite galaxies are all classified as Starburst galaxies in the [\ion{S}{2}] diagram. One composite galaxy--NSA 259478--is a Seyfert in the [\ion{O}{1}] diagnostic. None of the 5 BPT composite galaxies are cross-listed as SED AGN as all have extremely low or no AGN contribution to their SED models.

We also estimated BH masses for objects with broad H$\alpha$ emission. We found 12 of the variable galaxies to have a broad H$\alpha$ component (25\%), with BH masses ranging from 5$\times$10$^{\text{5}}$--2$\times$10$^{\text{8}}$ (see Table~\ref{tab:results_table}). We note that broad emission lines were observed in galaxies with stellar masses $>10^{10}\;\text{M}_{\odot}$. One possibility is that broad emission from lower-mass BHs is generally too faint to detect in SDSS observations, and could be detected in deeper, higher-resolution spectroscopy \citep{Baldassare_2015ApJ...809L..14B}. Alternatively, broad emission line production could be inefficient for BHs below $\sim10^{5}\;\text{M}_{\odot}$ \citep{Chakravorty_et_al_2014}.

%established introductory work to investigate if broad line emissions are in existence for the full range of BH masses. Their derived result was consistent with observation containing low-mass BHs, in that, once a they set a limit to the radius of the broad emission line region of the AGN in their model, the equivalent widths of the spectroscopic broad emission lines depended on BH mass. This relation then takes a steep drop-off as they approach lower BH mass, such that for masses below $10^5 M_{\odot}$, broad emissions are not found. They deduce that low-mass BHs have an inaptitude to produce observable broad emissions. The study of \cite{Baldassare2016ApJ...829...57B} showed that supernovae can be the main source of broad line emission in lower mass galaxies. These consideration will be remembered come the discussion of the BPT diagram diagnostics below.

\subsection{WISE Infrared Color Analysis}
\label{subsec: WISE_analysis}

\begin{figure}
    \centering
    \includegraphics[width=0.5\textwidth]{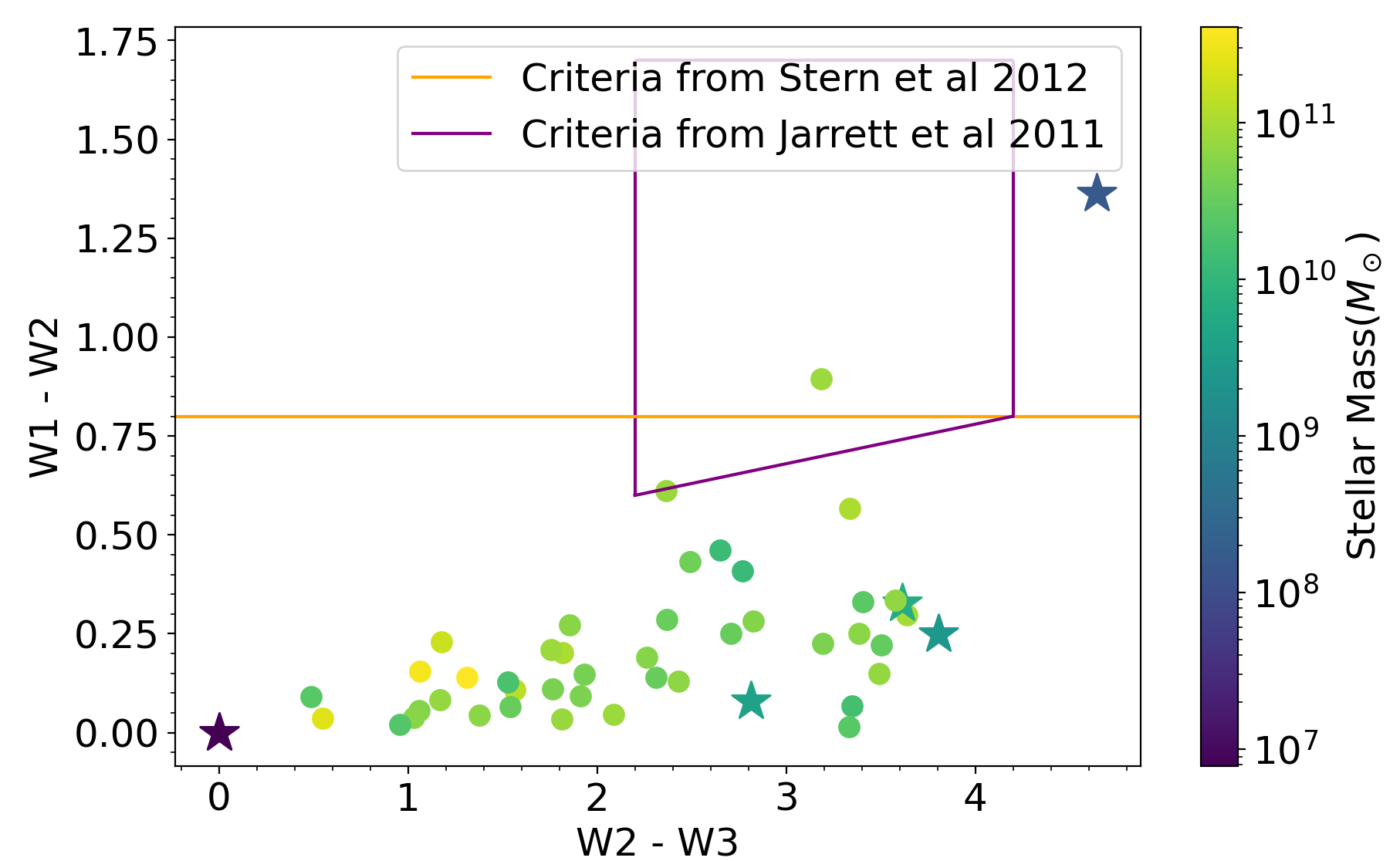}
    \caption{WISE color-color diagram. Maximum of 2 AGNs are selected by this method. Stars signify dwarf galaxies whose stellar masses are $\log M_{*} < 10.0 \text{M}_\odot$.  }
    \label{fig:WISE_diagram}
\end{figure}

We also explore the IR properties of this sample. The WISE AllWISE catalog \citep{WISE_Wright2010AJ....140.1868W} was used to collect the IR photometry in 3.4, 4.6, and 12$\mu$m (W1, W2, and W3 respectively). magnitudes for our sample. Photometry was available for 46/48 of our variable galaxies. 
Figure \ref{fig:WISE_diagram} shows a WISE color-color diagram for these 46 galaxies, where the color of each point is set by the stellar mass of the galaxy. AGN selection criteria are taken from \cite{Stern_2012} and \cite{Jarrett_2011}. 

Objects above the threshold of $W1-W2 >$ 0.80 met the \cite{Stern_2012} criterion for IR AGN. This value was set based on the high reliability and completeness of its WISE AGN selection in the COSMOS field. The criteria of \cite{Jarrett_2011} sets an interior box where within the selected IR AGN would lie. This box, originally defined in \cite{Lacy_2004ApJS..154..166L}, was set through evaluating the fraction of mid-IR AGN in the Spitzer Space Telescope First Look Survey. In examining Figure \ref{fig:WISE_diagram}, only two objects appear to picked as AGNs by the Stern criteria and one by the Jarrett criteria, as shown in the middle columns of Table \ref{tab:class_table}. 

NSA 28616 is selected by both IR criteria and NSA 631637 is only selected by the criteria of \cite{Jarrett_2011}. Interestingly, the vast majority of our highly variable sources are not selected by WISE color-color diagnostics. NSA 28616 is also selected as a BPT AGN based on log([\ion{O}{3}]/\text{H}$\beta$) vs log([\ion{N}{2}]/\text{H}$\alpha$). In addition, NSA 28616 has significant AGN contribution to its overall SED ($f_{\text{AGN}}$ = 0.90). NSA 631637 did not have SDSS spectroscopy. NSA 631637 is a dwarf galaxy, having a mass of $10^{8.19} \text{M}_{\odot}$ with a large AGN contribution ($f_{\text{AGN}}$ = 0.93) to its SED. Both WISE AGN are categorized as SED AGN (see Section \ref{subsec: SED_Analysis}.

\subsection{SED Analysis}
\label{subsec: SED_Analysis}

We constructed SEDs for each galaxy and modeled them following the methodology of \cite{Carroll_2021}. \cite{Carroll_2021} coadds these low-resolution templates using a non-negative linear least squares approach to construct SEDs. A simple $\chi ^2$ minimization is used to determine the normalization of each template needed to construct the full model for each object. In the last columns of Table \ref{tab:class_table}, percent AGN ($\% _{\text{AGN}}$) describes how many the percent of realizations that contained any AGN contribution during the fitting process, with $f_{\text{AGN}}$ giving the fraction contribution of the AGN template at $15\mu m$. We parameterize nuclear obscuration with color excess ($E(B-V)_{\text{AGN}}$) applied only to the AGN component. The latter two columns were then used to identify objects as SED AGN based on our selection criteria, as well as classifying obscured AGNs. We defined obscured AGNs as sources with $E(B-V)$, value greater than 0.15. Signatures of nuclear obscuration are visible in the SED decomposition as attenuation of the AGN component at shorter wavelengths as wavelength decreases.

We cross-match the spectroscopic and IR analysis with the AGN fractions. Of our 48 variable galaxies, twelve objects' do not require any AGN contribution. NSA 64266, 259880, and 649438 present interesting cases: they have $0.0$ for both their fractional AGN contribution and excess color values, but significant uncertainties on this $E(B-V)$ number. This demonstrates that small changes in their photometry cause large variations in their overall SED. We therefore lack the confidence to further consider the SEDs of these sources.

\begin{figure}
    %\centering
    %\plotone{\includegraphics[scale=0.45\textwidth]{SED_confident_AGN.jpeg}}%\quad
    \plotone{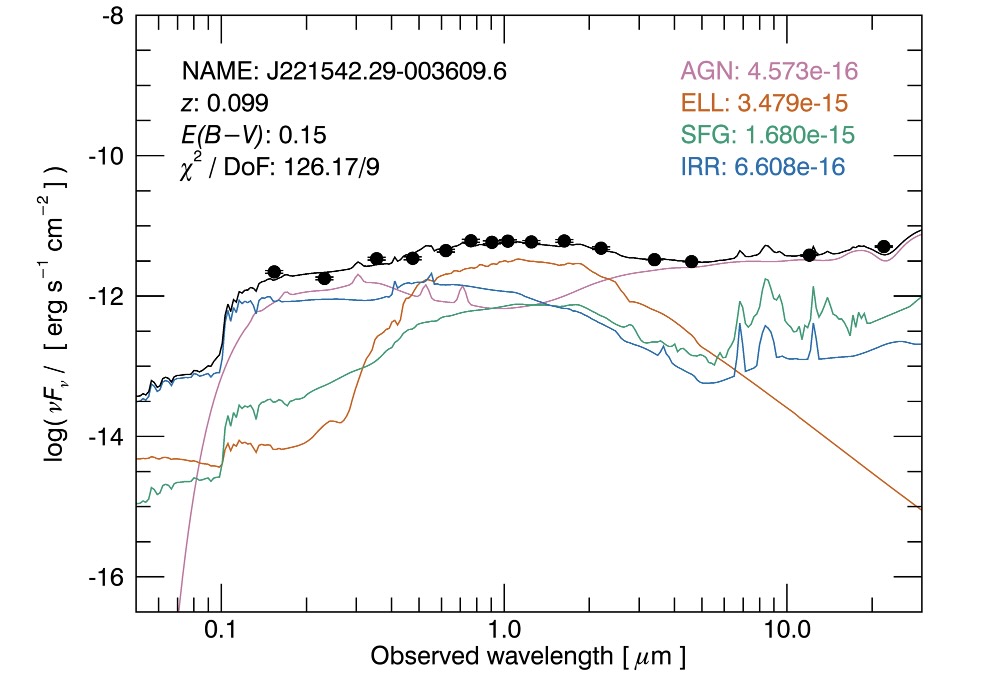}
    \plotone{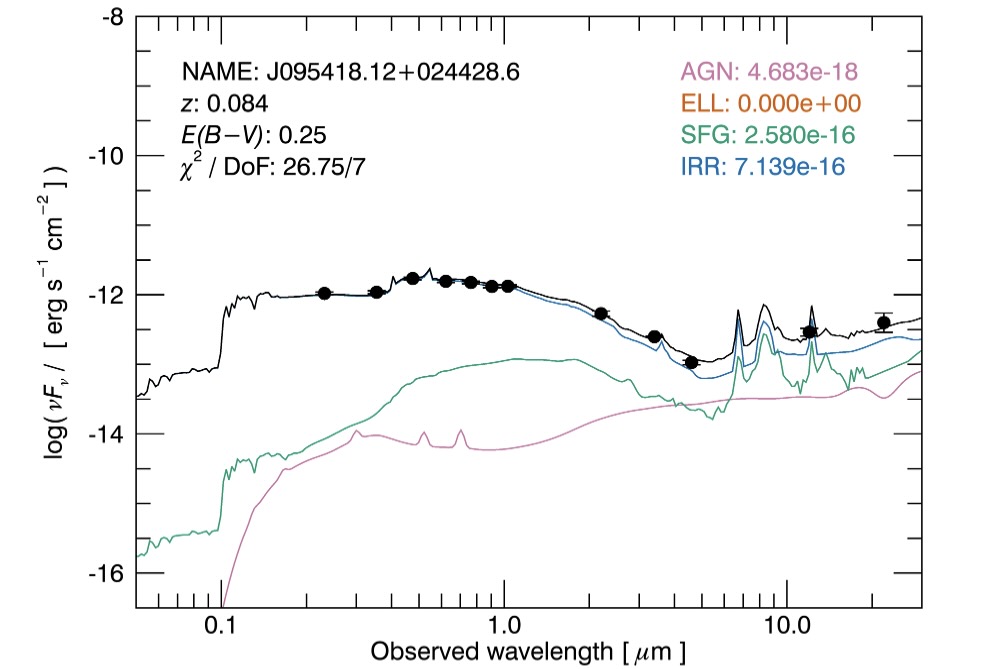}
    \plotone{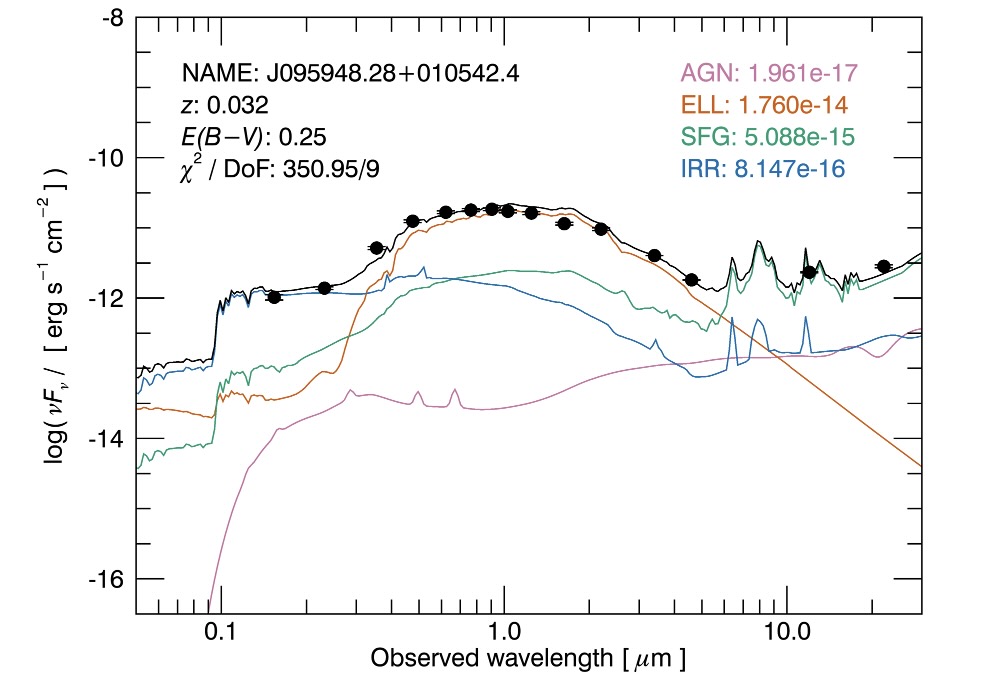}
    %\hspace{1em}% Space between image A and B
    %\plotone{\includegraphics[scale=0.45\textwidth]{SED_Dwarf.jpeg}}
    %\plotone{\includegraphics[scale=0.45\textwidth]{SED_composite.jpeg}}
    \caption{\emph{Top:} NSA 28616. This object is a BPT AGN that is categorized as a Syfert through the other spectroscopic diagnostics. \emph{Middle:} NSA 58917. This is a dwarf galaxy with a mass of $10^{9.39} \text{M}_{\odot}$. This model gives evidence for star-formation diluting AGN emission \emph{Bottom:} NSA 64234. This BPT Composite galaxy is in the starburst regime of the other spectroscopic diagnostics. Signatures of star-formation and AGN obscuration are present in this model.  }
    \label{fig:SEDs}
\end{figure}

Examples of SED models are given in Figure \ref{fig:SEDs}. These three models give a representation of sub-populations developed by this analysis. The top panel shows the SED of NSA 28616. With a high AGN fraction ($f_{\text{AGN}}$ = 0.90) and narrow lines placing it in the AGN region of the BPT diagram, this is a secure UV-variable AGN. The SED model of NSA 58917 has AGN fraction of 17.3$\%$ and only 3.1$\%$ of iterations contains an AGN template. This dwarf galaxy's SED contains high contribution of star-forming emission from the SFG and IRR templates. A BPT Composite galaxy SED is given in the final panel for NSA 64234. All three of these examples are also obscured AGNs with $E(B-V)>0.15$. 

\begin{figure}
    %\centering
    \includegraphics[scale=0.6]{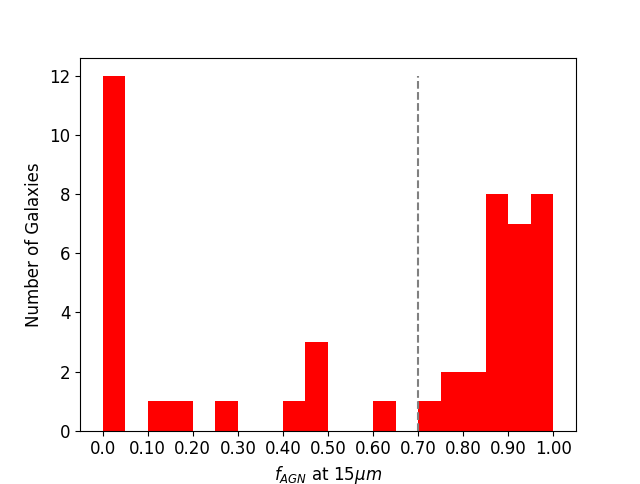}
    %\plotone{histo_AGN_fract.png}
    \caption{Histogram of AGN fraction at $15\mu m$ for our variable galaxy population.  }
    \label{fig:AGN_fract_Histogram}
\end{figure}

Using a cut of $f_{\text{AGN}} \geq$ 0.70, we identified 28 objects as SED AGN (see Figure \ref{fig:AGN_fract_Histogram}). Six of our 8 BPT AGNs are classified as SED AGNs. Four of the SED AGNs were classified as Starburst galaxies on the standard BPT diagram, with two of these objects having conflicting classifications. NSA 63970 falls in the Seyfert regime of [\ion{S}{2}]/$\text{H}\alpha$ vs [\ion{O}{3}]/$\text{H}\beta$ and NSA 583470 lies in the Seyfert regime of the [\ion{O}{1}]/$\text{H}\alpha$ vs [\ion{O}{3}]/$\text{H}\beta$ plot. These are interesting case studies; further study of host galaxy properties may yield information on their contrasting optical spectroscopic classifications. The rest of the objects in this subsample either lacked emission line spectra or did not have available SDSS spectroscopy. 

Of the 28 SED AGNs, 11 were identified as highly obscured, with $E(B-V)_{\text{AGN}} \geq$ 0.15 (see Table \ref{tab:results_table}). Of these 11 obscured AGNs, two have also been identified as dwarf galaxies. NSA 28616 and 64129 were identified as BPT AGNs as well, each with $E_{B-V} = 0.15$.

Of the 8 BPT AGNs (NSA 28616, 64059, 64129, 64272, 64286, 83201, 83204, and 259895), all but 64129 and 83204 had AGN contribution in every iteration of the model fitting (ie $\% _{AGN} =100 $). Only NSA 64059 and 64286 did not have a significant AGN fraction to their SEDs ($f_{\text{AGN}} <$ 0.70). We also note that NSA 64059 and 83204 were categorized as starburst galaxies in the [\ion{O}{1}]/$\text{H}\alpha$ vs [\ion{O}{3}]/$\text{H}\beta$ plot of Figure \ref{fig:BPT_diagram}.

%Figure \ref{fig:Venn Diagrams} shows a diagram conveying the cross categorization of the AGN selected through each method and the low-mass galaxies' place within these groups. 
Figure 11 depicts how our sample of 48 variable AGNs are cross-categorized using the different selection criteria. The bottom panel of Figure 11 focuses on the six select variable dwarf galaxies and their categorizations from our SED and IR color selection methods.
\input{table2}

%%%%%%%%%%%%%%%%%%%%%%%%%%%%%%%%%%%%%%%%%%%%%
\section{Discussion}
\label{sec:Discussion}

From the parent sampe of 1819 NSA galaxies, we identify 48 high-probability candidate AGNs based on their UV variability ($2.6\%$). Of these sources, 6 ($0.33\%$ of the parent population) are variable low-mass galaxies with masses $\leq 10^{10} \text{M}_{\odot}$. Of this variable population of 48 AGN candidates, only 8 were identified as AGNs by optical spectroscopic diagnostics alone. Only 2 objects were selected as AGNs based on their IR colors. SED modeling provided a more detailed look at the possible components of each variable object. Using the fractional AGN SED contribution at 15$\mu$m, a conservative cut of $f_{\text{AGN}} \geq 0.70$ identifies 28 objects. Twelve objects have no AGN contribution to their SEDs at all and 8 sources have some AGN contribution, but below the $f_{\text{AGN}}=0.70$ threshold. This selection criteria provided our initial threshold for AGN identification. 

\subsection{Importance of Multiwavelength Photometry}
\label{subsec: Importance_Multi_Wavelength} 
Optical, IR, and X-ray surveys identify AGNs with different BH and host galaxy properties. Each spectrum and selection method has its own benefits and selection biases. This emphasizes the important of comparing selection techniques across wavelength regimes to distinguish the processes at play within each galaxy.

Selection and categorization of AGNs through their optical emission uses particular spectroscopic signatures to identify candidates. Searching for AGNs based on photoionized emission line ratios and broad $H\alpha$ emission are directly tied to the quality of the observed spectrum for each object. As one of the most prominent optical signatures of AGN activity, broad $H\alpha$ becomes harder to detect for low-luminosity and low-mass galaxies \citep{Greene_2004, Reines:2013pia}. Host galaxy properties also determine whether narrow line signatures are detectable. This is particularly relevant for low-mass galaxies (see Section \ref{subsec: Dwarf Galaxies}).  

In contrast, IR color-color diagnostics are sensitive to the obscuring material surrounding the AGN. As this wavelength regime can be dominated by dust heated by star-formation within the galaxy, IR selection techniques are heavily influenced by both processes. 

The decomposition an SED into component parts allows us to estimate the contribution of different physical processes to the total energy budget of the galaxy. BH accretion is taken into account in the AGN template, while ongoing star-formation and starbusts are represented in the SFG and IRR templates, and finally with older stellar populations characterized in the ELL template. 

\begin{figure}
    \centering
    \subfigure{\includegraphics[scale=0.40]{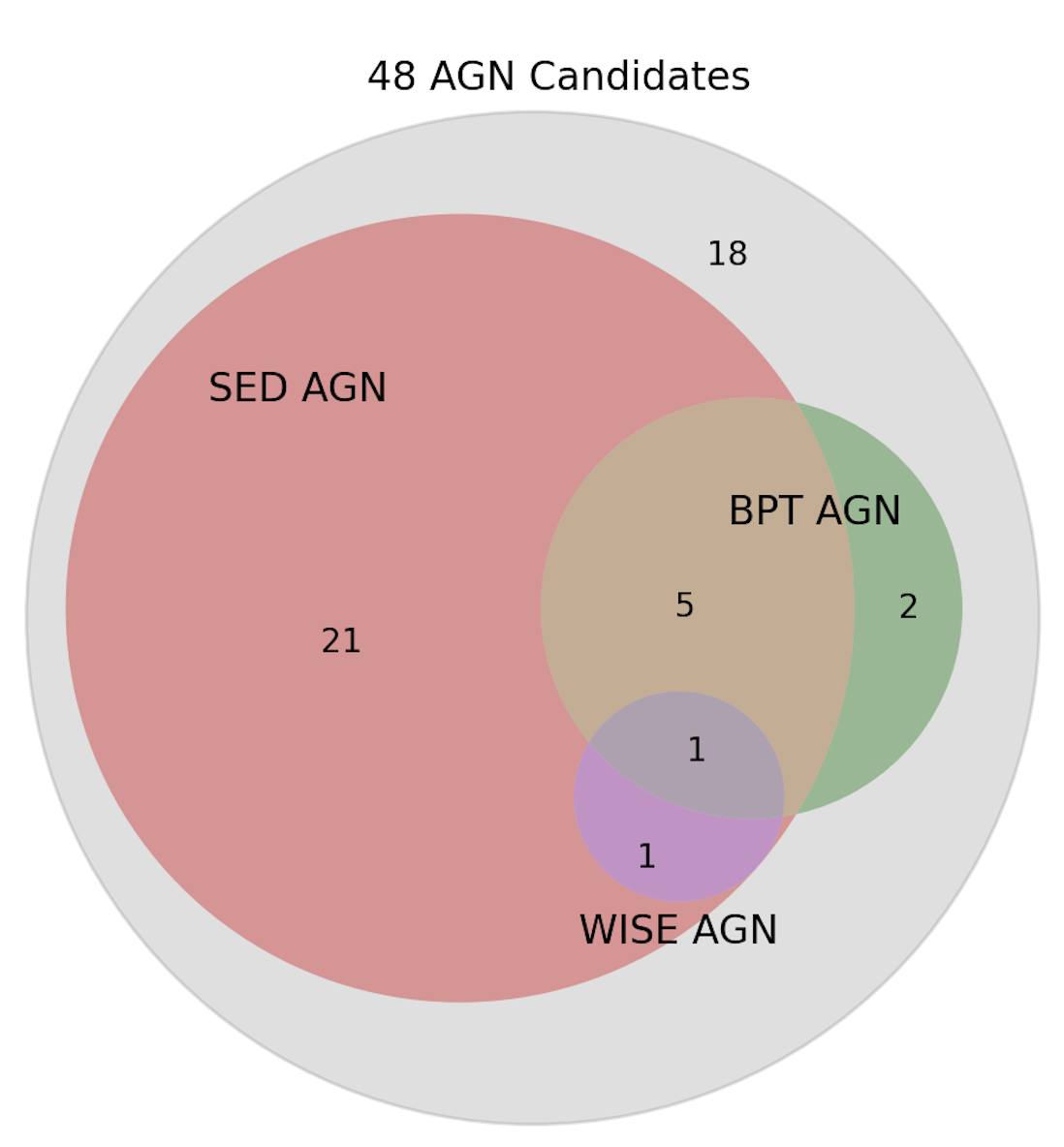}}%\quad
    \hspace{1em}% Space between image A and B
    \subfigure{\includegraphics[scale=0.34]{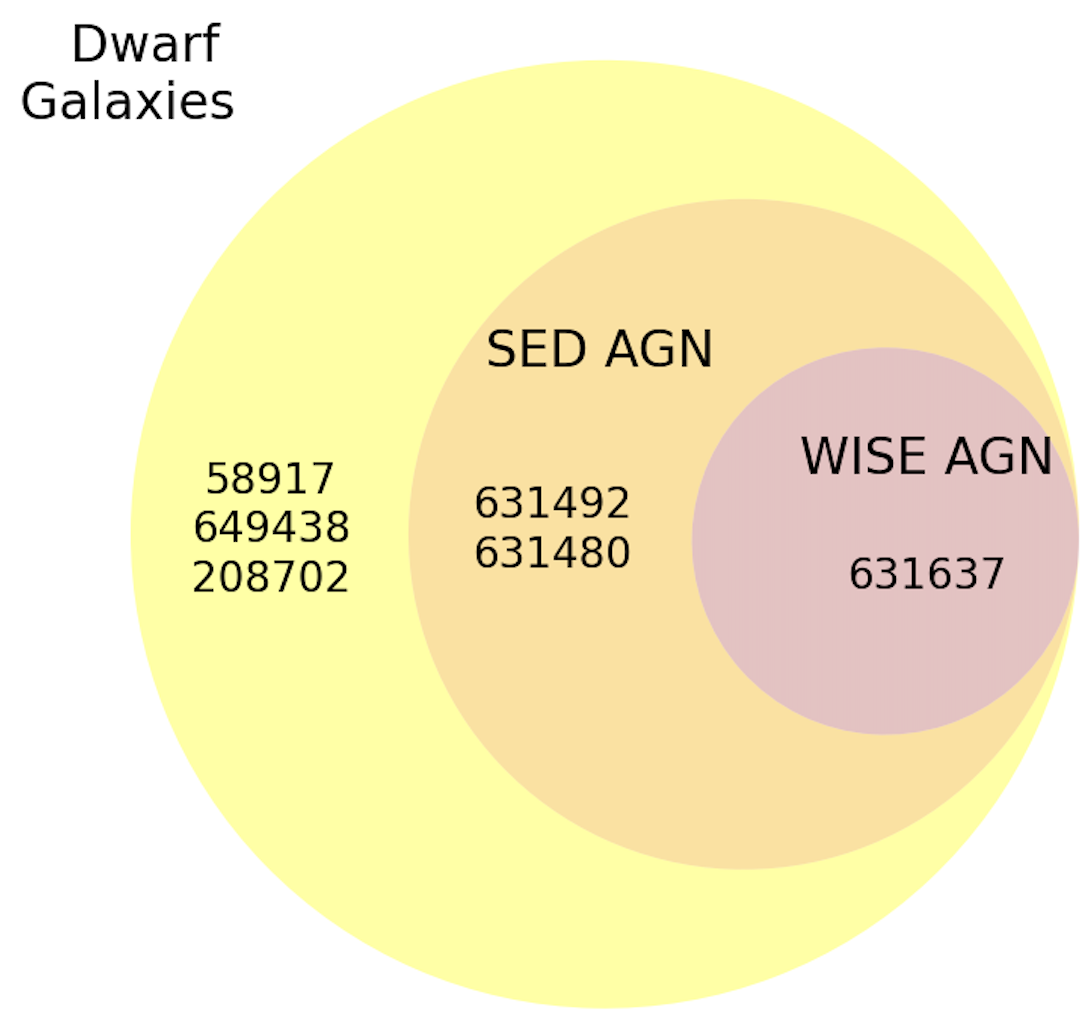}}
    \caption{\emph{Top:} Venn diagram of the AGN candidates cross-categorizations for the full population of variability selected candidates. A total of 28 of the 48 candidates met the SED AGN cut(given in the salmon circle). 8 candidates were identified as BPT AGNs(as shown in green), with 6 of these galaxies also selected as SED AGNs(shown in brown). Within the SED AGN population, the WISE cuts identified 2 candidates(given in pink), one of whom is cross categorized amongst all three of these methods(as shown in dark grey). 18 of the 48 AGN candidates had no cross categorization with these other methods (as shown in light grey). }\emph{Bottom:} Venn diagram of how the low-mass galaxy population is classified across different techniques. Note that none of these objects were selected as BPT AGNs. The six dwarf galaxies select have their NSA ID numbers listed in the sub population they fit in. The outer yellow circle represents dwarfs galaxies selected with variability, the orange SED AGN ring lists those dwarf galaxies that also made the SED AGN cut, and the pink WISE AGN circle represents the dwarf galaxy that was also identified as an SED AGN and a WISE AGN. 
    \label{fig:Venn Diagrams}
\end{figure}

Using a multiwavelength approach, we have compiled diagnostic techniques from the UV to IR, viewing AGN activity from its nuclear accretion and obscuring torus. Demonstrated by Figure \ref{fig:Venn Diagrams}, as some candidates would only be characterized as AGNs by one technique. Here we have demonstrated the importance of a multiwavelength approach when investigating active galaxies. Follow up studies will incorporate X-ray data to focus on the structure of the accretion disks (e.g., \citealt{Greene_X-ray_IMBH_2007ApJ...656...84G, Desriches_X-Ray_IMBH_2009ApJ...698.1515D}).

\subsection{Low-mass galaxies}
\label{subsec: Dwarf Galaxies}
Six galaxies of the selected variable population have stellar masses less than $10^{10} \text{M}_\odot$(NSA 58917, 208702, 631480, 631492, 631637, and 649438). The AGN fraction at $15\mu m$ splits this group into either virtually none ($f_{\text{AGN}}<20\%$, NSA 58917, 208702, and 649438) or significant ($f_{\text{AGN}}>90\%$, NSA 631480, 631492, 631637) AGN contribution. Only NSA 589117 and 208702 have emission line spectra for testing optical spectroscopic diagnostics: NSA 589117 was categorized as starburst and NSA 208702 is categorized as a composite galaxy in the standard BPT diagram. The lack of SDSS spectra for the remaining four low-mass galaxies and the results for NSA 589117 and 208702 demonstrate that spectroscopic selection would have missed these sources, whereas our use of UV variability included them in our population. 

$12.5\%$ of our candidates are dwarf galaxies. $30.68\%$ of the parent sample of 1819 sources are dwarf galaxies, of which we selected $1.1\%$ as active dwarf candidates. \cite{Reines:2013pia} uses an earlier version of the NSA, \texttt{v0\_1\_2,} to initially collect 44,594 galaxies with stellar masses $M_{*}<3 * 10^9 M_\Sun$, which represents $30.72\%$ of he entire catalog. From this, their optical selection identifies $~0.5\%$ as active candidates.

\cite{Cann_2019} discussed the  pitfalls of the BPT diagram classifications, particularly in confirming AGNs who host IMBHs. Through their models, they conclude that these standard optical diagnostics do not adequately apply to BHs with masses lower than $\sim10^4\;\text{M}_{\odot}$. For higher mass IMBHs, further theoretical investigation would ideally determine the uncertainties within these diagnostics. Optically, lower-mass AGNs have weaker line emission ratios than their more massive counterparts. 

It has been shown that AGNs in dwarf galaxies are not reliably selected using the WISE color diagnostics. \cite{Hainline_2016} found that a majority dwarf galaxies (which have masses in range of $10^7 - 10^9 \text{M}_{\odot}$) in the AGN and composite regions of the BPT diagram were below the AGN selection criteria on the WISE diagram. IR color selection for dwarf galaxies is limited as the host galaxy can are that the host galaxy can dominate the IR over the AGN and that high star-formation rates can heat dust enough to produce $W1-W2$  values within the AGN range \citep{Hainline_2016}. With these biases against low-mass galaxies, WISE color selection is not an individually reliable diagnostic to identify AGNs.

\subsection{SDSS Catalog Matching}
We crossmatched our 48 variable galaxies with the SDSS DR7 broad-line AGN catalog \citep{Liu2019ApJS..243...21L}. We find that $7$ of the $48$ variable galaxies are listed in that catalog. Six of these (NSA $64129, 64145, 64272, 64286, 83201, 28616$) are selected as BPT AGNs and one (NSA 65145) is not. 

Additionally, cross-matching with the SDSS Quasar Catalog DR16Q \citep{SDSS_Quasar_Catalog_DR16__2020ApJS..250....8L} reveals that only NSA $583470$ is listed in this catalog. This object was placed in the Starburst region of the BPT diagram but meets our criteria for SED AGN without obscuration. Using [\ion{O}{1}]/$\text{H}\alpha$ vs [\ion{O}{3}]/$\text{H}\beta$, this object was categorized as a Seyfert AGN. 
%%%%%%%%%%%%%%%%%%%%%%%%%%%%%%%%%%%%%%%%%%%%%
\section{Conclusions}
\label{sec:Conclusion}
In this study, we use the GALEX TDS NUV observations to search for AGNs via UV variability. We then compared our results to other AGNs selection techniques. Below, we summarize our main findings. 
\begin{itemize}
    \item We identify 48 high-probability AGN candidates from a parent sample of 1819 nearby galaxies in the NASA-Sloan Atlas. 
    \item Using optical spectroscopic diagnostics, $8$ of the 48 UV-variable galaxies are categorized as AGNs. Of the 6 low-mass galaxies in the variable population, only one has optical spectroscopic evidence for AGN activity.
    \item Only 2 of the UV-variable galaxies were selected as AGNs using the WISE IR color cuts of \cite{Jarrett_2011} and \cite{Stern_2012}. 
    \item Based on SED modeling, we find that 28 of our galaxies have a high ($f_{\text{AGN}}\geq 0.70$) AGN fractional contribution to their SEDs at 15$\mu$m. This group includes 11 obscured AGNs with color excess $E(B-V)$ $\geq 0.15$ and three dwarf galaxies.  
\end{itemize}
Through the UV-variability selection presented in this work, we have confidently identified a sample of elusive AGN candidates. Many would be missed by optical spectroscopic diagnostics or IR selection alone. In particular, AGNs in dwarf galaxies, as well as obscured AGNs, would have been missed through spectroscopic and WISE IR selection. If WISE IR color was used as the only selection technique, we would have only identified $\sim 4\%$ of the candidates selected through variability. Spectroscopic selection would have found $\sim 17\%$ of our candidates. This result further endorses variability as a selection method to detect otherwise elusive AGNs. 

Follow-up studies in the X-ray can help confirm the AGN nature of these objects. In future work, we plan to study this sample with Chandra X-ray Observatory data, searching for bright, broadband X-ray emission indicative of AGN activity. Additionally we will measure the ratio of X-ray--to--UV luminosity ($\alpha_{\rm OX}$). This value will be used in the relation between $\alpha_{\rm OX}$ and luminosity at $2500 \AA$. This relationship holds well for quasars, and deviations from this relationship can indicate changes in absorption, temperature, and accretion rate. Successful X-ray follow-up will allow us to investigate additional properties of AGNs, such as their disk structure and nature of the obscuring material.

Current X-ray observatories are capable of detecting the expected sources in this sample, especially for the higher-mass galaxies. For low-mass galaxies, high angular resolution X-ray observations are particularly important for pinpointing the position of the source within the galaxy and distinguishing between emission from an AGN and X-ray binaries. At the distance of our sample, Chandra observations of a few tens of kiloseconds are sufficient to detect sources down to $L_{\rm 0.5-7 keV}=10^{40} \rm{erg\;s^{-1}}$, at which point there is increased risk of contamination from X-ray binaries. Early search of the Chandra Archival Database finds observations for six of our candidates; these will be the subject of a forthcoming paper. We note that highly obscured candidates may lack X-ray emission in Chandra, XMM-Newton, and NuSTAR fields \citep{Lambrides_2020,Carroll_2021}. For our highly obscured sources, investigation of their obscuration in the X-ray may lead to new and interesting results. 

Radio emission would provide a complementary confirmation of the presence of an AGN and add additional constraints to SED studies. However, only a fraction of AGN are radio loud, and radio emission from low-mass AGN is expected to be relatively faint. There may also be confusion between star-formation related radio emission and AGN activity.

Finally, deeper, higher-resolution optical spectroscopy would aid in the search for faint, broad H$\alpha$ emission and measurements of stellar velocity dispersions to estimate BH masses. 
%%%%%%%%%%%%%%%%%%%%%%%%%%%%%%%%%%%%%%%%%%%%%

\begin{acknowledgments}

We thank Alexander Messick for helpful comments. We also thank the anonymous referee for their insightful thoughts and suggestions which have improved this manuscript. 

This work is based on observations made with the NASA Galaxy Evolution Explorer.
GALEX is operated for NASA by the California Institute of Technology under NASA contract NAS5-98034.

Funding for the SDSS and SDSS-II has been provided by the Alfred P. Sloan Foundation, the Participating Institutions, the National Science Foundation, the U.S. Department of Energy, the National Aeronautics and Space Administration, the Japanese Monbukagakusho, the Max Planck Society, and the Higher Education Funding Council for England. The SDSS Web Site is http://www.sdss.org/.

The SDSS is managed by the Astrophysical Research Consortium for the Participating Institutions. The Participating Institutions are the American Museum of Natural History, Astrophysical Institute Potsdam, University of Basel, University of Cambridge, Case Western Reserve University, University of Chicago, Drexel University, Fermilab, the Institute for Advanced Study, the Japan Participation Group, Johns Hopkins University, the Joint Institute for Nuclear Astrophysics, the Kavli Institute for Particle Astrophysics and Cosmology, the Korean Scientist Group, the Chinese Academy of Sciences (LAMOST), Los Alamos National Laboratory, the Max-Planck-Institute for Astronomy (MPIA), the Max-Planck-Institute for Astrophysics (MPA), New Mexico State University, Ohio State University, University of Pittsburgh, University of Portsmouth, Princeton University, the United States Naval Observatory, and the University of Washington.
\end{acknowledgments}

\vspace{5mm}
\facilities{GALEX \cite{inital_GALEX_paper}; SDSS \cite{SDSS_spect_2013AJ....146...32S}; UKIRT \cite{UKIRT_Lawrence2007MNRAS.379.1599L}; unWISE \cite{unWISE_Lang2014AJ....147..108L}; WISE \cite{WISE_Wright2010AJ....140.1868W}; 2MASS \cite{2MASS_2006AJ....131.1163S}  }

\software{\cite{astropy_2018AJ....156..123T}}

\bibliography{GALEX_AGN_REFS}{}
\bibliographystyle{aasjournal}

\appendix
Below are shown the light curves of all $48$ high standard deviation galaxies of the population selected by the way of Section \ref{subsubsec: galaxy_selection}. Error bars are set by the equation of the fit line of Figure \ref{fig:star_std_template}.

\begin{figure*}[hbt!]
    \centering
    \subfigure{\includegraphics[scale = 0.45]{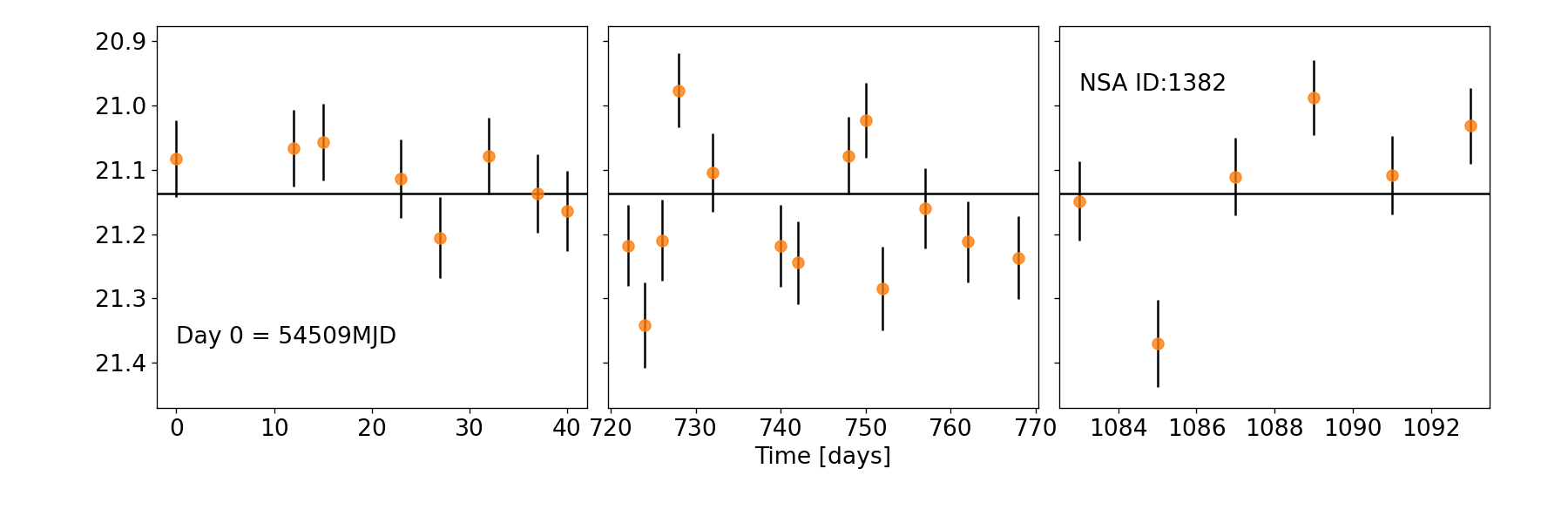} }
    \hspace{0em}% Space between images
    \subfigure{\includegraphics[scale = 0.45]{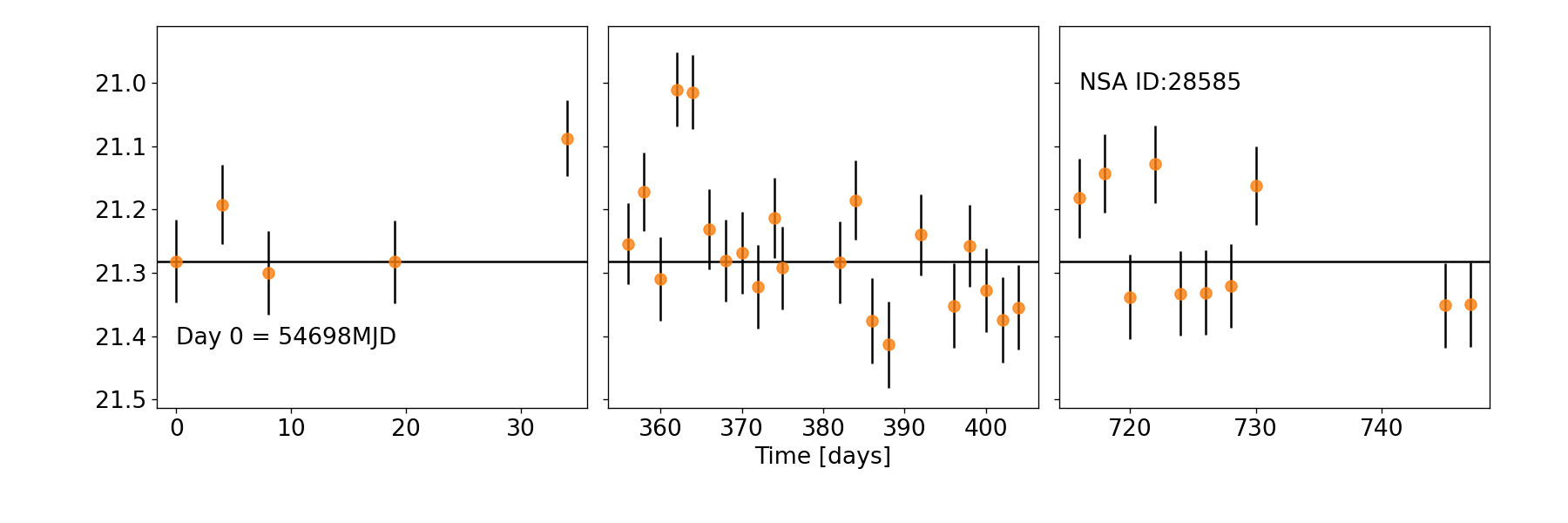} }
    \subfigure{\includegraphics[scale = 0.45]{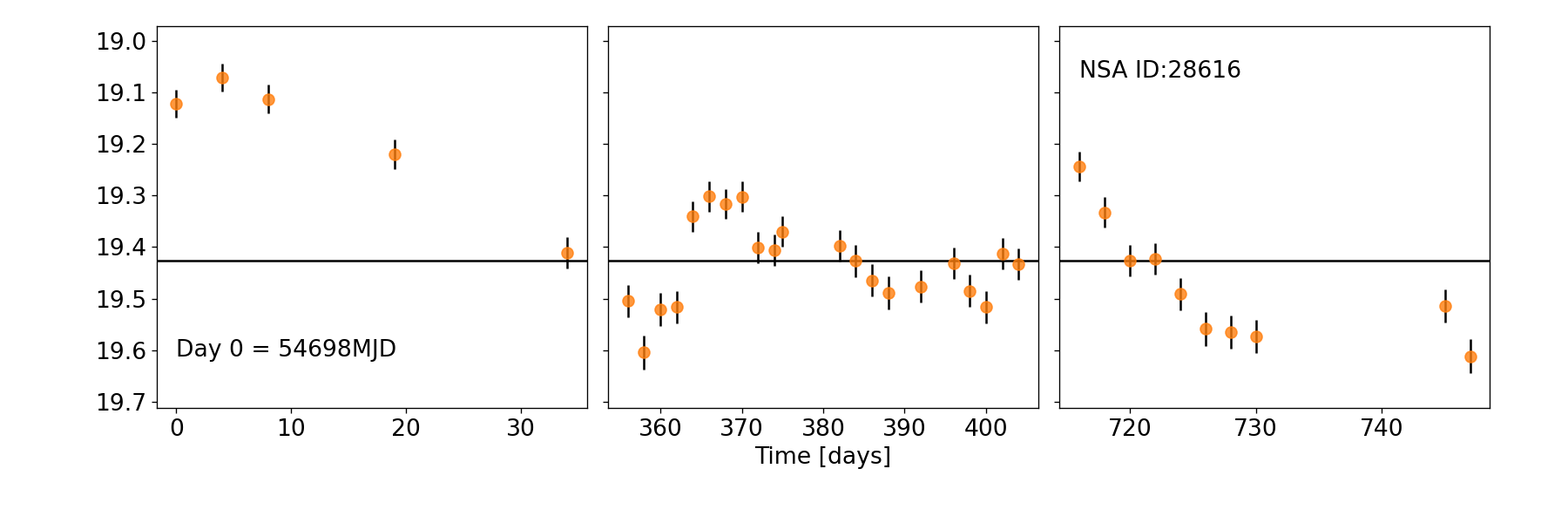} }
    \label{fig:Light_curves_set1}
\end{figure*}

\begin{figure*}
    \centering
    \subfigure{\includegraphics[scale = 0.45]{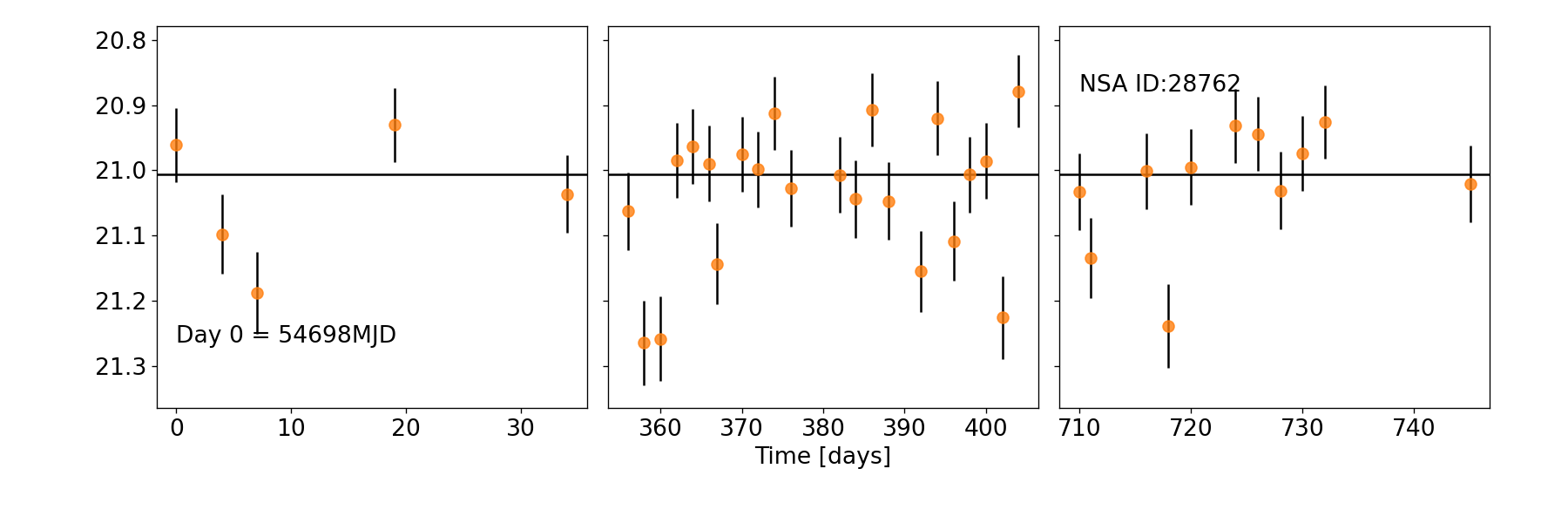} }
    \subfigure{\includegraphics[scale = 0.45]{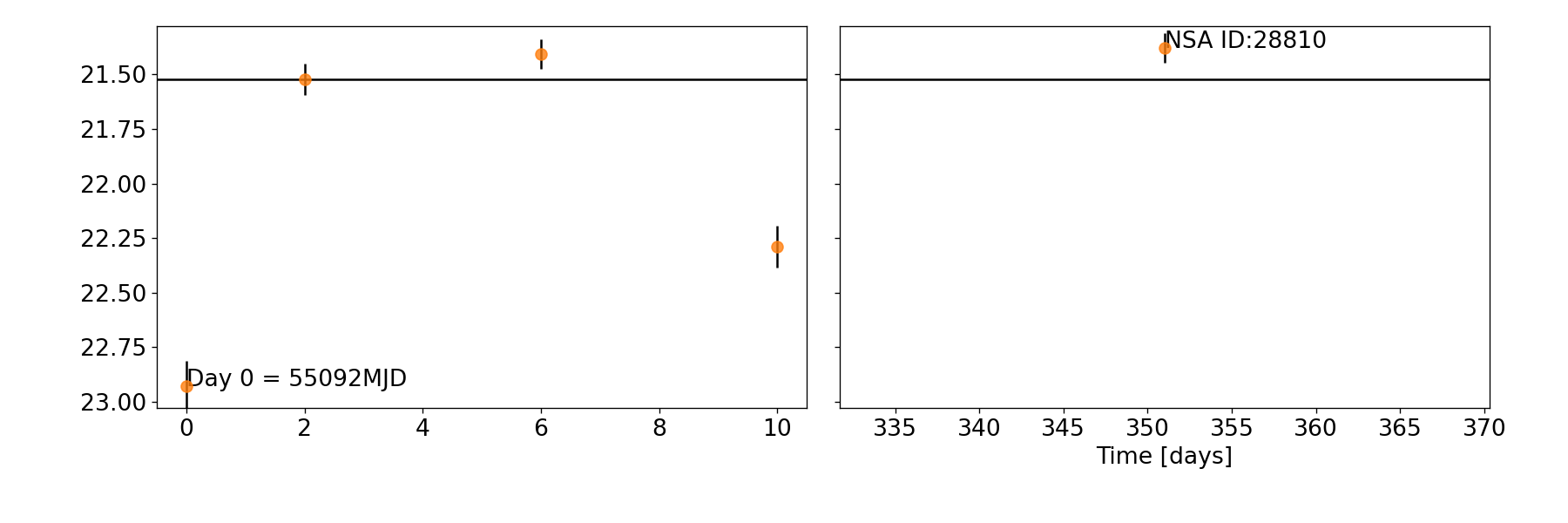} }
    \subfigure{\includegraphics[scale = 0.45]{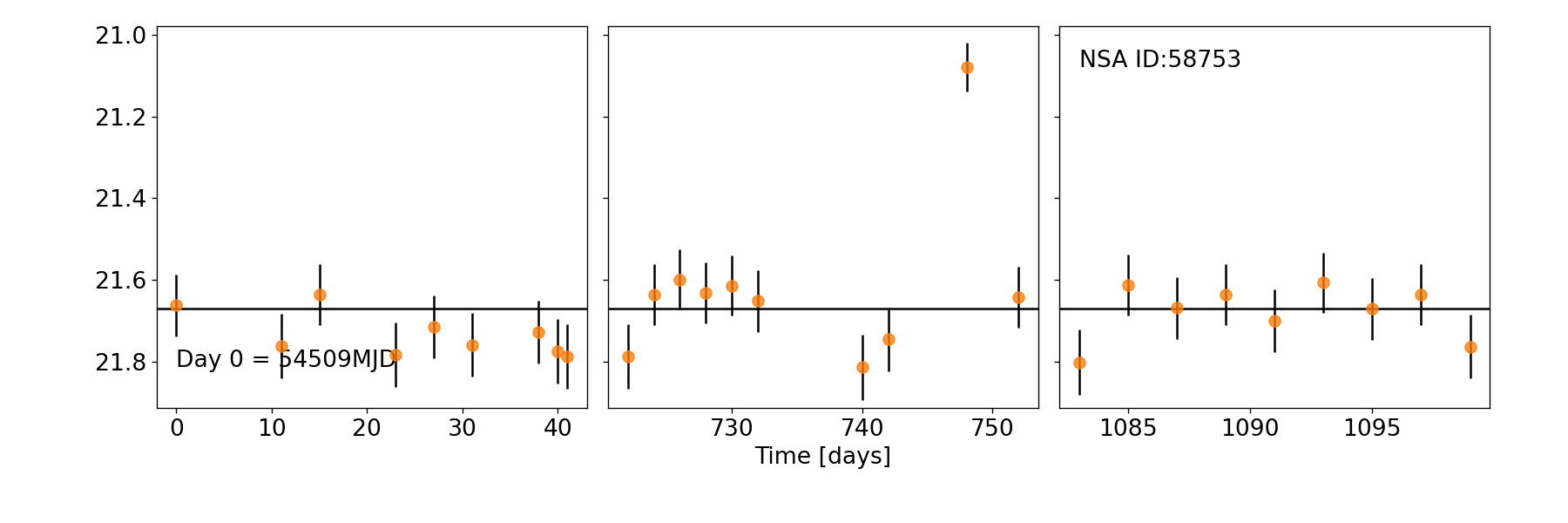} }
    \subfigure{\includegraphics[scale = 0.45]{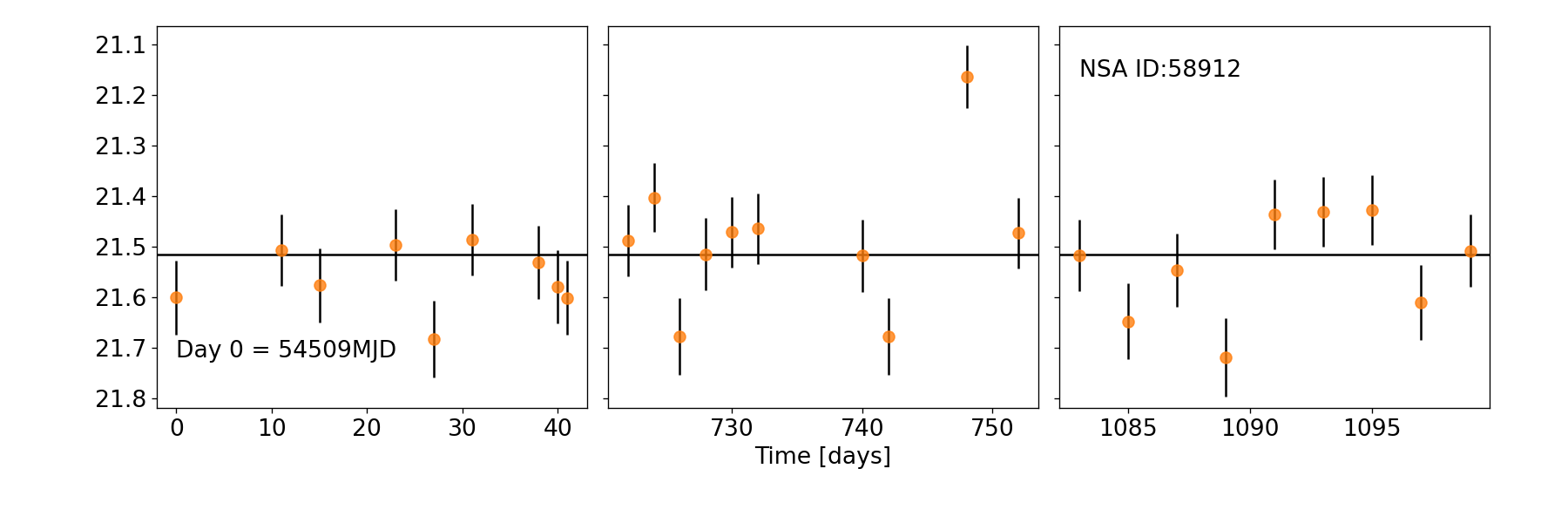} }
    \label{fig:Light_curves_set2}
\end{figure*}

\begin{figure*}
    \centering
    \subfigure{\includegraphics[scale = 0.45]{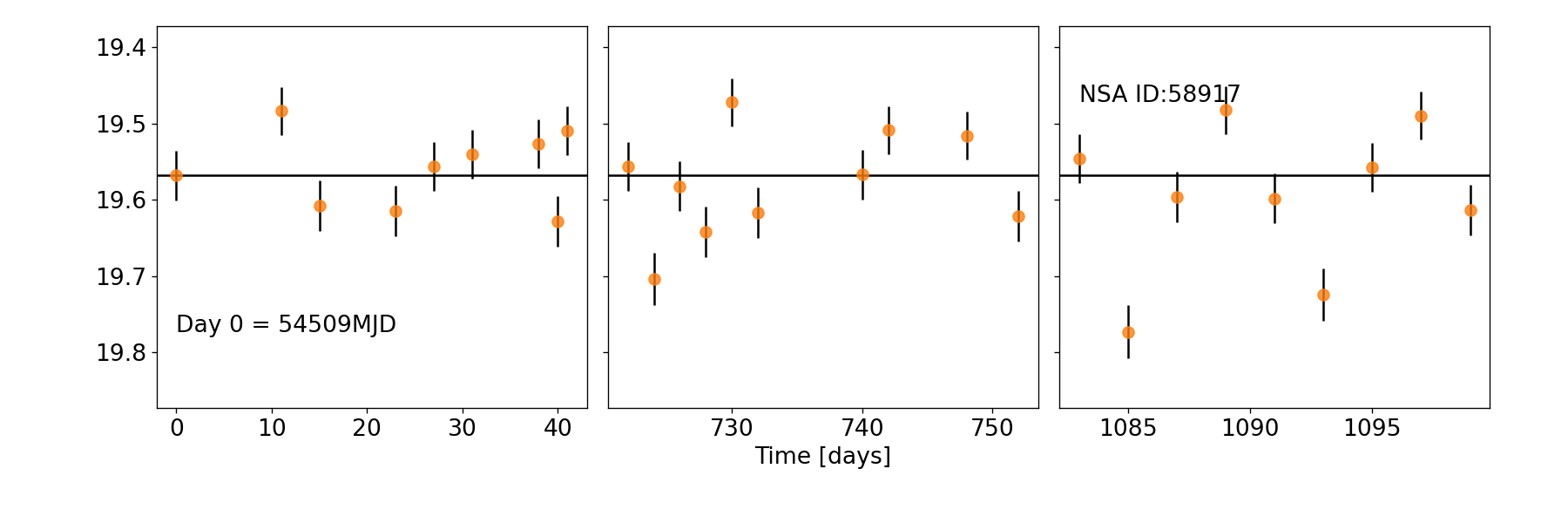} }
    \subfigure{\includegraphics[scale = 0.45]{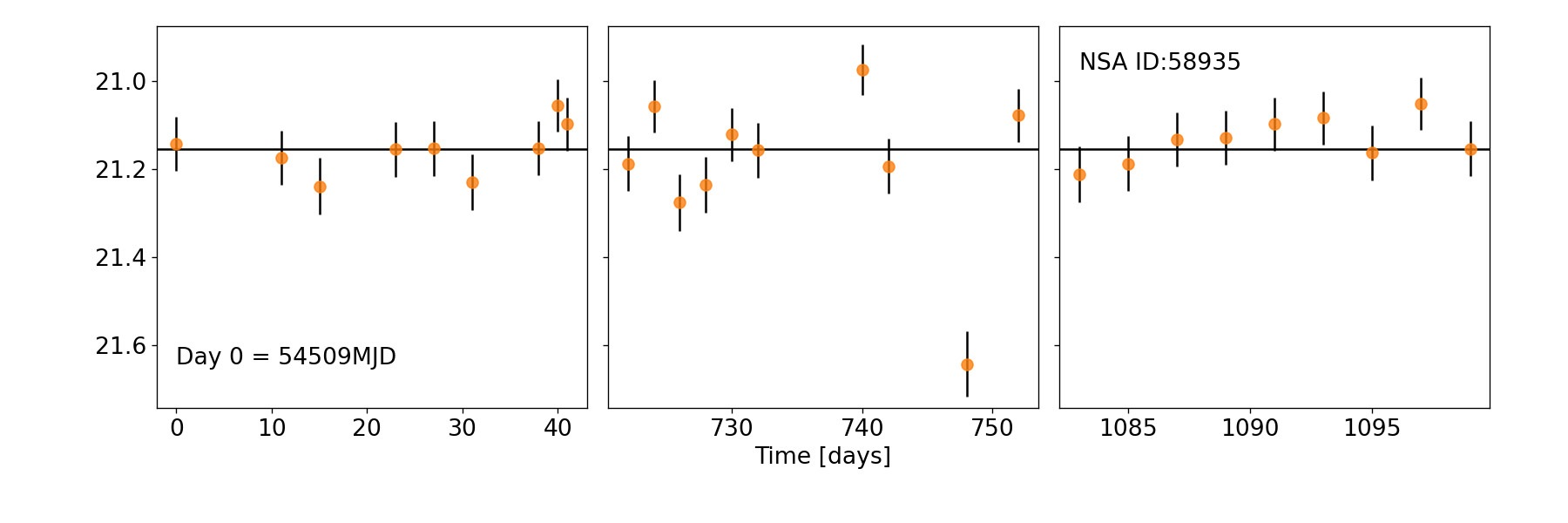} }
    \subfigure{\includegraphics[scale = 0.45]{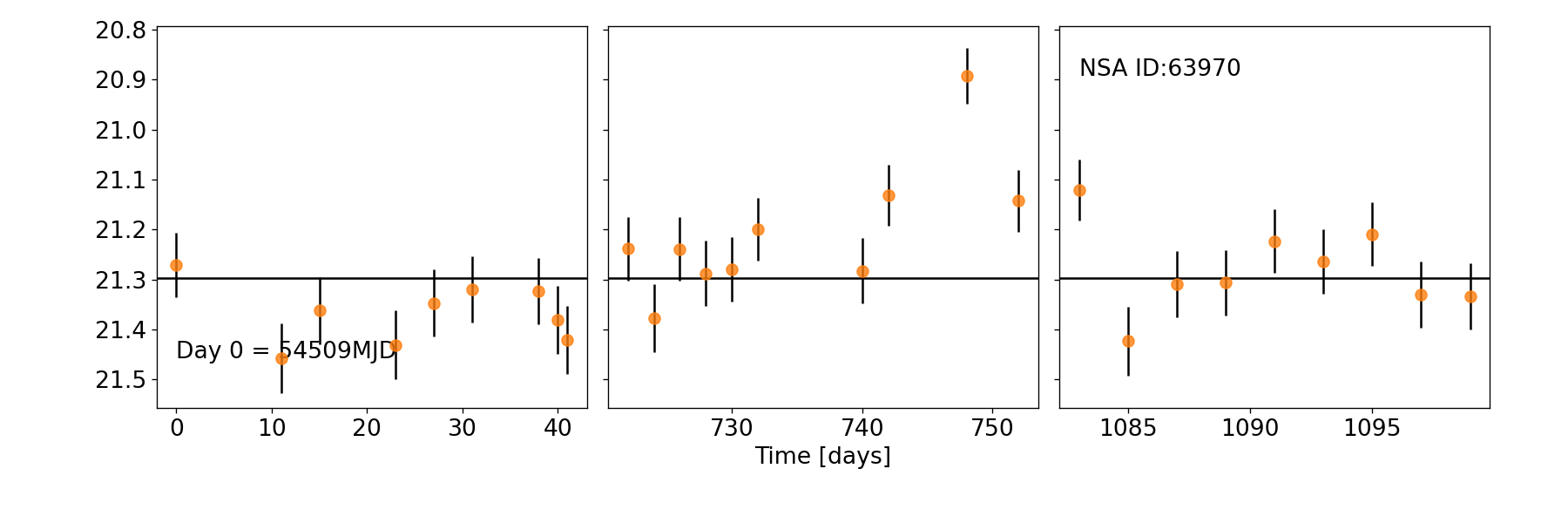} }
    \subfigure{\includegraphics[scale = 0.45]{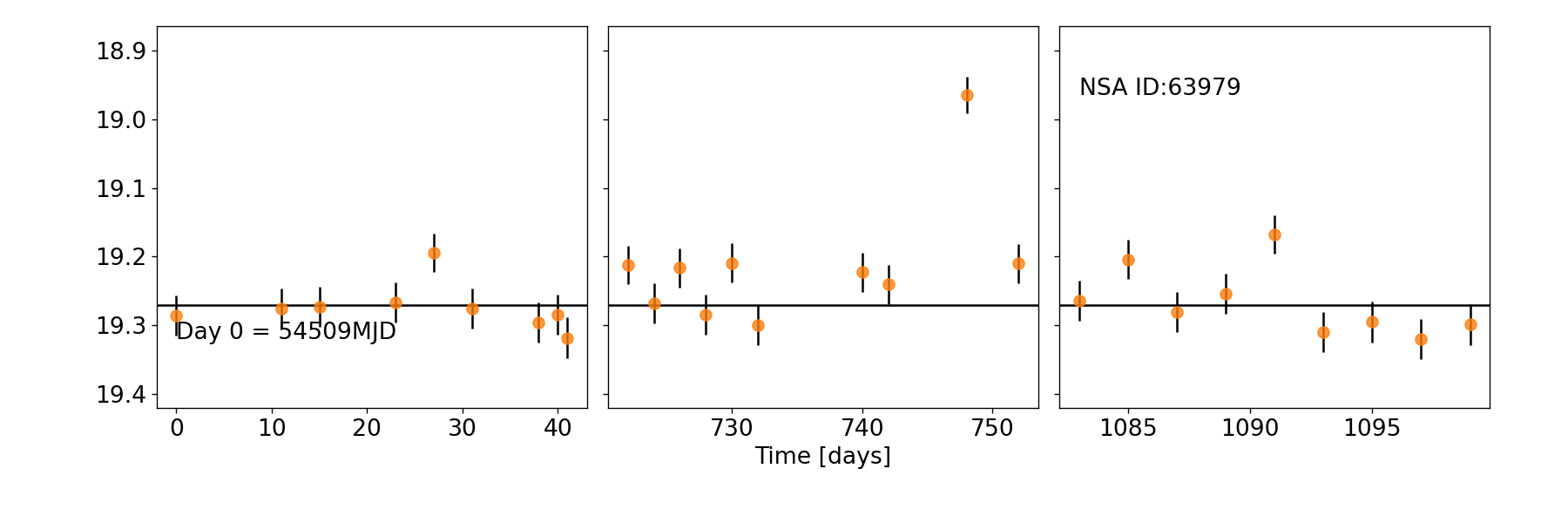} }
    \label{fig:Light_curves_set3}
\end{figure*}

\begin{figure*}
    \centering
    \subfigure{\includegraphics[scale = 0.45]{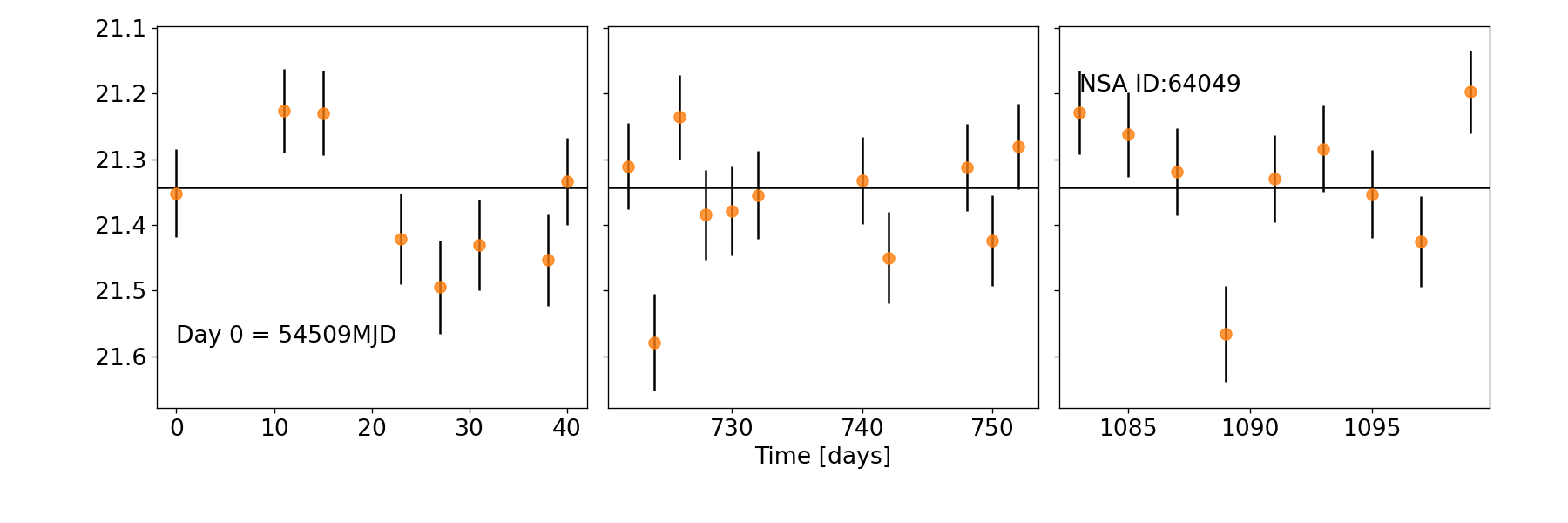} }
    \subfigure{\includegraphics[scale = 0.45]{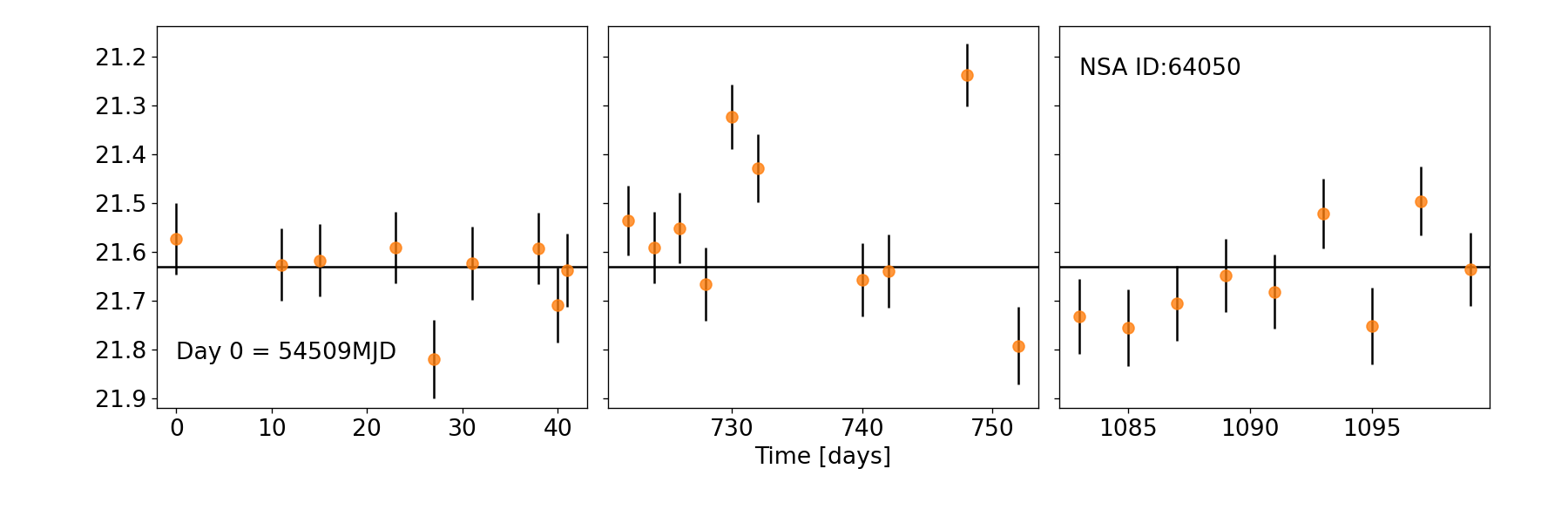} }
    \subfigure{\includegraphics[scale = 0.45]{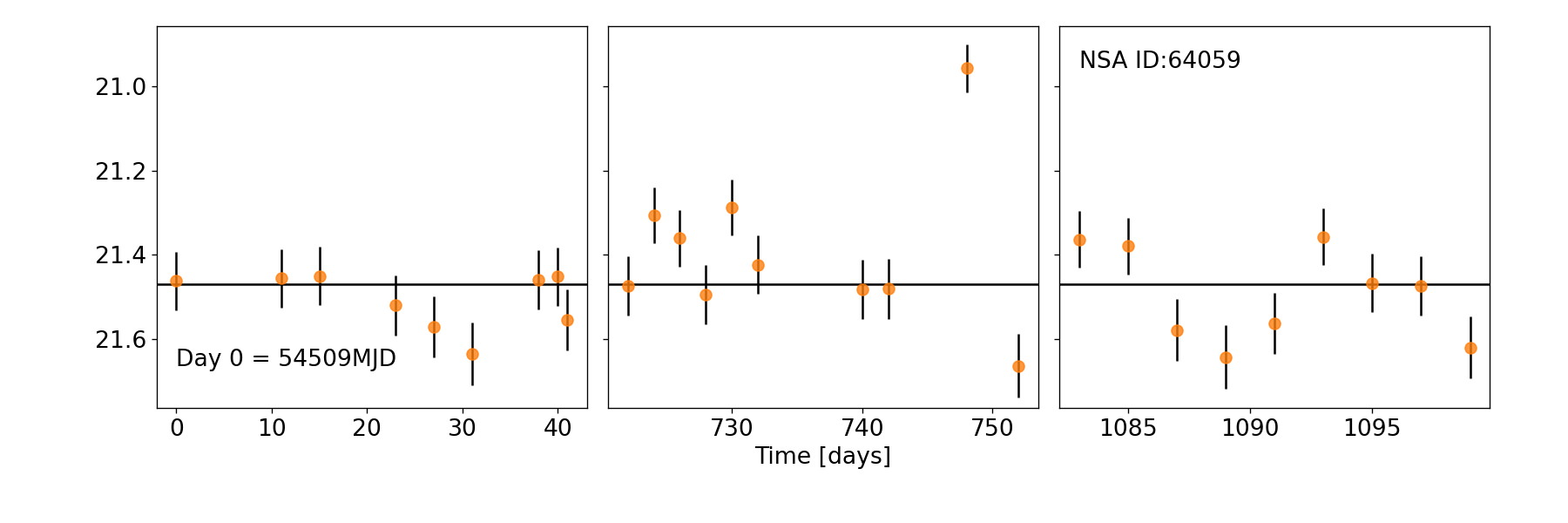} }
    \subfigure{\includegraphics[scale = 0.45]{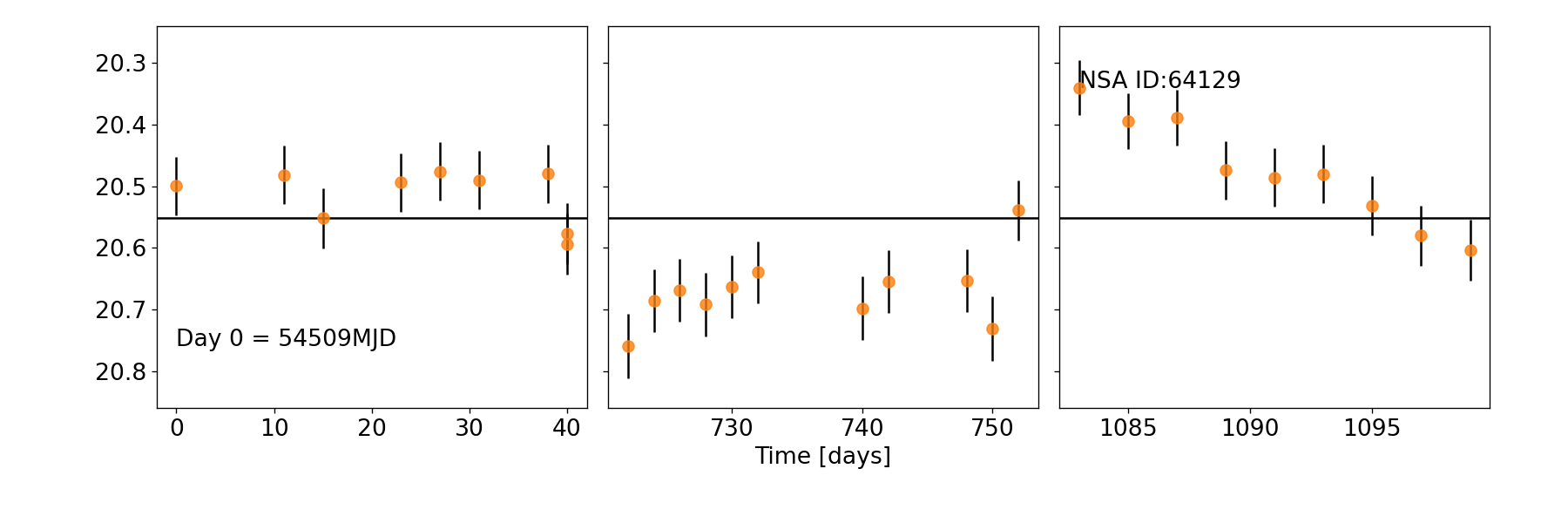} }
    \label{fig:Light_curves_set4}
\end{figure*}

\begin{figure*}
    \centering
    \subfigure{\includegraphics[scale = 0.45]{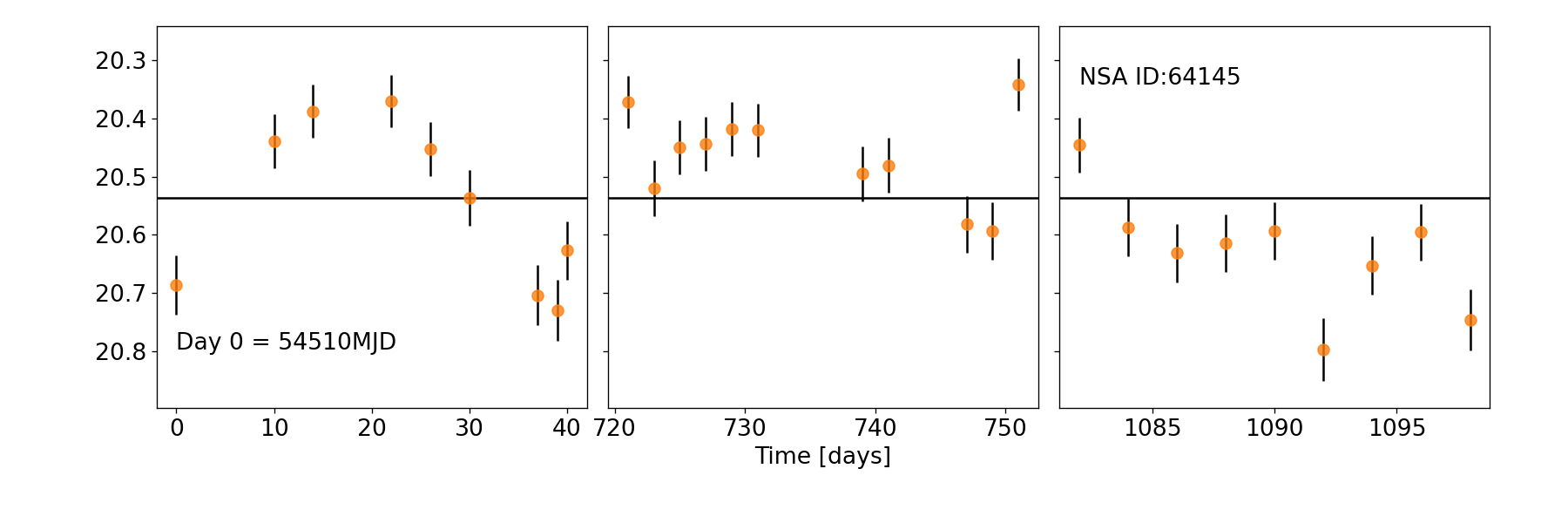} }
    \subfigure{\includegraphics[scale = 0.45]{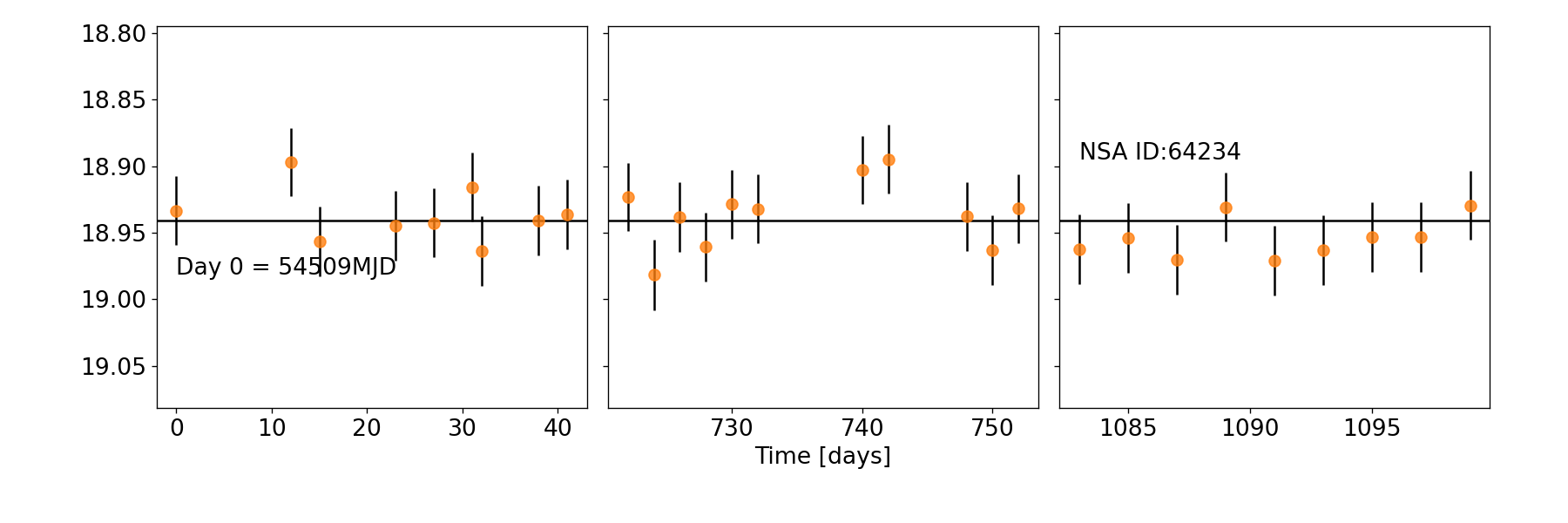} }
    \subfigure{\includegraphics[scale = 0.45]{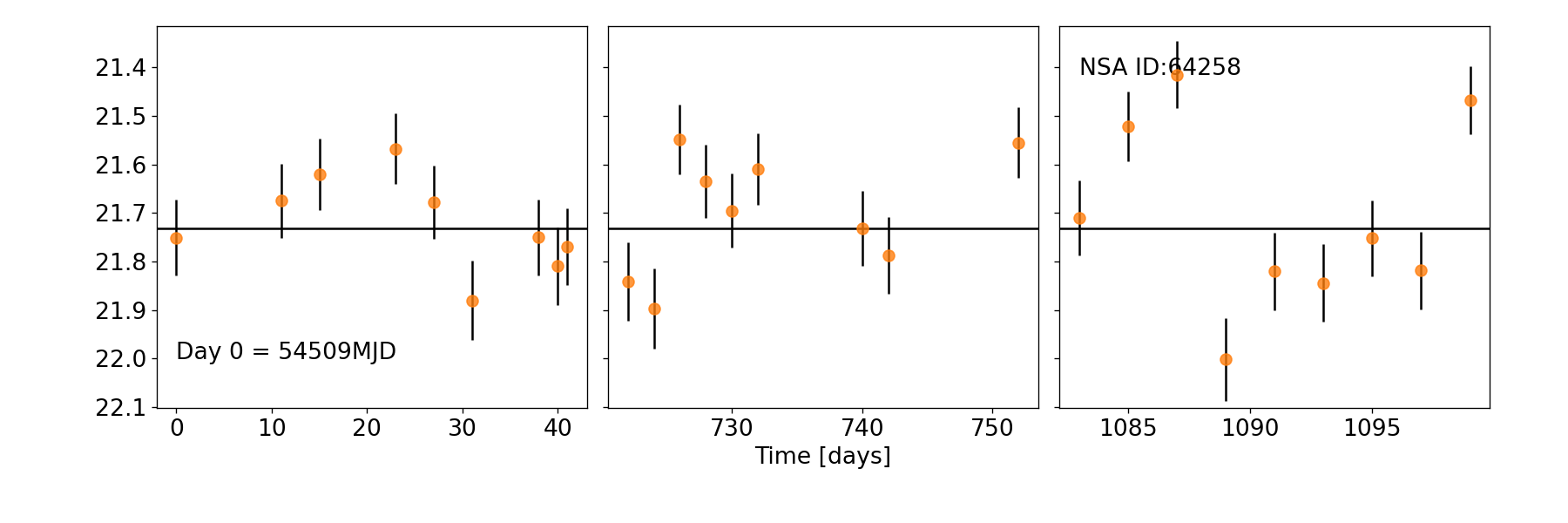} }
    \subfigure{\includegraphics[scale = 0.45]{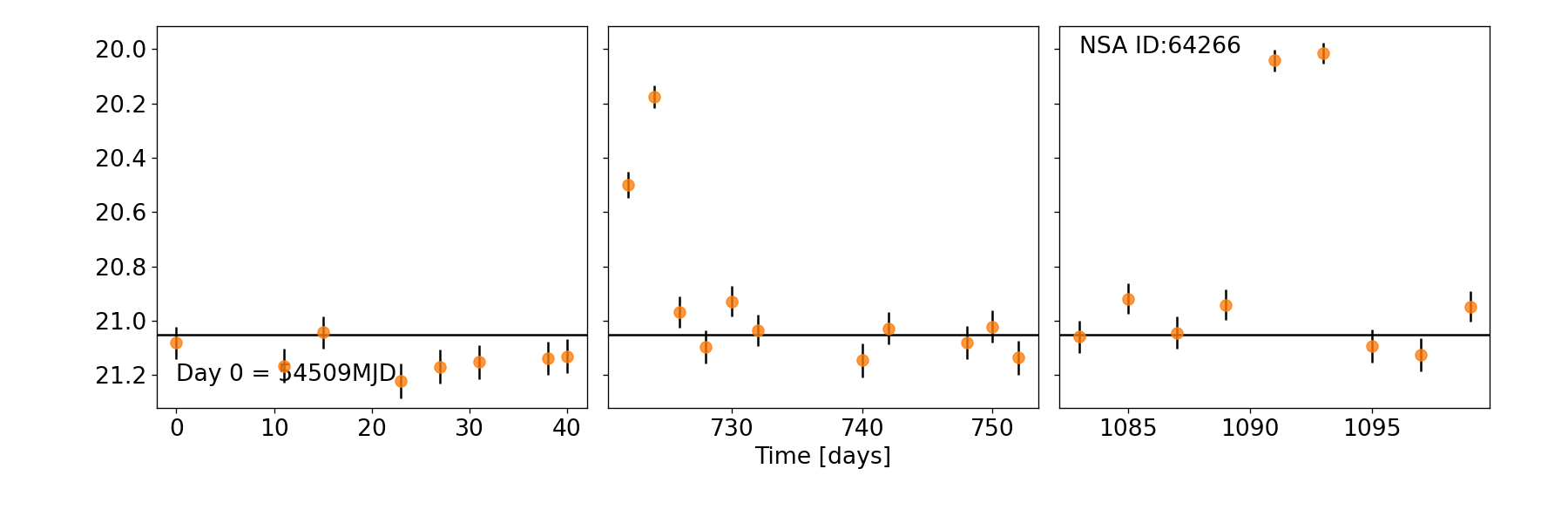} }
    \label{fig:Light_curves_set5}
\end{figure*}

\begin{figure*}
    \centering
    \subfigure{\includegraphics[scale = 0.45]{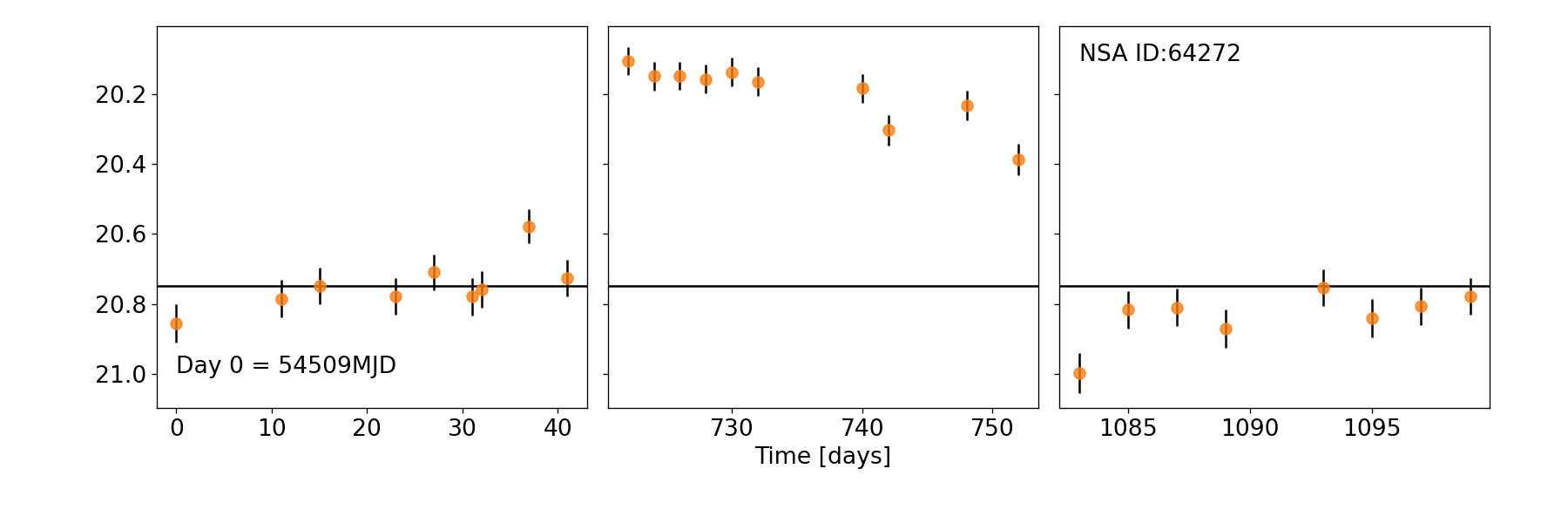} }
    \subfigure{\includegraphics[scale = 0.45]{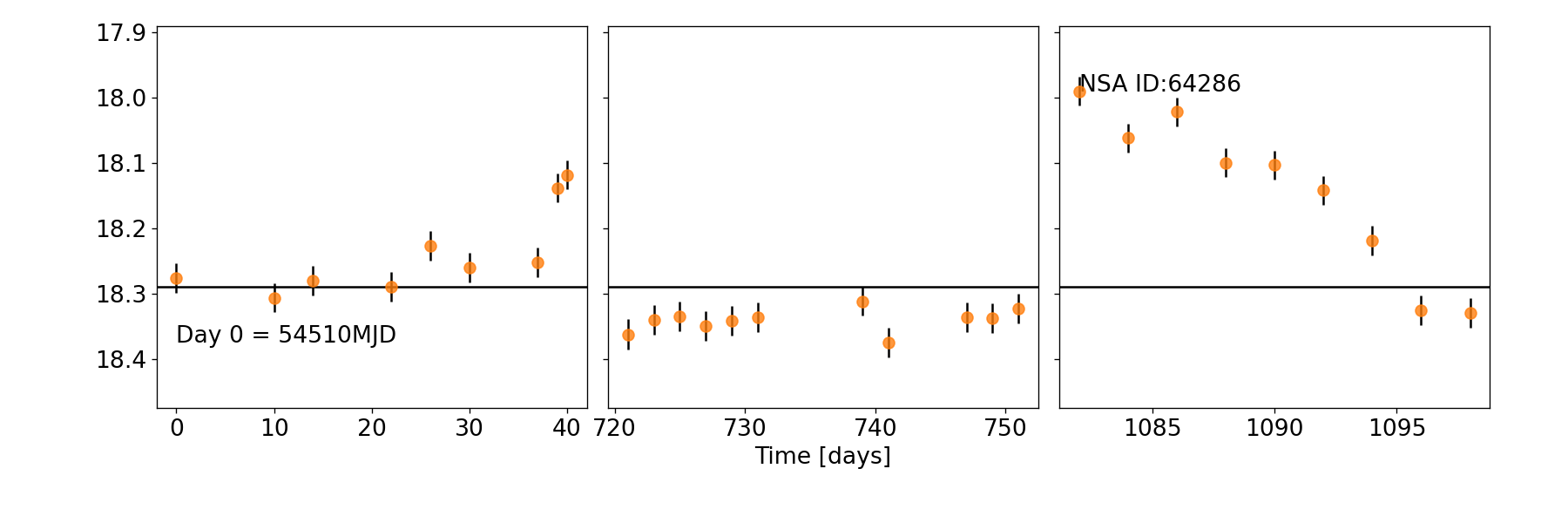} }
    \subfigure{\includegraphics[scale = 0.45]{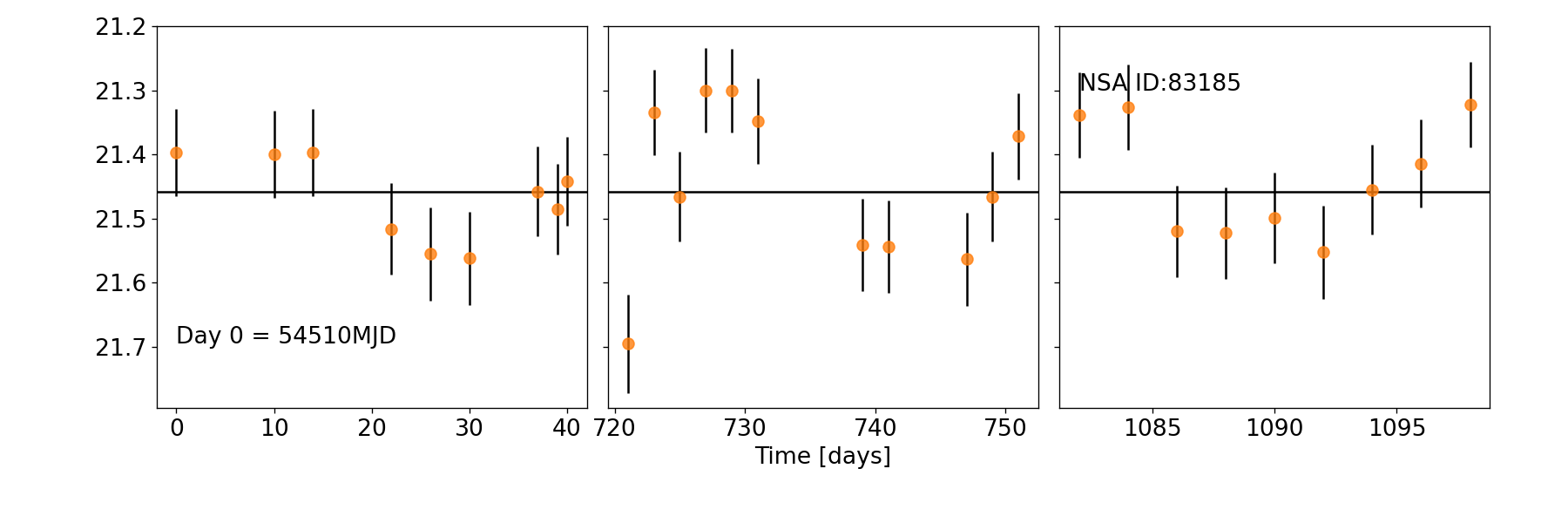} }
    \subfigure{\includegraphics[scale = 0.45]{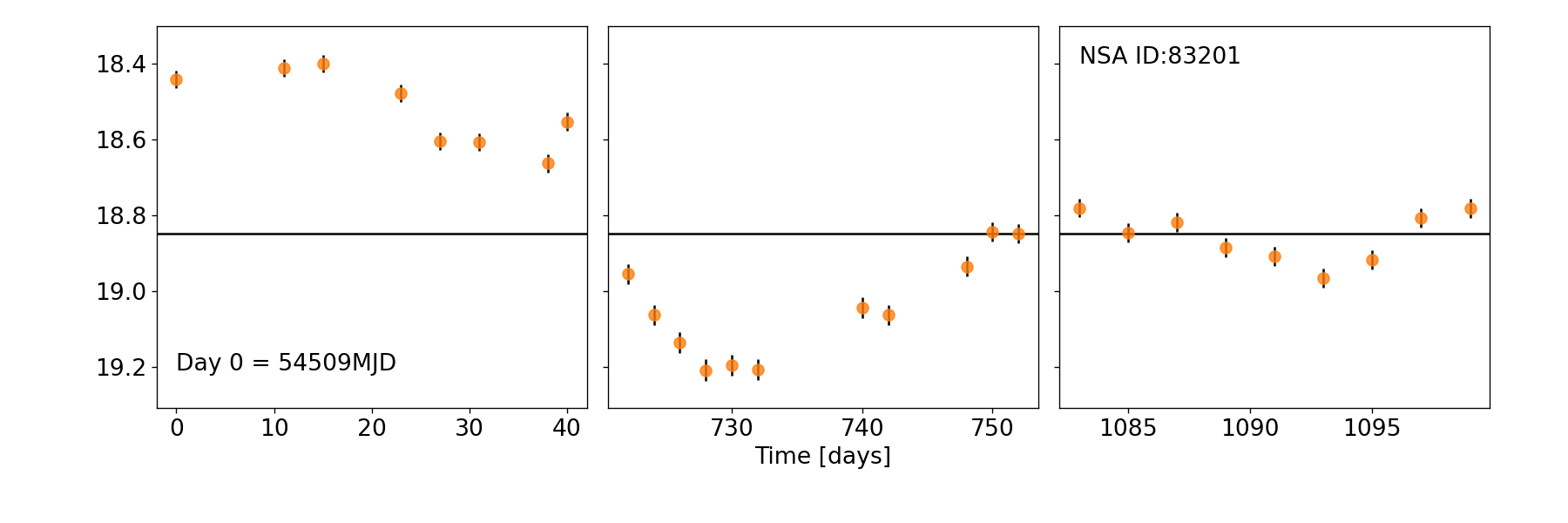} }
    \label{fig:Light_curves_set6}
\end{figure*}

\begin{figure*}
    \centering
    \subfigure{\includegraphics[scale = 0.45]{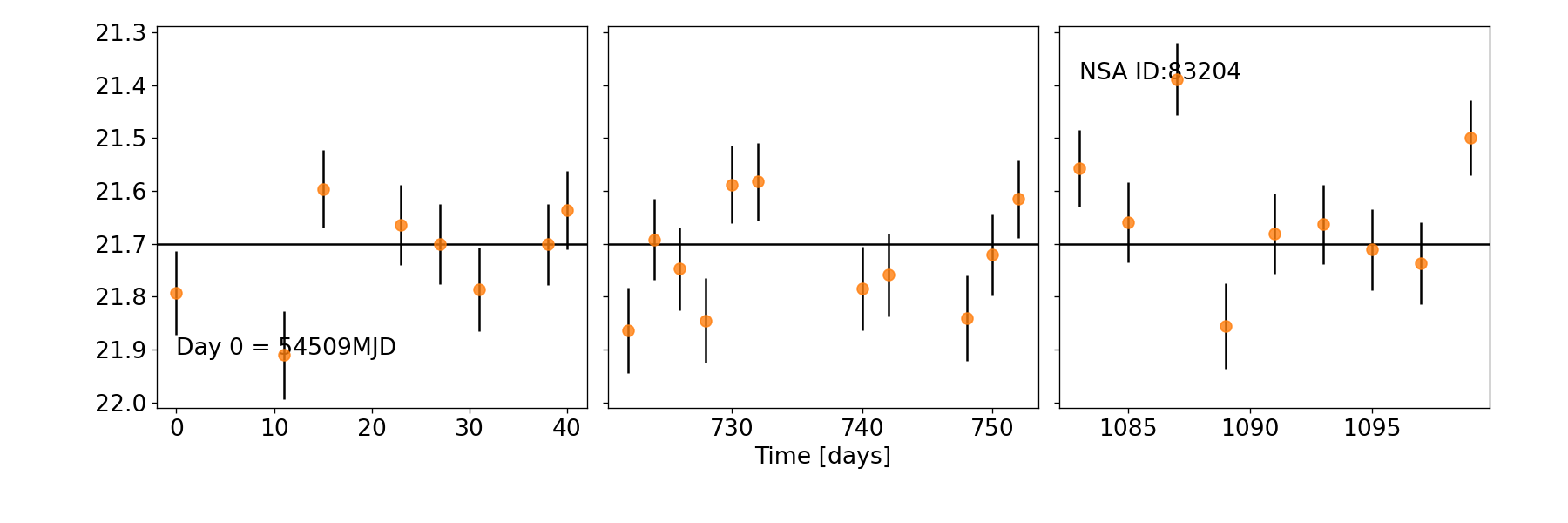} }
    \subfigure{\includegraphics[scale = 0.45]{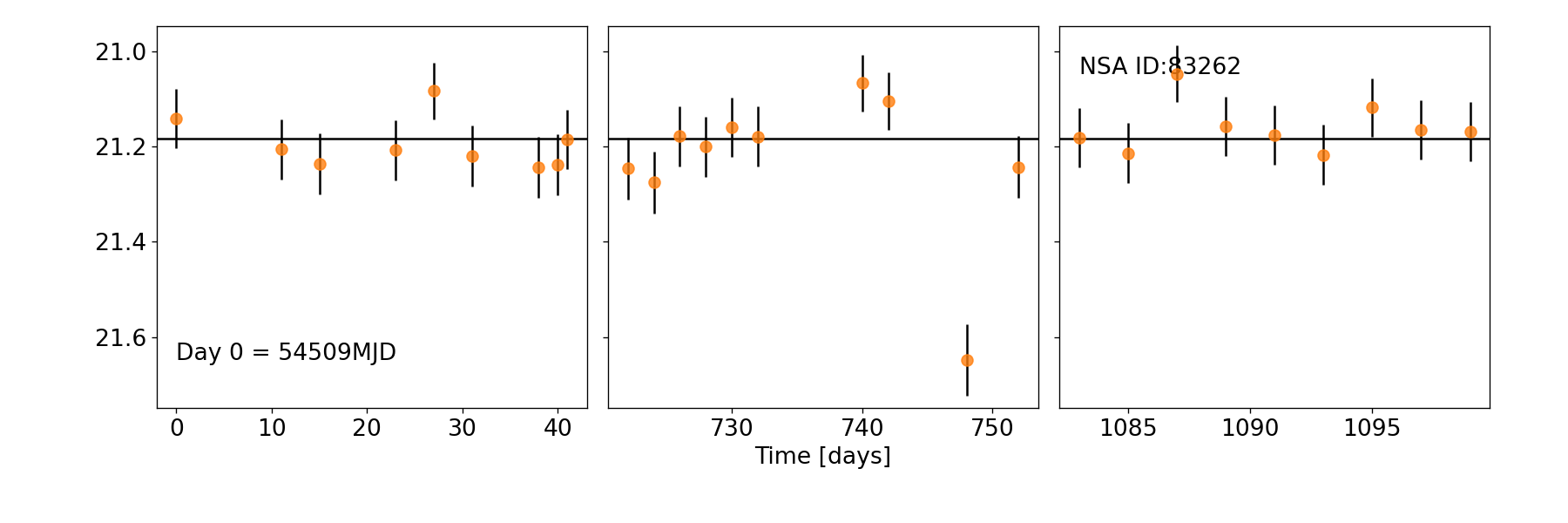} }
    \subfigure{\includegraphics[scale = 0.45]{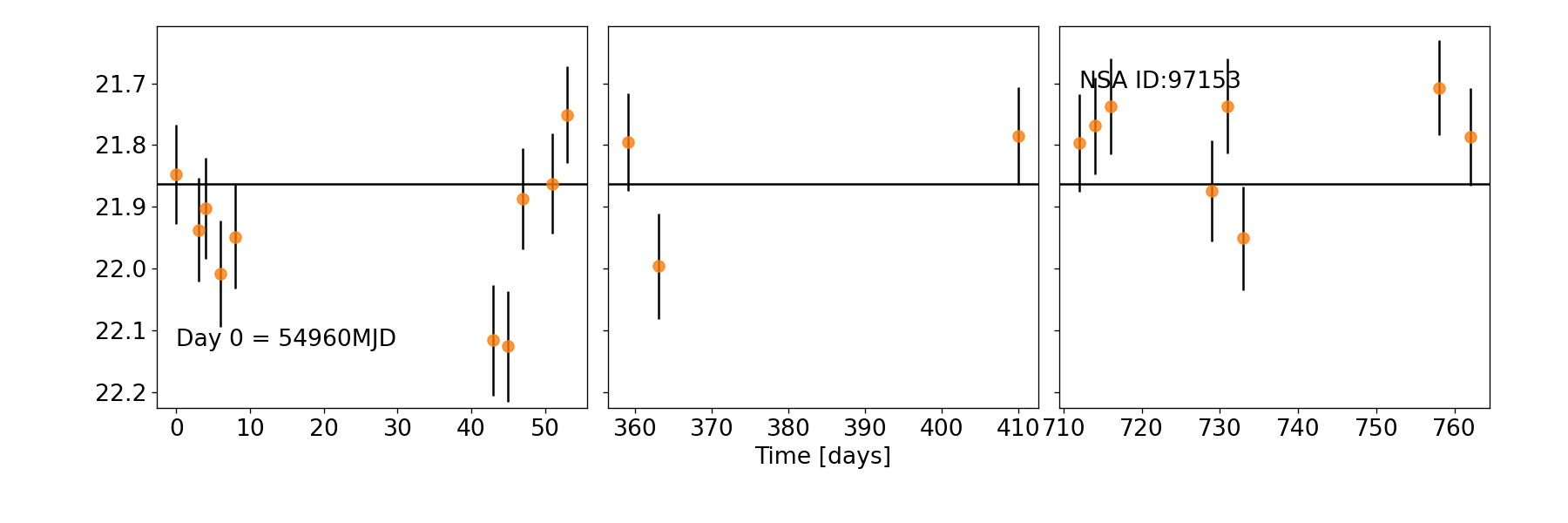} }
    \subfigure{\includegraphics[scale = 0.45]{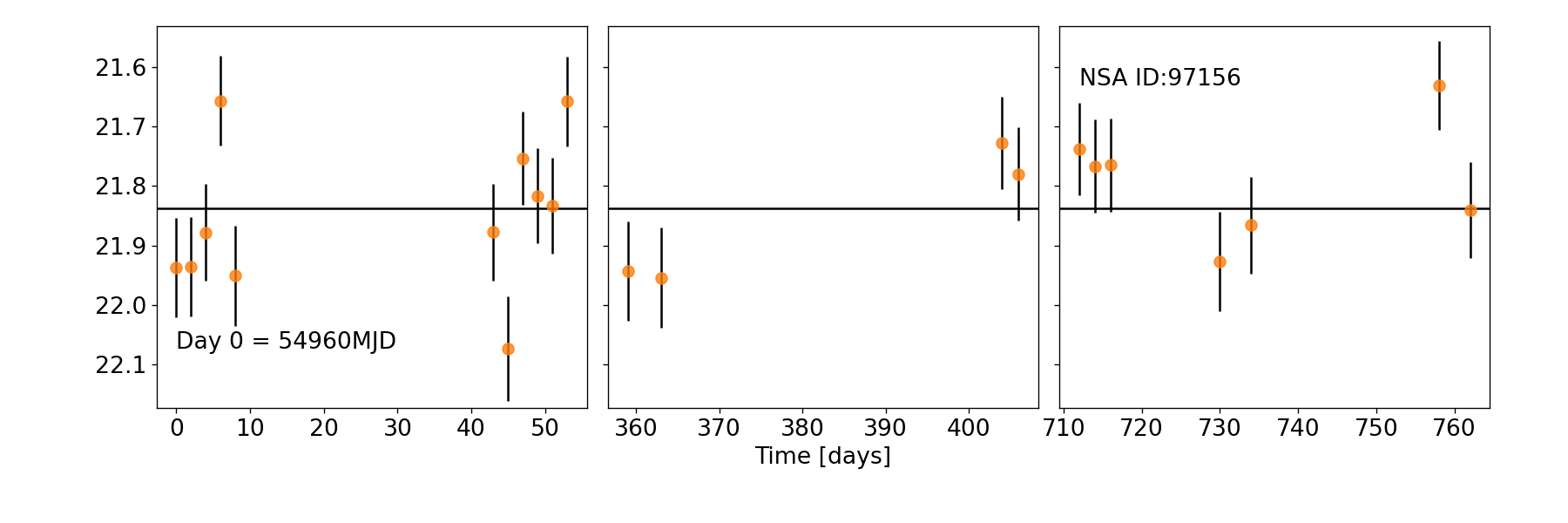} }
    \label{fig:Light_curves_set7}
\end{figure*}

\begin{figure*}
    \centering
    \subfigure{\includegraphics[scale = 0.45]{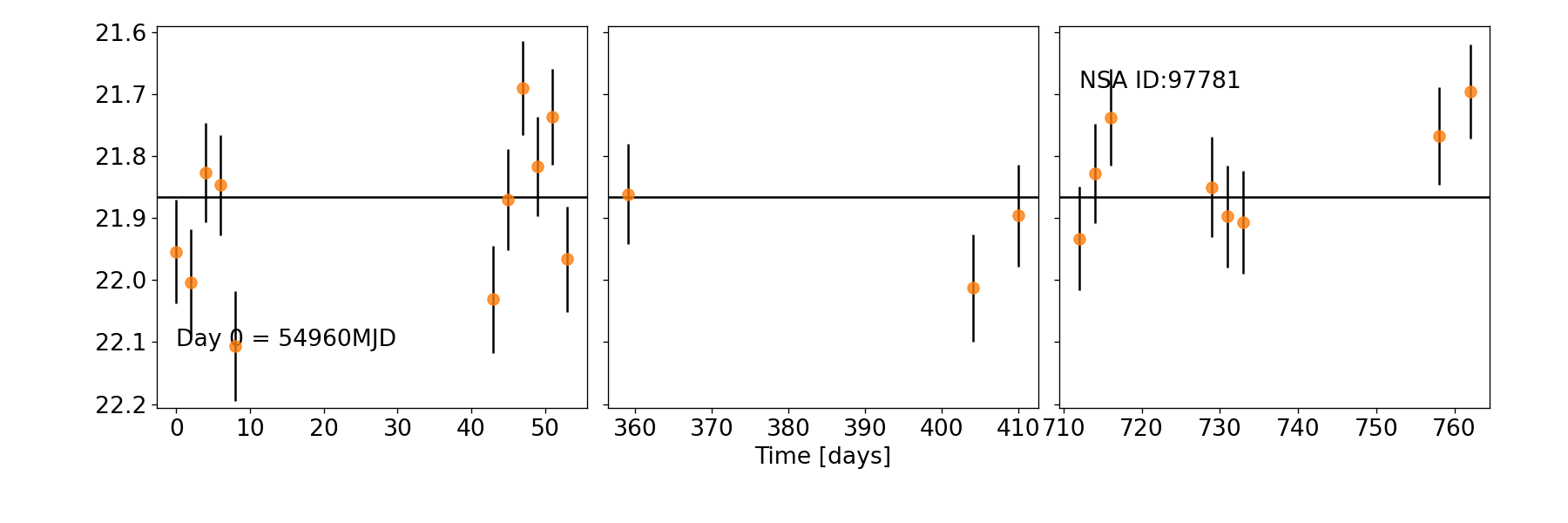} }
    \subfigure{\includegraphics[scale = 0.45]{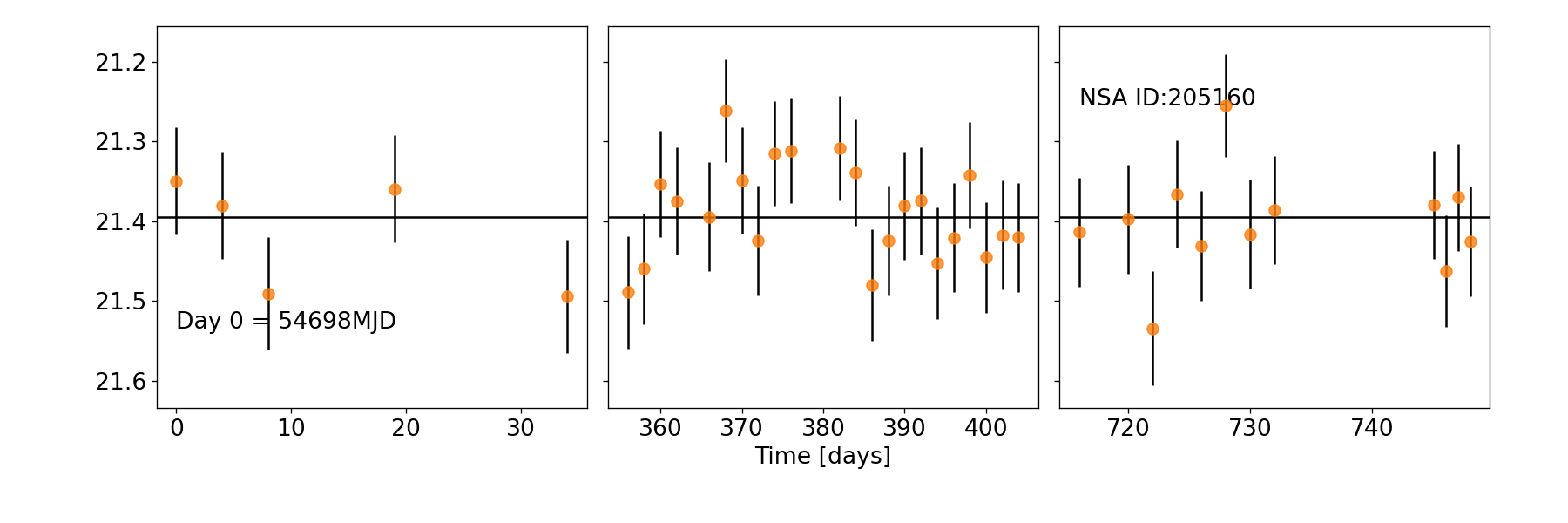} }
    \subfigure{\includegraphics[scale = 0.45]{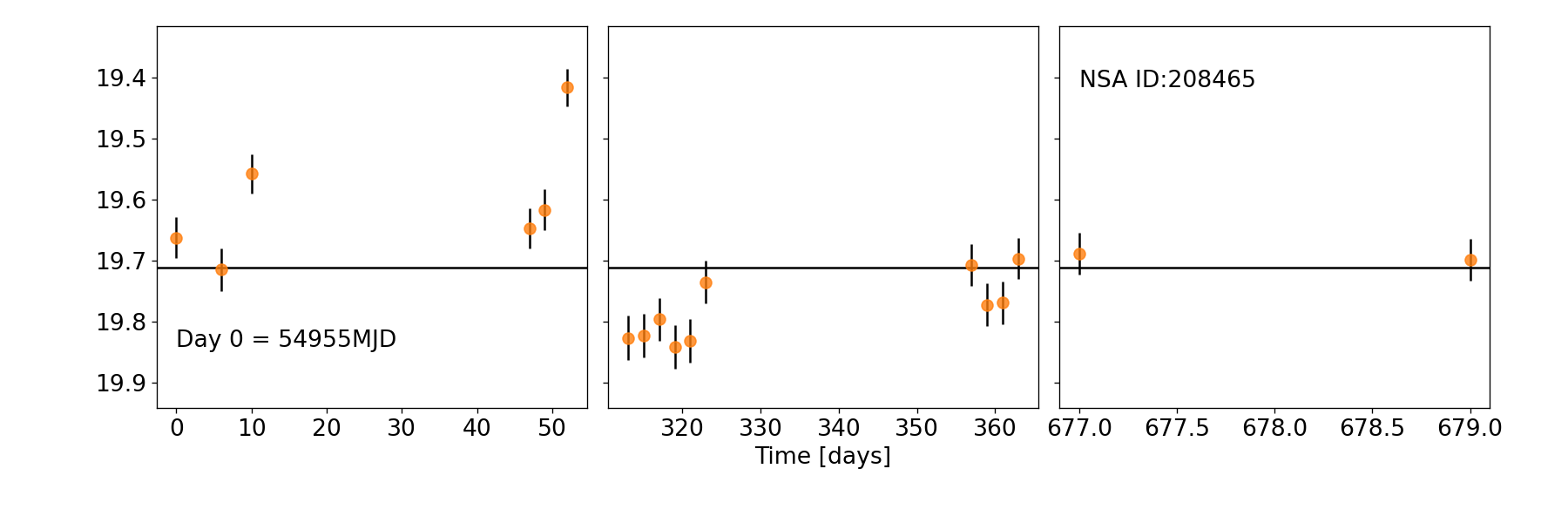} }
    \subfigure{\includegraphics[scale = 0.45]{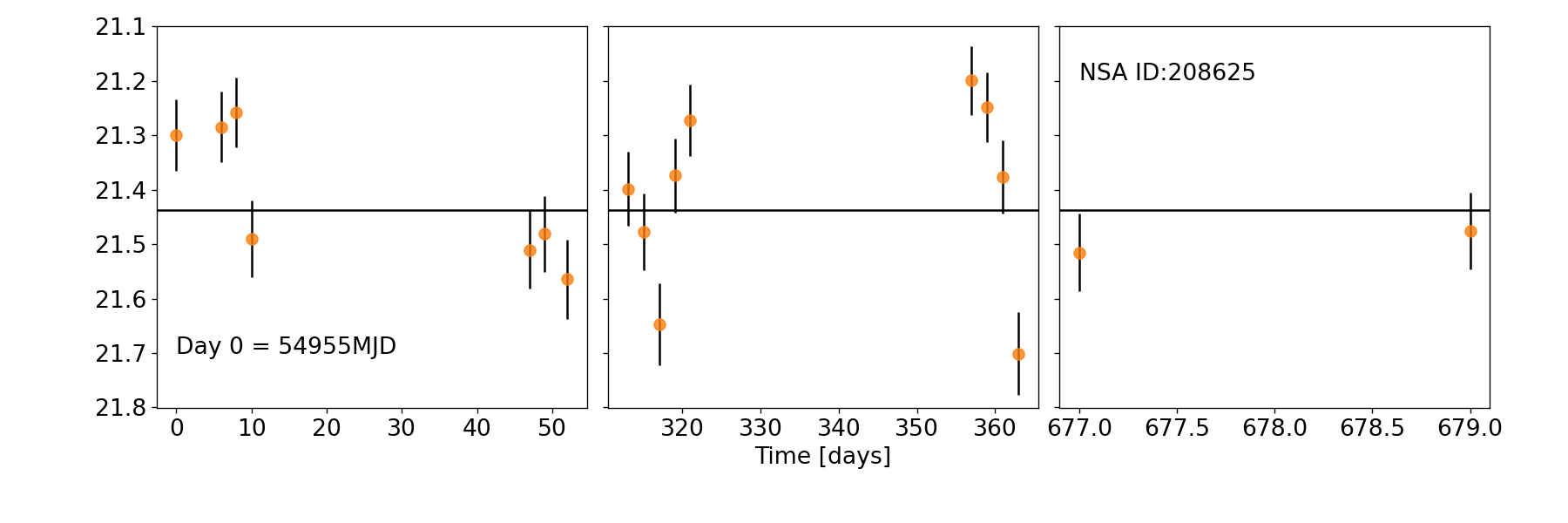} }
    \label{fig:Light_curves_set8}
\end{figure*}

\begin{figure*}
    \centering
    \subfigure{\includegraphics[scale = 0.45]{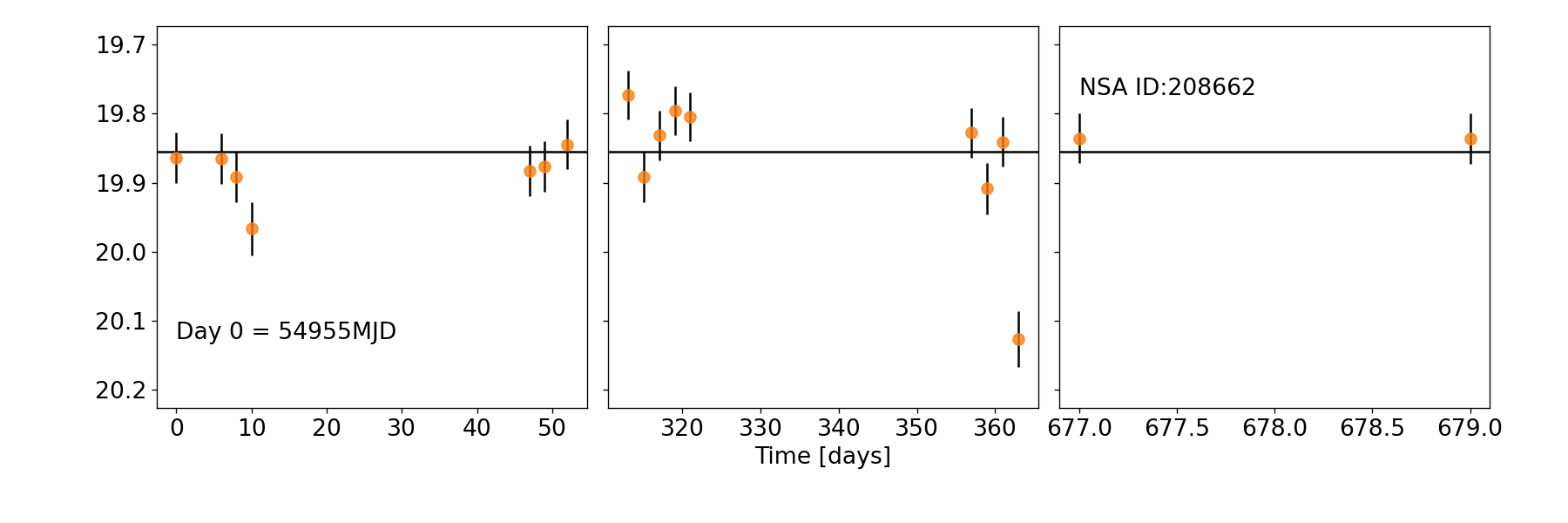} }
    \subfigure{\includegraphics[scale = 0.45]{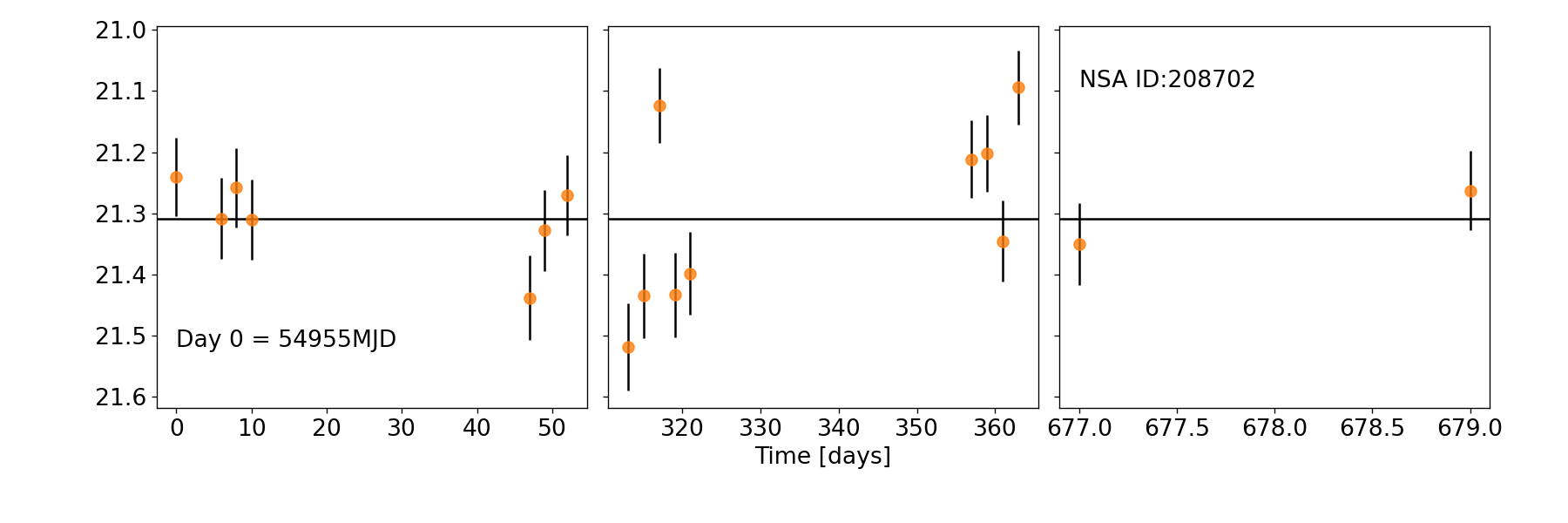} }
    \subfigure{\includegraphics[scale = 0.45]{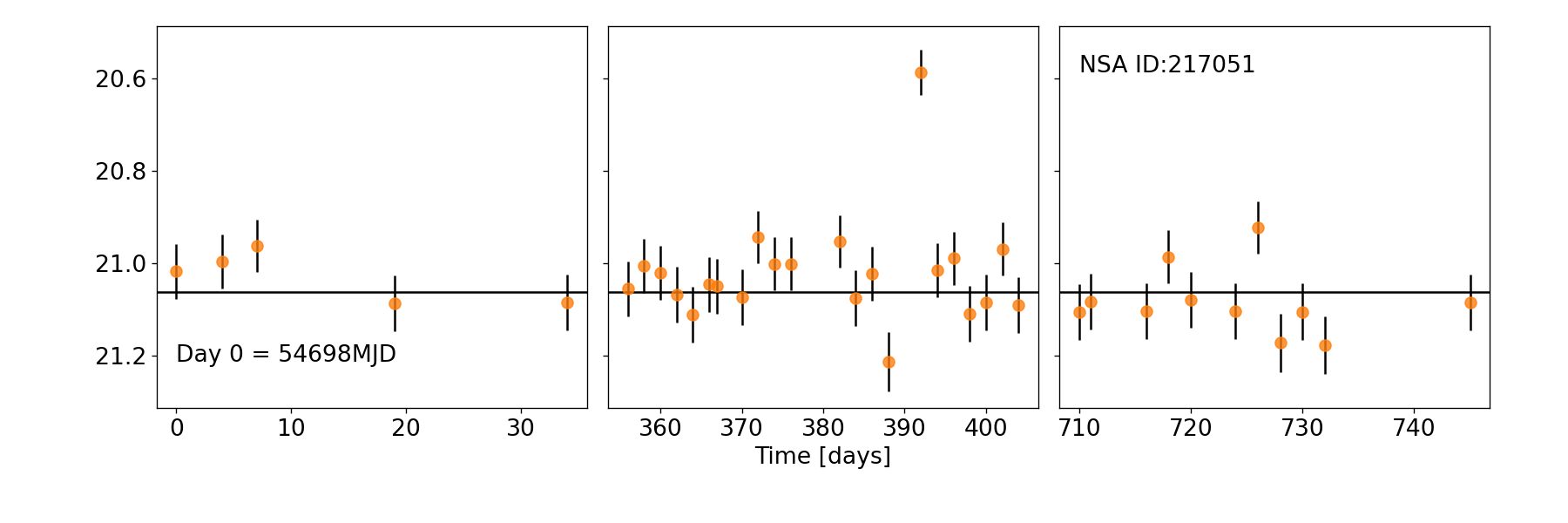} }
    \subfigure{\includegraphics[scale = 0.45]{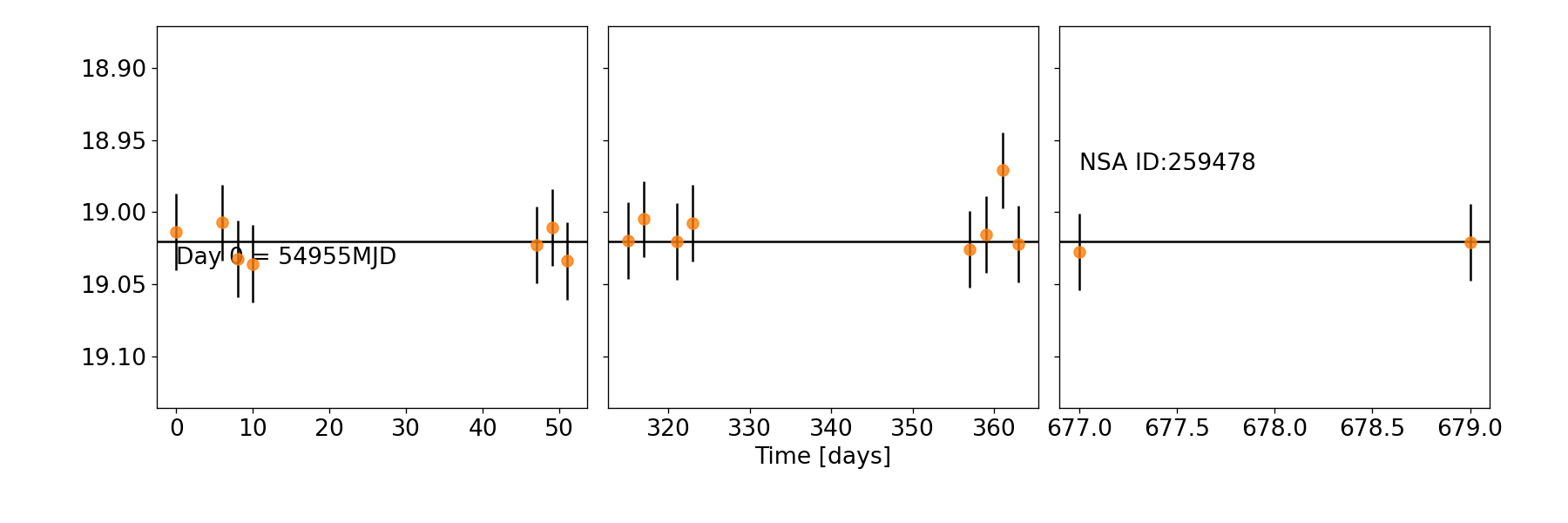} }
    \label{fig:Light_curves_set9}
\end{figure*}

\begin{figure*}
    \centering
    \subfigure{\includegraphics[scale = 0.45]{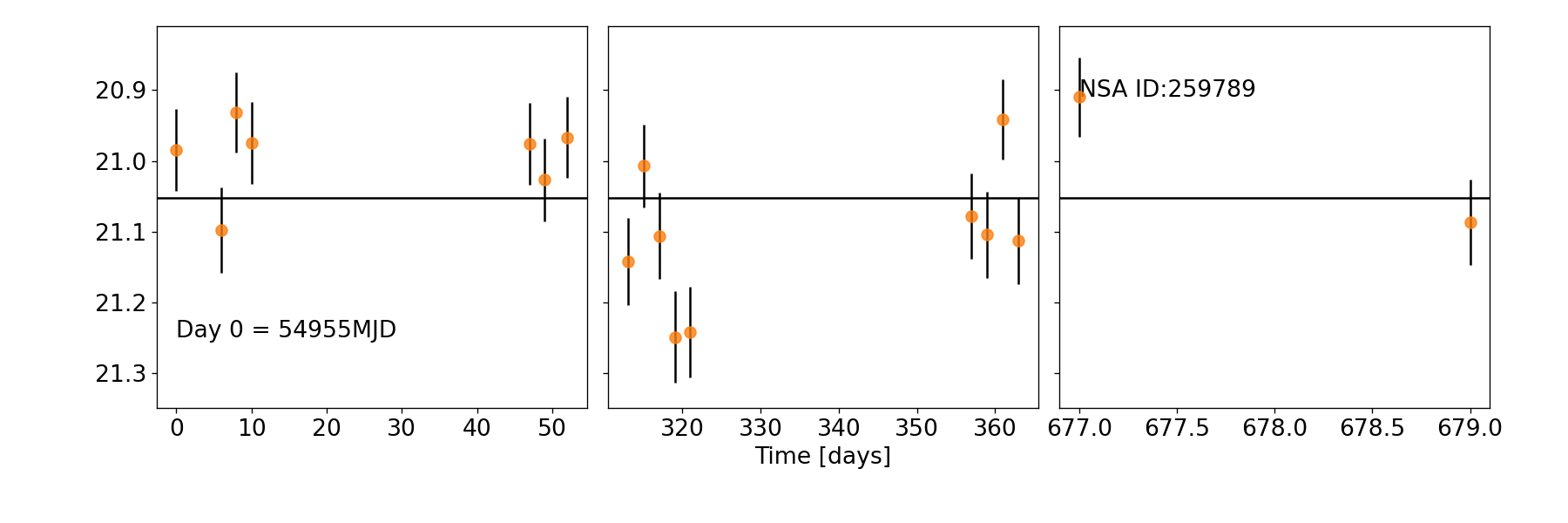} }
    \subfigure{\includegraphics[scale = 0.45]{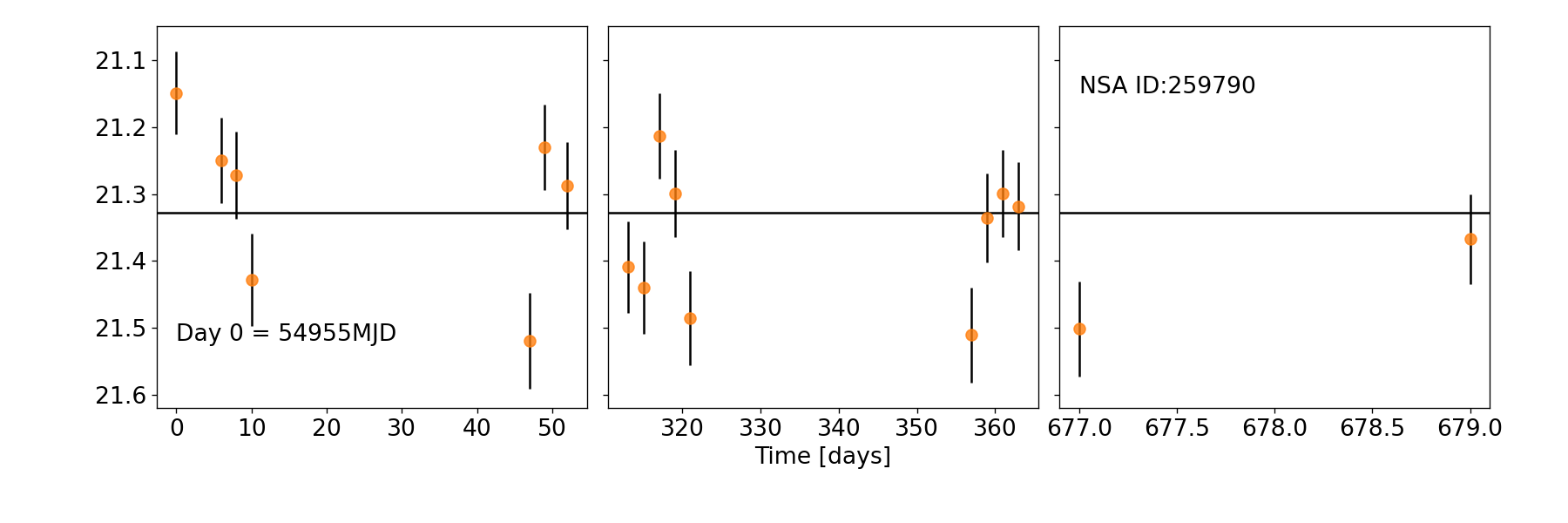} }
    \subfigure{\includegraphics[scale = 0.45]{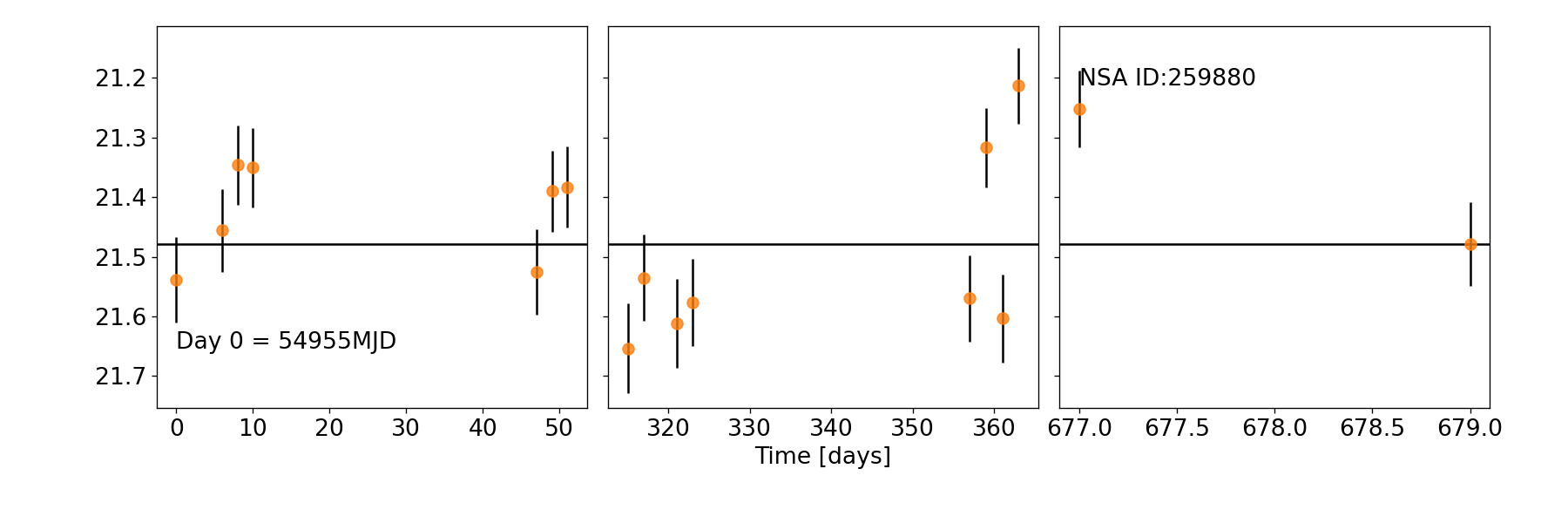} }
    \subfigure{\includegraphics[scale = 0.45]{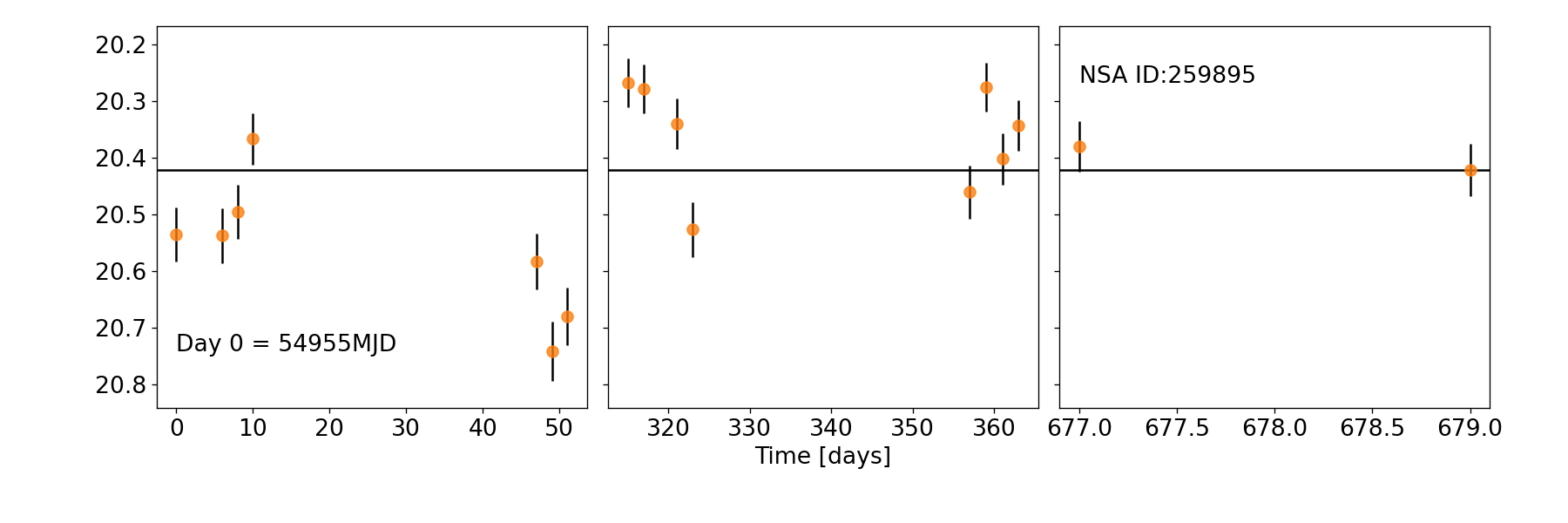} }
    \label{fig:Light_curves_set10}
\end{figure*}

\begin{figure*}
    \centering
    \subfigure{\includegraphics[scale = 0.45]{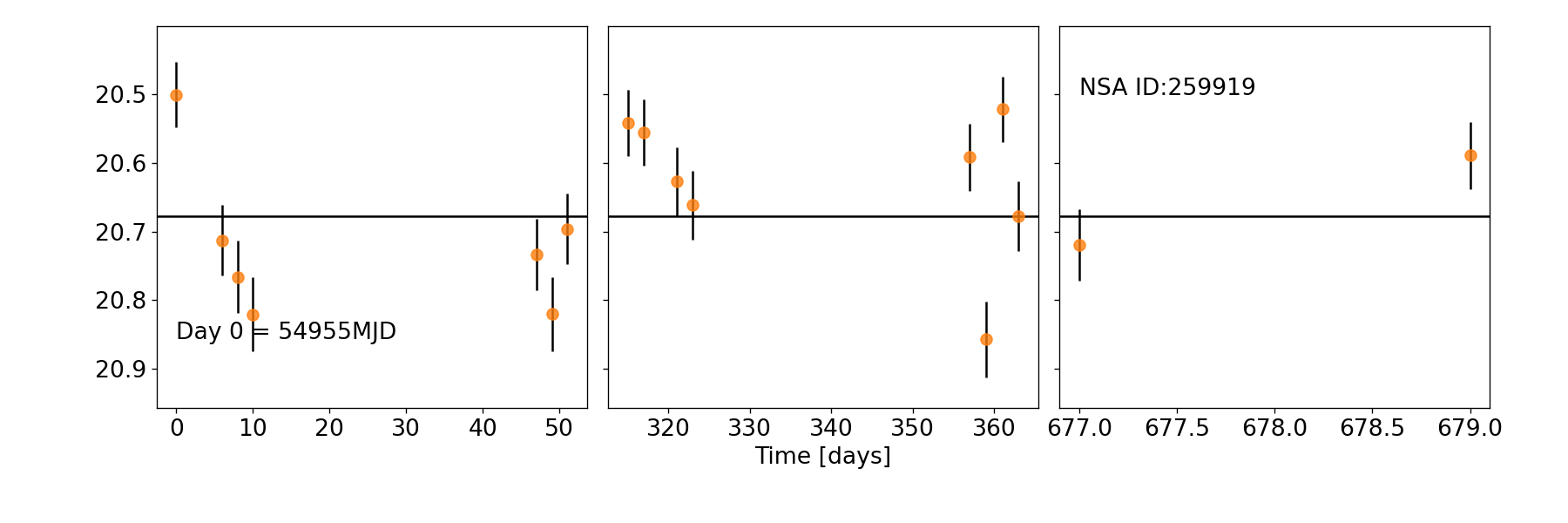} }
    \subfigure{\includegraphics[scale = 0.45]{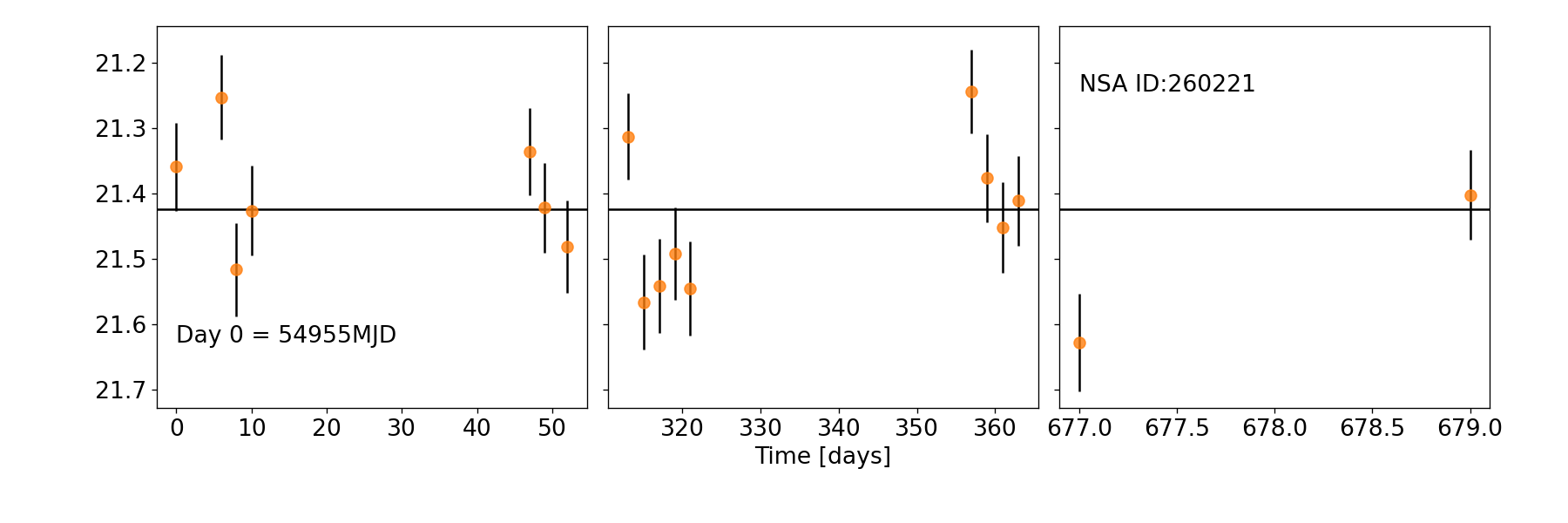} }
    \subfigure{\includegraphics[scale = 0.45]{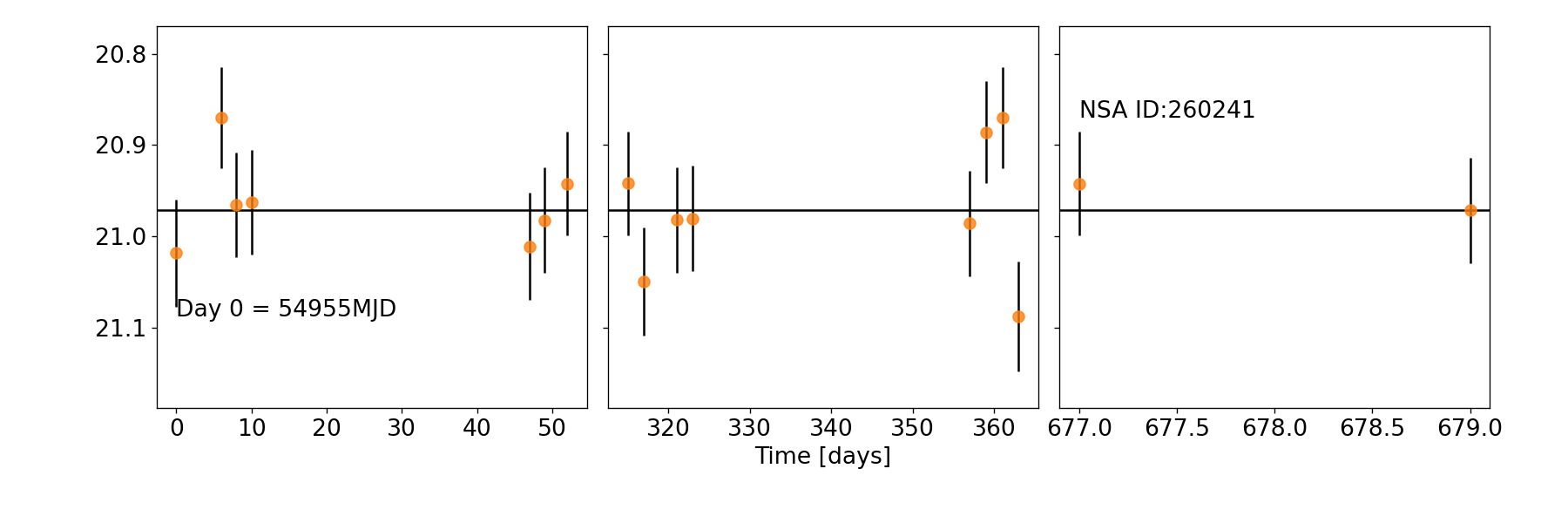} }
    \subfigure{\includegraphics[scale = 0.45]{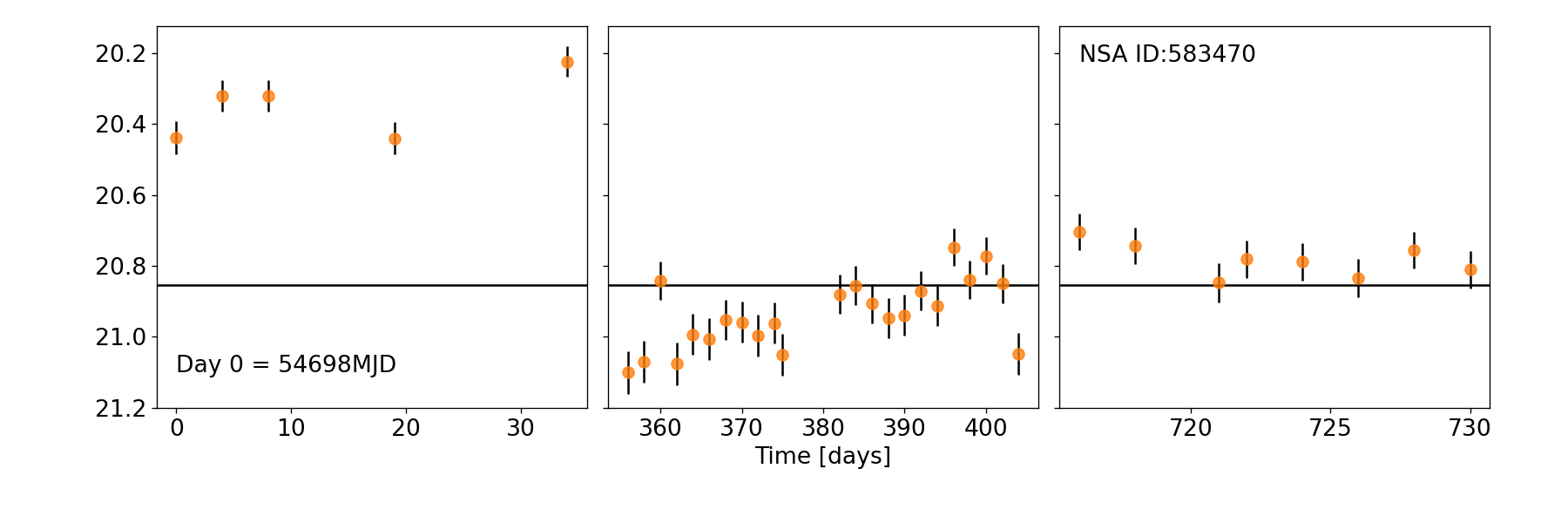} }
    \label{fig:Light_curves_set11}
\end{figure*}

\begin{figure*}
    \centering
    \subfigure{\includegraphics[scale = 0.45]{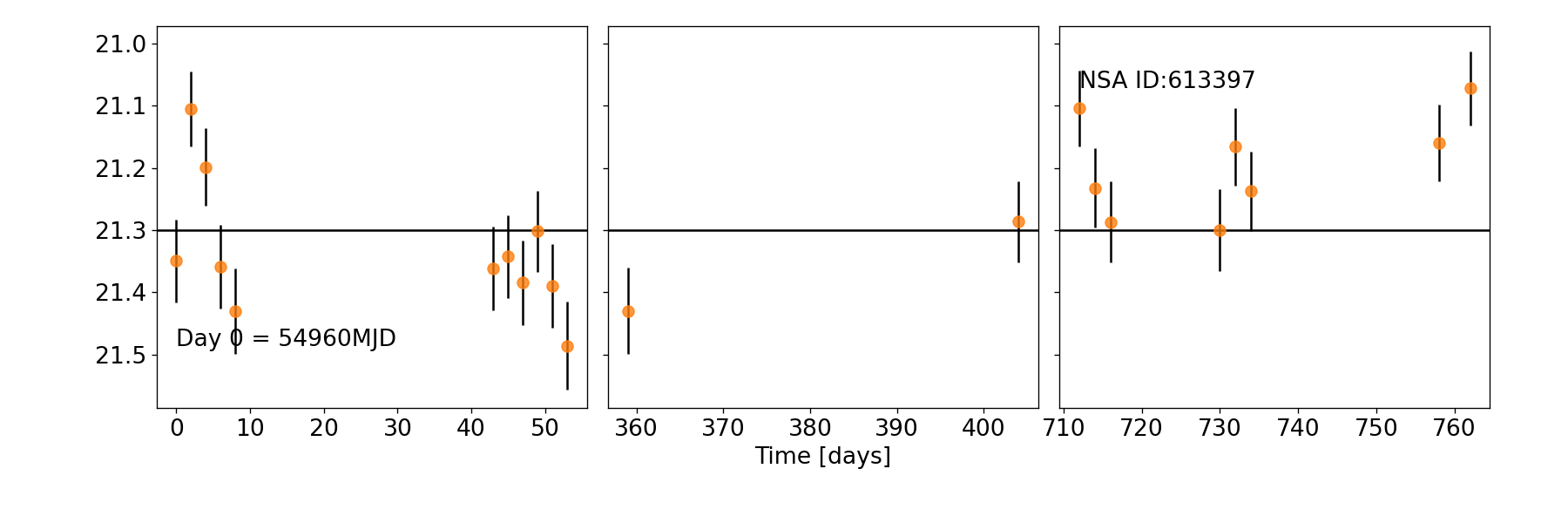} }
    \subfigure{\includegraphics[scale = 0.45]{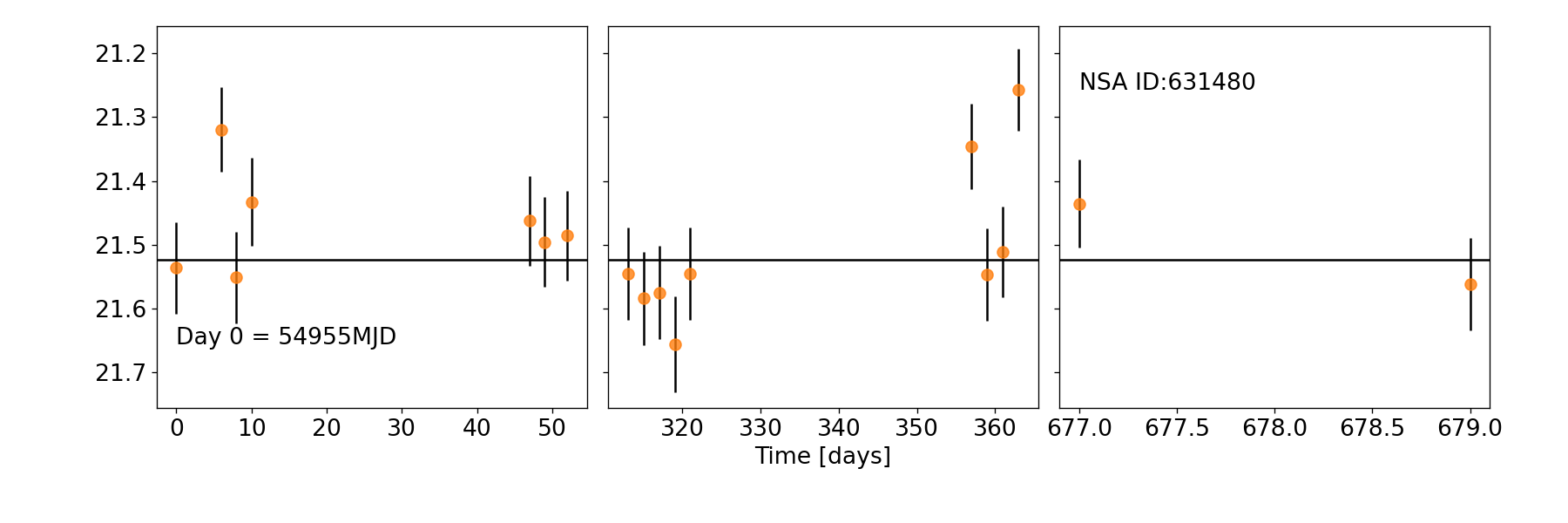} }
    \subfigure{\includegraphics[scale = 0.45]{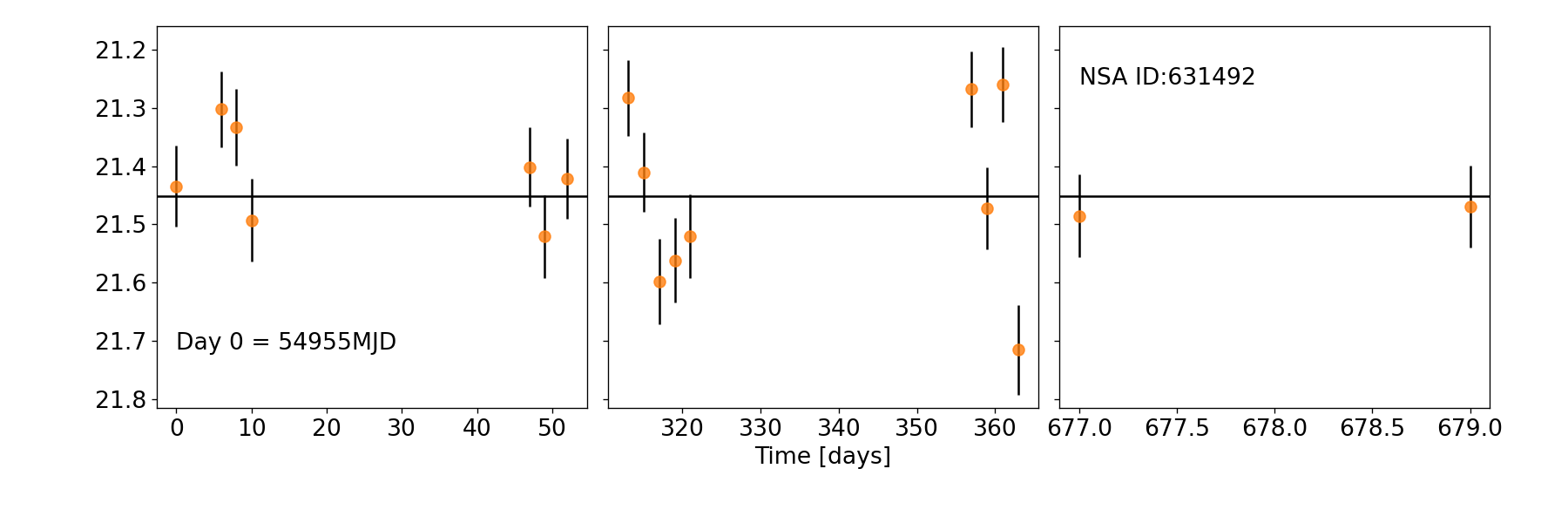} }
    \subfigure{\includegraphics[scale = 0.45]{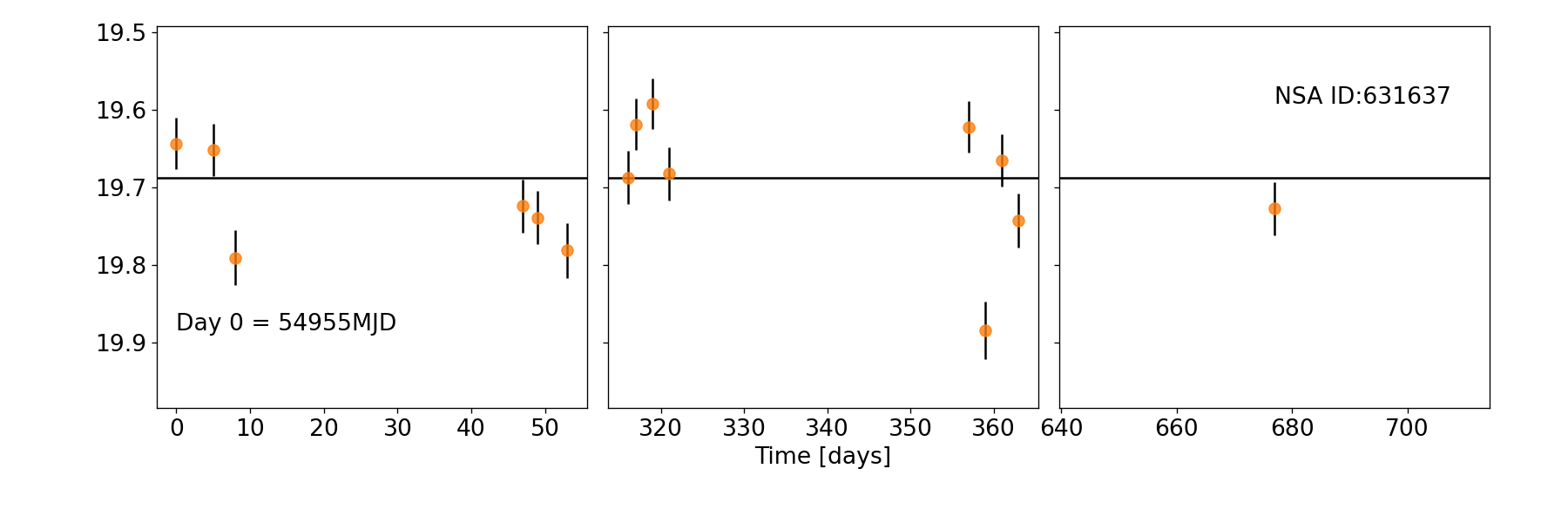} }
    \label{fig:Light_curves_set12}
\end{figure*}

\begin{figure*}
    \centering
    \subfigure{\includegraphics[scale = 0.45]{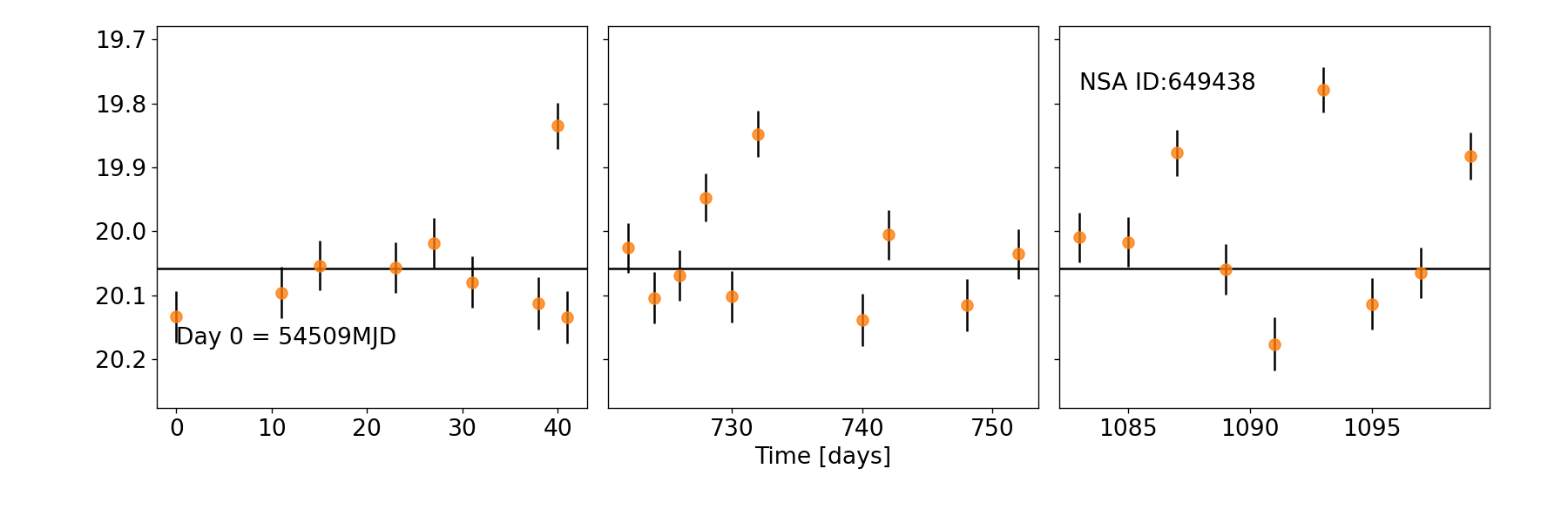} }
    \label{fig:Light_curves_set13}
\end{figure*}

\end{document}

%% file: table1.tex
\startlongtable
\begin{deluxetable*}{rrrrrrrrr}
\tablecolumns{9}
\tablecaption{Sample properties.\label{tab:results_table}}
\tablehead{\colhead{NSA ID} &  \colhead{R.A.} & \colhead{Decl.} & \colhead{$\log M_{*}$} & \colhead{\emph{z}} & \colhead{NUV Mag} & \colhead{$\sigma$} & \colhead{$F_{var}$} & \colhead{$\log M_{\text{BH}}$} \\
\colhead{} &  \colhead{[deg]} & \colhead{[deg]} & \colhead{[$M_\Sun$]} & \colhead{} & \colhead{} & \colhead{} & \colhead{} & \colhead{[$M_\Sun$]}}
\startdata
1382 &  150.8071007 & 0.9166644 & 11.13 & 0.0966 & 21.15 & 0.097 & 0.0035 & \nodata \\  
28585 &  334.1202944 & -1.2454316 & 10.41 & 0.1162 & 21.26 & 0.0974 & 0.0034 & \nodata\\  
28616 &  333.9262375 & -0.6027033 & 10.89 & 0.0991 & 19.41 & 0.1321 & 0.0066 & \nodata \\   
28762 &  333.3791149 & 0.8221411 & 10.77 & 0.0982 & 21.03 & 0.1016 & 0.0039 & \nodata \\  
28810 &  334.0740128 & 0.1650286 & 11.24 & 0.1394 & 21.45 & 0.0982 & 0.0032 & $8.29^{+0.90}_{-2.41} $ \\  
58753 &  149.0006157 & 1.7933316 & 10.87 & 0.0981 & 21.68 & 0.1333 & 0.0051 & \nodata \\  
58912 &  148.5312013 & 2.2277921 & 10.79 & 0.0698 & 21.53 & 0.1094 & 0.0038 & \nodata \\  
58917 &  148.5755379 & 2.7412974 & 9.39 & 0.0845 & 19.58 & 0.0724 & 0.0033 & \nodata \\  
58935 &  148.8960094 & 2.7014311 & 11.0 & 0.1309 & 21.16 & 0.1136 & 0.0045 & $7.09^{+0.40}_{-0.63} $ \\   
63970 &  148.821601 & 1.7666729 & 10.81 & 0.1003 & 21.28 & 0.115 & 0.0045 & \nodata \\ 
63979 &  148.8004955 & 1.7608818 & 10.19 & 0.0349 & 19.25 & 0.0679 & 0.0032 & \nodata \\ 
64049 &  149.1378657 & 2.7037899 & 10.83 & 0.1283 & 21.36 & 0.0958 & 0.0032 & \nodata \\  
64050 &  149.0904962 & 2.640145 & 10.91 & 0.1299 & 21.61 & 0.1271 & 0.0048 & \nodata \\ 
64059 &  148.5897165 & 1.9534373 & 10.49 & 0.1 & 21.46 & 0.1371 & 0.0055 & \nodata \\  
64129 &  149.7027968 & 2.8786719 & 10.53 & 0.0783 & 20.56 & 0.1063 & 0.0046 & $5.75^{+0.02}_{-0.02} $ \\  
64145 &  150.0656545 & 2.9292917 & 10.59 & 0.103 & 20.54 & 0.1228 & 0.0055 & \nodata \\ 
64234 &  149.951175 & 1.0951285 & 10.38 & 0.0323 & 19.02 & 0.0691 & 0.0034 & \nodata \\  
64258 &  149.4748494 & 1.9582107 & 10.74 & 0.1084 & 21.75 & 0.1137 & 0.0038 & \nodata \\ 
64266 &  149.9034921 & 2.9396754 & 10.68 & 0.0805 & 20.95 & 0.3258 & 0.0153 & \nodata \\  
64272 &  149.7438977 & 2.2497568 & 10.89 & 0.1329 & 20.57 & 0.2974 & 0.0143 & $7.87^{+0.05}_{-0.05} $ \\  
64286 &  150.5293341 & 3.0577013 & 10.67 & 0.0234 & 18.25 & 0.1106 & 0.0059 & $7.35^{+0.03}_{-0.03} $\\ 
83185 &  150.0678076 & 3.3380463 & 10.63 & 0.1033 & 21.45 & 0.0968 & 0.0031 & \nodata \\   
83201 &  149.8727195 & 3.1837508 & 11.02 & 0.1244 & 18.82 & 0.2395 & 0.0127 & $7.18^{+0.03}_{-0.03} $ \\ 
83204 &  149.8440779 & 3.1908699 & 10.9 & 0.1246 & 21.7 & 0.1145 & 0.0039 & \nodata \\  
83262 &  149.2909323 & 2.5633208 & 10.8 & 0.1263 & 21.2 & 0.1033 & 0.0039 & $6.22^{+0.55}_{-0.45} $ \\  
97153 &  241.366954 & 54.9073605 & 10.35 & 0.0624 & 21.87 & 0.1179 & 0.0039 & \nodata \\  
97156 &  241.3026668 & 54.9307658 & 10.26 & 0.0642 & 21.83 & 0.1118 & 0.0036 & \nodata \\  
97781 &  241.9093182 & 53.7136762 & 10.39 & 0.0661 & 21.87 & 0.1084 & 0.0033 & \nodata \\
205160 & 334.1350066 & 0.4734242 & 10.52 & 0.1396 & 21.44 & 0.0981 & 0.0032 & \nodata \\  
208465 & 212.9559958 & 52.8167127 & 11.61 & 0.0765 & 19.71 & 0.105 & 0.005 & $6.28^{+0.54}_{-1.17} $ \\  
208625 & 213.7651102 & 52.0761692 & 10.85 & 0.0732 & 21.42 & 0.1388 & 0.0056 & \nodata \\  
208662 & 213.9684008 & 52.0524492 & 10.98 & 0.1122 & 19.87 & 0.076 & 0.0034 & $5.93^{+0.45}_{-1.39} $\\  
208702 & 214.8480654 & 52.0831737 & 9.61 & 0.0448 & 21.31 & 0.1093 & 0.0041 & \nodata \\  
217051 & 333.8808418 & 1.0297436 & 10.77 & 0.0614 & 21.04 & 0.0989 & 0.0038 & \nodata \\ 
259478 & 212.194614 & 53.2740888 & 10.84 & 0.0784 & 19.15 & 0.0726 & 0.0035 & \nodata \\  
259789 & 214.6720162 & 52.871023 & 10.76 & 0.1063 & 21.05 & 0.0969 & 0.0036 & \nodata \\  
259790 & 214.7703074 & 53.047393 & 11.38 & 0.0818 & 21.35 & 0.1091 & 0.004 & \nodata \\ 
259880 & 212.9356623 & 54.1584797 & 10.65 & 0.0762 & 21.46 & 0.1292 & 0.0051 & \nodata \\  
259895 & 212.1853626 & 53.8578921 & 10.11 & 0.0834 & 20.45 & 0.1349 & 0.0062 & $6.50^{+0.02}_{-0.02} $ \\ 
259919 & 211.9709118 & 53.6142803 & 10.52 & 0.0781 & 20.67 & 0.107 & 0.0046 & \nodata \\  
260221 & 214.2461774 & 52.7304623 & 10.73 & 0.0742 & 21.43 & 0.1039 & 0.0036 & \nodata \\  
260241 & 214.0785351 & 53.8415265 & 10.51 & 0.0743 & 21.03 & 0.1083 & 0.0043 & \nodata \\  
583470 & 332.4441817 & -0.6396369 & 10.08 & 0.0889 & 20.82 & 0.2173 & 0.0101 & $6.37^{+0.01}_{-0.01} $ \\  
613397 & 243.2130022 & 54.4802705 & 11.54 & 0.1274 & 21.28 & 0.1146 & 0.0044 & $5.92^{+0.57}_{-1.41} $ \\  
631480 & 214.2188319 & 52.5807769 & 7.74 & 0.0442 & 21.49 & 0.0988 & 0.0032 & \nodata \\  
631492 & 214.3279483 & 52.5141623 & 6.89 & 0.0119 & 21.44 & 0.1192 & 0.0045 & \nodata \\  
631637 & 215.6606315 & 54.2344583 & 8.19 & 0.0206 & 19.7 & 0.075 & 0.0034 & \nodata \\ 
649438 & 149.3216589 & 2.1478328 & 9.8 & 0.0862 & 20.03 & 0.1011 & 0.0047 & \nodata \\  
\enddata
\tablecomments{Table displaying the characteristic quantities for the selected variable population. No BH mass estimates were given for those spectrum with H$\alpha$ absorption features, poor quality spectrum, or those lacking any broadline H$\alpha$ to be fit. Equation \ref{eq:Mass_BH} has an intrinsic {$\sim$}0.4 dex uncertainty \citep{Baldassare_2015ApJ...809L..14B, Burke21_MNRAS}. The quoted uncertainties are derived from the bootstrapping method.}

\end{deluxetable*}

%% file: table2.tex
\startlongtable
\begin{deluxetable*}{rrrrrrrrr}
\tablecolumns{9}
\tablecaption{Object Classifications.\label{tab:class_table}}
    \tablehead{\colhead{NSA ID} & \colhead{BPT Class} & \colhead{[SII]--[OIII]} & \colhead{[OI]--[OIII]} & \colhead{Stern IR} & \colhead{Jarrett IR} & \colhead{$\%_{\text{AGN}}$} & \colhead{$f_{\text{AGN}}$} & \colhead{$E(B-V)$} }
\startdata
1382 & Absorption & \nodata & \nodata & \nodata & \nodata & 100 & $0.92 \pm 0.02$  & $0.0413 \pm 0.0193$ \\ 
28585 & Starburst & Starburst & Starburst & \nodata & \nodata & 99.8 & $0.40 \pm 0.10$ & $0.0 \pm 0.0486$ \\ 
28616 & AGN & Seyfert & Seyfert & AGN & AGN & 100 & $0.90 \pm 0.01$ & $0.1512 \pm 0.0015$\\
28762 & Absorption & Seyfert & LINER & \nodata & \nodata & 65.2 & $0.0 \pm 0.11$ & $0.0 \pm 0.6080$ \\ 
28810 & \nodata & \nodata & \nodata & \nodata & \nodata & 97.8 & $0.96 \pm 0.09$ & $0.0260 \pm 0.0003$ \\ 
58753 & Absorption & \nodata & \nodata & \nodata & \nodata & 25.7 & $0.0 \pm 0.06$ & $0.0 \pm 0.0$ \\ 
58912 & Absorption & Starburst & LINER & \nodata & \nodata & 99.9 & $0.88 \pm 0.016$ & $0.0 \pm 0.0172$ \\ 
58917 & Starburst & Starburst & Starburst & \nodata & \nodata & 3.1 & $0.17 \pm 0.12$ & $0.2548 \pm 0.0386$\\ 
58935 & Absorption & \nodata & \nodata & \nodata & \nodata & 98.4 & $0.79 \pm 0.11$ & $0.2162 \pm 0.0730$ \\ 
63970 & Starburst & Seyfert & Starburst & \nodata & \nodata & 99.6 & $0.97 \pm 0.04$ & $0.0259 \pm 0.2136$ \\ 
63979 & Composite & Starburst & Starburst & \nodata & \nodata & 0 & $0.0 \pm 0.0 $ & $0.0 \pm 0.0$ \\ 
64049 & \nodata & \nodata & \nodata & \nodata & \nodata & 100 & $0.62 \pm 0.02$ & $0.2981 \pm 0.0545$ \\ 
64050 & Absorption & \nodata & \nodata & \nodata & \nodata & 8.6 & $0.0 \pm 0.02$ & $0.0 \pm 0.0$ \\ 
64059 & AGN & Seyfert & Starburst & \nodata & \nodata & 100 & $0.49 \pm 0.04$ & $0.0778 \pm 0.0008$ \\ 
64129 & AGN & Starburst & Seyfert & \nodata & \nodata & 100 & $0.88 \pm 0.02$ & $0.1512 \pm 0.0517$ \\ 
64145 & Absorption & \nodata & \nodata & \nodata & \nodata & 100 & $0.90 \pm 0.02$ & $0.1512 \pm 0.0580$ \\ 
64234 & Composite & Starburst & Starburst & \nodata & \nodata & 49.8 & $0.13 \pm 0.02$ & $0.2548 \pm 0.0026$ \\ 
64258 & Starburst & \nodata & \nodata & \nodata & \nodata & 100 & $0.99 \pm 0.01$ & $0.2548 \pm 0.0026$ \\ 
64266 & None & \nodata & \nodata & \nodata & \nodata & 45.1 & $0.0 \pm 0.09$ & $0.0 \pm 21.9349$ \\ 
64272 & AGN & LINER & LINER & \nodata & \nodata & 100 & $0.99 \pm 0.01$ & $0.1239 \pm 0.0012$ \\ 
64286 & AGN & Seyfert & Seyfert & \nodata & \nodata & 100 & $0.47 \pm 0.05$ & $0.0259 \pm 0.8638$ \\ 
83185 & None & \nodata & \nodata & \nodata & \nodata & 89.9 & $0.72 \pm 0.12$ & $0.4012 \pm 0.0486$\\ 
83201 & AGN & Seyfert & Seyfert & \nodata & \nodata & 100 & $0.81 \pm 0.02$ & $0.0  \pm 0.0$\\ 
83204 & AGN & LINER & Starburst & \nodata & \nodata & 94.1 & $0.96 \pm 0.08$ & $0.0259 \pm 0.0172$ \\ 
83262 & None & \nodata & \nodata & \nodata & \nodata & 0 & $0.0\pm 0.0$ & $0.0 \pm 0.0$ \\ 
97153 & Absorption & Starburst & LINER & \nodata & \nodata & 11.4 & $0.0 \pm 0.05$ & $0.0 \pm 0.0$ \\ 
97156 & Absorption & \nodata & \nodata & \nodata & \nodata & 51.9 & $0.0 \pm 0.11$ & $0.0 \pm 0.0$ \\ 
97781 & Starburst & Starburst & \nodata & \nodata & \nodata & 100 & $0.91 \pm 0.01$ & $0.0122 \pm 0.0137$\\ 
205160 & None & \nodata & \nodata & \nodata & \nodata & 99.9 & $0.85 \pm 0.17$ & $0.0 \pm 0.0654$\\ 
208465 & Absorption & \nodata & \nodata & \nodata & \nodata & 100 & $0.95 \pm 0.01$ & $0.1239 \pm 0.0012$ \\ 
208625 & None & \nodata & \nodata & \nodata & \nodata & 100 & $0.94 \pm 0.01$ & $0.0585 \pm 0.0006$ \\ 
208662 & Composite & Starburst & \nodata & \nodata & \nodata & 100 & $0.29 \pm 0.01$ & $0.0 \pm 0.0$ \\ 
208702 & Composite & Starburst & Starburst & \nodata & \nodata & 31.3 & $0.0 \pm 0.24$ & $0.0 \pm 0.0$ \\ 
217051 & Absorption & \nodata & \nodata & \nodata & \nodata & 100 & $0.93 \pm 0.01$ & $0.0 \pm 0.0154$ \\ 
259478 & Composite & Starburst & Seyfert & \nodata & \nodata & 0.3 & $0.0 \pm 0.01$ & $0.0 \pm 0.6862$\\ 
259789 & None & \nodata & \nodata & \nodata & \nodata & 87.5 & $0.47 \pm 0.23$ & $0.6079 \pm 0.1456$ \\ 
259790 & Absorption & Starburst & Starburst & \nodata & \nodata & 100 & $0.89 \pm 0.03$ & $0.2162 \pm 0.0386$ \\ 
259880 & Absorption & \nodata & \nodata & \nodata & \nodata & 7.8 & $0.0 \pm 0.07$ & $0.0 \pm 6.8213$\\ 
259895 & AGN & Seyfert & Seyfert & \nodata &  \nodata& 100 & $0.78 \pm 0.01$ & $0.2981 \pm 0.2194$ \\ 
259919 & Absorption & \nodata & Starburst & \nodata & \nodata & 70.9 & $0.89 \pm 0.05$ & $50.0187 \pm 0.5002$\\ 
260221 & Absorption & Starburst & \nodata & \nodata & \nodata & 92.7 & $0.87 \pm 0.05$ &$0.0 \pm 0.0$ \\ 
260241 & None & \nodata & \nodata & \nodata & \nodata & 99.9 & $0.94 \pm 0.07$ & $0.0 \pm 0.0$ \\ 
583470 & Starburst & Starburst & Seyfert & \nodata & \nodata & 100 & $0.88 \pm 0.02$ & $0.0 \pm 0.0122$ \\ 
613397 & Absorption & \nodata & Starburst & \nodata & \nodata & 100 & $0.90 \pm 0.01$ & $0.0585 \pm 0.0006$\\ 
631480 & \nodata & \nodata & \nodata & \nodata & \nodata & 100 & $0.99 \pm 0.04$ & $0.0 \pm 0.0$ \\ 
631492 & \nodata & \nodata & \nodata & \nodata & \nodata & 100 & $0.99 \pm 0.01$ & $2.4119 \pm 0.0241$  \\ 
631637 & \nodata & \nodata & \nodata & AGN & \nodata & 100 & $0.93 \pm 0.01$ & $22.2872 \pm 0.2229$ \\ 
649438 & \nodata & \nodata & \nodata & \nodata & \nodata & 6.2 & $0.0 \pm 0.02$ & $0.0 \pm 10.3080$ \\ 
\enddata

\tablecomments{Table of classifications for the spectroscopic, WISE IR, and SED modeling Analysis. A BPT Class of "Absorption" was given for those galaxies not shown on the standard BPT diagram who contain absorption features, a signature of older stellar populations. A BPT Class of "None" indicates a lack of emission lines for use. The BPT Class column is blank where there was no available spectrum. Column \%$_{\text{AGN}}$ depicts the percentage of realizations which contain AGN contribution to their SED, while $f_{\text{AGN}}$ represents the fractional AGN contribution at 15 um.}

\end{deluxetable*}

%% file: MAIN.bbl
\begin{thebibliography}{}
\expandafter\ifx\csname natexlab\endcsname\relax\def\natexlab#1{#1}\fi
\providecommand{\url}[1]{\href{#1}{#1}}
\providecommand{\dodoi}[1]{doi:~\href{http://doi.org/#1}{\nolinkurl{#1}}}
\providecommand{\doeprint}[1]{\href{http://ascl.net/#1}{\nolinkurl{http://ascl.net/#1}}}
\providecommand{\doarXiv}[1]{\href{https://arxiv.org/abs/#1}{\nolinkurl{https://arxiv.org/abs/#1}}}

\bibitem[{{Abazajian} {et~al.}(2009){Abazajian}, {Adelman-McCarthy},
  {Ag{\"u}eros}, {Allam}, {Allende Prieto}, {An}, {Anderson}, {Anderson},
  {Annis}, {Bahcall}, {Bailer-Jones}, {Barentine}, {Bassett}, {Becker},
  {Beers}, {Bell}, {Belokurov}, {Berlind}, {Berman}, {Bernardi}, {Bickerton},
  {Bizyaev}, {Blakeslee}, {Blanton}, {Bochanski}, {Boroski}, {Brewington},
  {Brinchmann}, {Brinkmann}, {Brunner}, {Budav{\'a}ri}, {Carey}, {Carliles},
  {Carr}, {Castander}, {Cinabro}, {Connolly}, {Csabai}, {Cunha}, {Czarapata},
  {Davenport}, {de Haas}, {Dilday}, {Doi}, {Eisenstein}, {Evans}, {Evans},
  {Fan}, {Friedman}, {Frieman}, {Fukugita}, {G{\"a}nsicke}, {Gates},
  {Gillespie}, {Gilmore}, {Gonzalez}, {Gonzalez}, {Grebel}, {Gunn},
  {Gy{\"o}ry}, {Hall}, {Harding}, {Harris}, {Harvanek}, {Hawley}, {Hayes},
  {Heckman}, {Hendry}, {Hennessy}, {Hindsley}, {Hoblitt}, {Hogan}, {Hogg},
  {Holtzman}, {Hyde}, {Ichikawa}, {Ichikawa}, {Im}, {Ivezi{\'c}}, {Jester},
  {Jiang}, {Johnson}, {Jorgensen}, {Juri{\'c}}, {Kent}, {Kessler}, {Kleinman},
  {Knapp}, {Konishi}, {Kron}, {Krzesinski}, {Kuropatkin}, {Lampeitl},
  {Lebedeva}, {Lee}, {Lee}, {French Leger}, {L{\'e}pine}, {Li}, {Lima}, {Lin},
  {Long}, {Loomis}, {Loveday}, {Lupton}, {Magnier}, {Malanushenko},
  {Malanushenko}, {Mandelbaum}, {Margon}, {Marriner}, {Mart{\'\i}nez-Delgado},
  {Matsubara}, {McGehee}, {McKay}, {Meiksin}, {Morrison}, {Mullally}, {Munn},
  {Murphy}, {Nash}, {Nebot}, {Neilsen}, {Newberg}, {Newman}, {Nichol},
  {Nicinski}, {Nieto-Santisteban}, {Nitta}, {Okamura}, {Oravetz}, {Ostriker},
  {Owen}, {Padmanabhan}, {Pan}, {Park}, {Pauls}, {Peoples}, {Percival}, {Pier},
  {Pope}, {Pourbaix}, {Price}, {Purger}, {Quinn}, {Raddick}, {Re Fiorentin},
  {Richards}, {Richmond}, {Riess}, {Rix}, {Rockosi}, {Sako}, {Schlegel},
  {Schneider}, {Scholz}, {Schreiber}, {Schwope}, {Seljak}, {Sesar}, {Sheldon},
  {Shimasaku}, {Sibley}, {Simmons}, {Sivarani}, {Allyn Smith}, {Smith},
  {Smol{\v{c}}i{\'c}}, {Snedden}, {Stebbins}, {Steinmetz}, {Stoughton},
  {Strauss}, {SubbaRao}, {Suto}, {Szalay}, {Szapudi}, {Szkody}, {Tanaka},
  {Tegmark}, {Teodoro}, {Thakar}, {Tremonti}, {Tucker}, {Uomoto}, {Vanden
  Berk}, {Vandenberg}, {Vidrih}, {Vogeley}, {Voges}, {Vogt}, {Wadadekar},
  {Watters}, {Weinberg}, {West}, {White}, {Wilhite}, {Wonders}, {Yanny},
  {Yocum}, {York}, {Zehavi}, {Zibetti}, \& {Zucker}}]{abzajian2009}
{Abazajian}, K.~N., {Adelman-McCarthy}, J.~K., {Ag{\"u}eros}, M.~A., {et~al.}
  2009, \apjs, 182, 543, \dodoi{10.1088/0067-0049/182/2/543}

\bibitem[{{Ahumada} {et~al.}(2020){Ahumada}, {Prieto}, {Almeida}, {Anders},
  {Anderson}, {Andrews}, {Anguiano}, {Arcodia}, {Armengaud}, {Aubert}, {Avila},
  {Avila-Reese}, {Badenes}, {Balland}, {Barger}, {Barrera-Ballesteros}, {Basu},
  {Bautista}, {Beaton}, {Beers}, {Benavides}, {Bender}, {Bernardi}, {Bershady},
  {Beutler}, {Bidin}, {Bird}, {Bizyaev}, {Blanc}, {Blanton}, {Boquien},
  {Borissova}, {Bovy}, {Brandt}, {Brinkmann}, {Brownstein}, {Bundy}, {Bureau},
  {Burgasser}, {Burtin}, {Cano-D{\'\i}az}, {Capasso}, {Cappellari}, {Carrera},
  {Chabanier}, {Chaplin}, {Chapman}, {Cherinka}, {Chiappini}, {Doohyun Choi},
  {Chojnowski}, {Chung}, {Clerc}, {Coffey}, {Comerford}, {Comparat}, {da
  Costa}, {Cousinou}, {Covey}, {Crane}, {Cunha}, {Ilha}, {Dai}, {Damsted},
  {Darling}, {Davidson}, {Davies}, {Dawson}, {De}, {de la Macorra}, {De Lee},
  {Queiroz}, {Deconto Machado}, {de la Torre}, {Dell'Agli}, {du Mas des
  Bourboux}, {Diamond-Stanic}, {Dillon}, {Donor}, {Drory}, {Duckworth},
  {Dwelly}, {Ebelke}, {Eftekharzadeh}, {Davis Eigenbrot}, {Elsworth},
  {Eracleous}, {Erfanianfar}, {Escoffier}, {Fan}, {Farr},
  {Fern{\'a}ndez-Trincado}, {Feuillet}, {Finoguenov}, {Fofie},
  {Fraser-McKelvie}, {Frinchaboy}, {Fromenteau}, {Fu}, {Galbany}, {Garcia},
  {Garc{\'\i}a-Hern{\'a}ndez}, {Oehmichen}, {Ge}, {Maia}, {Geisler}, {Gelfand},
  {Goddy}, {Gonzalez-Perez}, {Grabowski}, {Green}, {Grier}, {Guo}, {Guy},
  {Harding}, {Hasselquist}, {Hawken}, {Hayes}, {Hearty}, {Hekker}, {Hogg},
  {Holtzman}, {Horta}, {Hou}, {Hsieh}, {Huber}, {Hunt}, {Chitham}, {Imig},
  {Jaber}, {Angel}, {Johnson}, {Jones}, {J{\"o}nsson}, {Jullo}, {Kim},
  {Kinemuchi}, {Kirkpatrick}, {Kite}, {Klaene}, {Kneib}, {Kollmeier}, {Kong},
  {Kounkel}, {Krishnarao}, {Lacerna}, {Lan}, {Lane}, {Law}, {Le Goff}, {Leung},
  {Lewis}, {Li}, {Lian}, {Lin}, {Long}, {Longa-Pe{\~n}a}, {Lundgren}, {Lyke},
  {Ted Mackereth}, {MacLeod}, {Majewski}, {Manchado}, {Maraston}, {Martini},
  {Masseron}, {Masters}, {Mathur}, {McDermid}, {Merloni}, {Merrifield},
  {M{\'e}sz{\'a}ros}, {Miglio}, {Minniti}, {Minsley}, {Miyaji}, {Mohammad},
  {Mosser}, {Mueller}, {Muna}, {Mu{\~n}oz-Guti{\'e}rrez}, {Myers}, {Nadathur},
  {Nair}, {Nandra}, {do Nascimento}, {Nevin}, {Newman}, {Nidever}, {Nitschelm},
  {Noterdaeme}, {O'Connell}, {Olmstead}, {Oravetz}, {Oravetz}, {Osorio},
  {Pace}, {Padilla}, {Palanque-Delabrouille}, {Palicio}, {Pan}, {Pan},
  {Parker}, {Paviot}, {Peirani}, {Ram{\'r}ez}, {Penny}, {Percival},
  {Perez-Fournon}, {P{\'e}rez-R{\`a}fols}, {Petitjean}, {Pieri},
  {Pinsonneault}, {Poovelil}, {Povick}, {Prakash}, {Price-Whelan}, {Raddick},
  {Raichoor}, {Ray}, {Rembold}, {Rezaie}, {Riffel}, {Riffel}, {Rix}, {Robin},
  {Roman-Lopes}, {Rom{\'a}n-Z{\'u}{\~n}iga}, {Rose}, {Ross}, {Rossi},
  {Rowlands}, {Rubin}, {Salvato}, {S{\'a}nchez}, {S{\'a}nchez-Menguiano},
  {S{\'a}nchez-Gallego}, {Sayres}, {Schaefer}, {Schiavon}, {Schimoia},
  {Schlafly}, {Schlegel}, {Schneider}, {Schultheis}, {Schwope}, {Seo},
  {Serenelli}, {Shafieloo}, {Shamsi}, {Shao}, {Shen}, {Shetrone}, {Shirley},
  {Aguirre}, {Simon}, {Skrutskie}, {Slosar}, {Smethurst}, {Sobeck}, {Sodi},
  {Souto}, {Stark}, {Stassun}, {Steinmetz}, {Stello}, {Stermer},
  {Storchi-Bergmann}, {Streblyanska}, {Stringfellow}, {Stutz}, {Su{\'a}rez},
  {Sun}, {Taghizadeh-Popp}, {Talbot}, {Tayar}, {Thakar}, {Theriault}, {Thomas},
  {Thomas}, {Tinker}, {Tojeiro}, {Toledo}, {Tremonti}, {Troup}, {Tuttle},
  {Unda-Sanzana}, {Valentini}, {Vargas-Gonz{\'a}lez}, {Vargas-Maga{\~n}a},
  {V{\'a}zquez-Mata}, {Vivek}, {Wake}, {Wang}, {Weaver}, {Weijmans}, {Wild},
  {Wilson}, {Wilson}, {Wolthuis}, {Wood-Vasey}, {Yan}, {Yang}, {Y{\`e}che},
  {Zamora}, {Zarrouk}, {Zasowski}, {Zhang}, {Zhao}, {Zhao}, {Zheng}, {Zheng},
  {Zhu}, \& {Zou}}]{SDSSDR16paper}
{Ahumada}, R., {Prieto}, C.~A., {Almeida}, A., {et~al.} 2020, \apjs, 249, 3,
  \dodoi{10.3847/1538-4365/ab929e}

\bibitem[{Assef {et~al.}(2010)Assef, Kochanek, Brodwin, Cool, Forman, Gonzalez,
  Hickox, Jones, Floc{\textquotesingle}h, Moustakas, Murray, \&
  Stern}]{Assef_2010}
Assef, R.~J., Kochanek, C.~S., Brodwin, M., {et~al.} 2010, The Astrophysical
  Journal, 713, 970, \dodoi{10.1088/0004-637x/713/2/970}

\bibitem[{Bai {et~al.}(2018)Bai, Liu, Wicker, Wang, Guo, Qin, He, Wang, Wu,
  Dong, Zhang, Hou, Wang, \& Cao}]{Bai_2018}
Bai, Y., Liu, J., Wicker, J., {et~al.} 2018, The Astrophysical Journal
  Supplement Series, 235, 16, \dodoi{10.3847/1538-4365/aaaab9}

\bibitem[{{Baldassare} {et~al.}(2018){Baldassare}, {Geha}, \&
  {Greene}}]{Baldassare2018ApJ..868..152}
{Baldassare}, V.~F., {Geha}, M., \& {Greene}, J. 2018, ApJ, 868, 152.
\newblock \doarXiv{1808.09578}

\bibitem[{{Baldassare} {et~al.}(2020){Baldassare}, {Geha}, \&
  {Greene}}]{Baldassare2020ApJ...896...10B}
---. 2020, \apj, 896, 10, \dodoi{10.3847/1538-4357/ab8936}

\bibitem[{{Baldassare} {et~al.}(2015){Baldassare}, {Reines}, {Gallo}, \&
  {Greene}}]{Baldassare_2015ApJ...809L..14B}
{Baldassare}, V.~F., {Reines}, A.~E., {Gallo}, E., \& {Greene}, J.~E. 2015,
  \apjl, 809, L14, \dodoi{10.1088/2041-8205/809/1/L14}

\bibitem[{{Baldwin} {et~al.}(1981){Baldwin}, {Phillips}, \&
  {Terlevich}}]{1981_BPT_Paper}
{Baldwin}, J.~A., {Phillips}, M.~M., \& {Terlevich}, R. 1981, \pasp, 93, 5,
  \dodoi{10.1086/130766}

\bibitem[{{Blanton} \& {Moustakas}(2009)}]{2009ARA&A..47..159B}
{Blanton}, M.~R., \& {Moustakas}, J. 2009, \araa, 47, 159,
  \dodoi{10.1146/annurev-astro-082708-101734}

\bibitem[{Burke {et~al.}(2021)Burke, Liu, Chen, Shen, \& Guo}]{Burke21_MNRAS}
Burke, C.~J., Liu, X., Chen, Y.-C., Shen, Y., \& Guo, H. 2021, Monthly Notices
  of the Royal Astronomical Society, 504, 543, \dodoi{10.1093/mnras/stab912}

\bibitem[{Cann {et~al.}(2019)Cann, Satyapal, Abel, Blecha, Mushotzky, Reynolds,
  \& Secrest}]{Cann_2019}
Cann, J.~M., Satyapal, S., Abel, N.~P., {et~al.} 2019, The Astrophysical
  Journal, 870, L2, \dodoi{10.3847/2041-8213/aaf88d}

\bibitem[{{Caplar} {et~al.}(2017){Caplar}, {Lilly}, \&
  {Trakhtenbrot}}]{Caplar2017ApJ...834..111C}
{Caplar}, N., {Lilly}, S.~J., \& {Trakhtenbrot}, B. 2017, \apj, 834, 111,
  \dodoi{10.3847/1538-4357/834/2/111}

\bibitem[{Carroll {et~al.}(2021)Carroll, Hickox, Masini, Lanz, Assef, Stern,
  Chen, \& Ananna}]{Carroll_2021}
Carroll, C.~M., Hickox, R.~C., Masini, A., {et~al.} 2021, The Astrophysical
  Journal, 908, 185, \dodoi{10.3847/1538-4357/abd185}

\bibitem[{Chakravorty {et~al.}(2013)Chakravorty, Elvis, \&
  Ferland}]{Chakravorty_et_al_2014}
Chakravorty, S., Elvis, M., \& Ferland, G. 2013, Monthly Notices of the Royal
  Astronomical Society, 437, 740, \dodoi{10.1093/mnras/stt1930}

\bibitem[{{Chambers} {et~al.}(2016){Chambers}, {Magnier}, {Metcalfe},
  {Flewelling}, {Huber}, {Waters}, {Denneau}, {Draper}, {Farrow}, {Finkbeiner},
  {Holmberg}, {Koppenhoefer}, {Price}, {Saglia}, {Schlafly}, {Smartt},
  {Sweeney}, {Wainscoat}, {Burgett}, {Grav}, {Heasley}, {Hodapp}, {Jedicke},
  {Kaiser}, {Kudritzki}, {Luppino}, {Lupton}, {Monet}, {Morgan}, {Onaka},
  {Stubbs}, {Tonry}, {Banados}, {Bell}, {Bender}, {Bernard}, {Botticella},
  {Casertano}, {Chastel}, {Chen}, {Chen}, {Cole}, {Deacon}, {Frenk},
  {Fitzsimmons}, {Gezari}, {Goessl}, {Goggia}, {Goldman}, {Grebel}, {Hambly},
  {Hasinger}, {Heavens}, {Heckman}, {Henderson}, {Henning}, {Holman}, {Hopp},
  {Ip}, {Isani}, {Keyes}, {Koekemoer}, {Kotak}, {Long}, {Lucey}, {Liu},
  {Martin}, {McLean}, {Morganson}, {Murphy}, {Nieto-Santisteban}, {Norberg},
  {Peacock}, {Pier}, {Postman}, {Primak}, {Rae}, {Rest}, {Riess}, {Riffeser},
  {Rix}, {Roser}, {Schilbach}, {Schultz}, {Scolnic}, {Szalay}, {Seitz},
  {Shiao}, {Small}, {Smith}, {Soderblom}, {Taylor}, {Thakar}, {Thiel},
  {Thilker}, {Urata}, {Valenti}, {Walter}, {Watters}, {Werner}, {White},
  {Wood-Vasey}, \& {Wyse}}]{Chambers2016arXiv161205560C}
{Chambers}, K.~C., {Magnier}, E.~A., {Metcalfe}, N., {et~al.} 2016, ArXiv
  e-prints.
\newblock \doarXiv{1612.05560}

\bibitem[{{Chauvenet}(1960)}]{1960mspa.book.....C}
{Chauvenet}, W. 1960, {A Manual of spherical and practical astronomy - Vol.1:
  Spherical astronomy; Vol.2: Theory and use of astronomical instruments.
  Method of least squares}

\bibitem[{{Desroches} {et~al.}(2009){Desroches}, {Greene}, \&
  {Ho}}]{Desriches_X-Ray_IMBH_2009ApJ...698.1515D}
{Desroches}, L.-B., {Greene}, J.~E., \& {Ho}, L.~C. 2009, \apj, 698, 1515,
  \dodoi{10.1088/0004-637X/698/2/1515}

\bibitem[{Dong {et~al.}(2012)Dong, Ho, Yuan, Wang, Fan, Zhou, \&
  Jiang}]{Dong_2012}
Dong, X.-B., Ho, L.~C., Yuan, W., {et~al.} 2012, The Astrophysical Journal,
  755, 167, \dodoi{10.1088/0004-637x/755/2/167}

\bibitem[{{Edelson} {et~al.}(1990){Edelson}, {Krolik}, \&
  {Pike}}]{1990ApJ...359...86E}
{Edelson}, R.~A., {Krolik}, J.~H., \& {Pike}, G.~F. 1990, \apj, 359, 86,
  \dodoi{10.1086/169036}

\bibitem[{{Ferrarese} \& {Merritt}(2000)}]{2000ApJ...539L...9F}
{Ferrarese}, L., \& {Merritt}, D. 2000, \apjl, 539, L9, \dodoi{10.1086/312838}

\bibitem[{{Gebhardt} {et~al.}(2000){Gebhardt}, {Bender}, {Bower}, {Dressler},
  {Faber}, {Filippenko}, {Green}, {Grillmair}, {Ho}, {Kormendy}, {Lauer},
  {Magorrian}, {Pinkney}, {Richstone}, \& {Tremaine}}]{2000ApJ...539L..13G}
{Gebhardt}, K., {Bender}, R., {Bower}, G., {et~al.} 2000, \apjl, 539, L13,
  \dodoi{10.1086/312840}

\bibitem[{{Geha} {et~al.}(2003){Geha}, {Alcock}, {Allsman}, {Alves}, {Axelrod},
  {Becker}, {Bennett}, {Cook}, {Drake}, {Freeman}, {Griest}, {Keller},
  {Lehner}, {Marshall}, {Minniti}, {Nelson}, {Peterson}, {Popowski}, {Pratt},
  {Quinn}, {Stubbs}, {Sutherland}, {Tomaney}, {Vandehei}, \&
  {Welch}}]{Geha2003AJ....125....1G}
{Geha}, M., {Alcock}, C., {Allsman}, R.~A., {et~al.} 2003, \aj, 125, 1,
  \dodoi{10.1086/344947}

\bibitem[{Gezari {et~al.}(2013)Gezari, Martin, Forster, Neill, Huber, Heckman,
  Bianchi, Morrissey, Neff, Seibert, Schiminovich, Wyder, Burgett, Chambers,
  Kaiser, Magnier, Price, \& Tonry}]{Gezari_2013}
Gezari, S., Martin, D.~C., Forster, K., {et~al.} 2013, The Astrophysical
  Journal, 766, 60, \dodoi{10.1088/0004-637x/766/1/60}

\bibitem[{{Graham} {et~al.}(2019){Graham}, {Kulkarni}, {Bellm}, {Adams},
  {Barbarino}, {Blagorodnova}, {Bodewits}, {Bolin}, {Brady}, {Cenko}, {Chang},
  {Coughlin}, {De}, {Eadie}, {Farnham}, {Feindt}, {Franckowiak}, {Fremling},
  {Gezari}, {Ghosh}, {Goldstein}, {Golkhou}, {Goobar}, {Ho}, {Huppenkothen},
  {Ivezi{\'c}}, {Jones}, {Juric}, {Kaplan}, {Kasliwal}, {Kelley}, {Kupfer},
  {Lee}, {Lin}, {Lunnan}, {Mahabal}, {Miller}, {Ngeow}, {Nugent}, {Ofek},
  {Prince}, {Rauch}, {van Roestel}, {Schulze}, {Singer}, {Sollerman}, {Taddia},
  {Yan}, {Ye}, {Yu}, {Barlow}, {Bauer}, {Beck}, {Belicki}, {Biswas}, {Brinnel},
  {Brooke}, {Bue}, {Bulla}, {Burruss}, {Connolly}, {Cromer}, {Cunningham},
  {Dekany}, {Delacroix}, {Desai}, {Duev}, {Feeney}, {Flynn}, {Frederick},
  {Gal-Yam}, {Giomi}, {Groom}, {Hacopians}, {Hale}, {Helou}, {Henning},
  {Hover}, {Hillenbrand}, {Howell}, {Hung}, {Imel}, {Ip}, {Jackson}, {Kaspi},
  {Kaye}, {Kowalski}, {Kramer}, {Kuhn}, {Landry}, {Laher}, {Mao}, {Masci},
  {Monkewitz}, {Murphy}, {Nordin}, {Patterson}, {Penprase}, {Porter},
  {Rebbapragada}, {Reiley}, {Riddle}, {Rigault}, {Rodriguez}, {Rusholme}, {van
  Santen}, {Shupe}, {Smith}, {Soumagnac}, {Stein}, {Surace}, {Szkody}, {Terek},
  {Van Sistine}, {van Velzen}, {Vestrand}, {Walters}, {Ward}, {Zhang}, \&
  {Zolkower}}]{Graham2019PASP..131g8001G}
{Graham}, M.~J., {Kulkarni}, S.~R., {Bellm}, E.~C., {et~al.} 2019, \pasp, 131,
  078001, \dodoi{10.1088/1538-3873/ab006c}

\bibitem[{Greene \& Ho(2004)}]{Greene_2004}
Greene, J.~E., \& Ho, L.~C. 2004, The Astrophysical Journal, 610, 722,
  \dodoi{10.1086/421719}

\bibitem[{{Greene} \& {Ho}(2005)}]{greene2005ApJ...630..122G}
{Greene}, J.~E., \& {Ho}, L.~C. 2005, \apj, 630, 122, \dodoi{10.1086/431897}

\bibitem[{{Greene} \& {Ho}(2007)}]{Greene_X-ray_IMBH_2007ApJ...656...84G}
---. 2007, \apj, 656, 84, \dodoi{10.1086/509064}

\bibitem[{Greene {et~al.}(2020)Greene, Strader, \&
  Ho}]{IMBH_Handbook_Greene_2020}
Greene, J.~E., Strader, J., \& Ho, L.~C. 2020, Annual Review of Astronomy and
  Astrophysics, 58, 257, \dodoi{10.1146/annurev-astro-032620-021835}

\bibitem[{{Groves} {et~al.}(2006){Groves}, {Heckman}, \&
  {Kauffmann}}]{groves2006MNRAS.371.1559G}
{Groves}, B.~A., {Heckman}, T.~M., \& {Kauffmann}, G. 2006, \mnras, 371, 1559,
  \dodoi{10.1111/j.1365-2966.2006.10812.x}

\bibitem[{Hainline {et~al.}(2016)Hainline, Reines, Greene, \&
  Stern}]{Hainline_2016}
Hainline, K.~N., Reines, A.~E., Greene, J.~E., \& Stern, D. 2016, The
  Astrophysical Journal, 832, 119, \dodoi{10.3847/0004-637x/832/2/119}

\bibitem[{Ho {et~al.}(1997)Ho, Filippenko, \& Sargent}]{Ho_1997}
Ho, L.~C., Filippenko, A.~V., \& Sargent, W. L.~W. 1997, The Astrophysical
  Journal Supplement Series, 112, 315, \dodoi{10.1086/313041}

\bibitem[{Hung {et~al.}(2016)Hung, Gezari, Jones, Kirshner, Chornock, Berger,
  Rest, Huber, Narayan, Scolnic, Waters, Wainscoat, Martin, Forster, \&
  Neill}]{Hung_2016}
Hung, T., Gezari, S., Jones, D.~O., {et~al.} 2016, The Astrophysical Journal,
  833, 226, \dodoi{10.3847/1538-4357/833/2/226}

\bibitem[{Jarrett {et~al.}(2011)Jarrett, Cohen, Masci, Wright, Stern, Benford,
  Blain, Carey, Cutri, Eisenhardt, Lonsdale, Mainzer, Marsh, Padgett, Petty,
  Ressler, Skrutskie, Stanford, Surace, Tsai, Wheelock, \& Yan}]{Jarrett_2011}
Jarrett, T.~H., Cohen, M., Masci, F., {et~al.} 2011, The Astrophysical Journal,
  735, 112, \dodoi{10.1088/0004-637x/735/2/112}

\bibitem[{Kauffmann {et~al.}(2003)Kauffmann, Heckman, Tremonti, Brinchmann,
  Charlot, White, Ridgway, Brinkmann, Fukugita, Hall, Ivezić, Richards, \&
  Schneider}]{Kauffmann_03_Host_Gals_of_AGN}
Kauffmann, G., Heckman, T.~M., Tremonti, C., {et~al.} 2003, Monthly Notices of
  the Royal Astronomical Society, 346, 1055,
  \dodoi{10.1111/j.1365-2966.2003.07154.x}

\bibitem[{Kewley {et~al.}(2001)Kewley, Dopita, Sutherland, Heisler, \&
  Trevena}]{Kewley_2001_Starburst_Galaxies}
Kewley, L., Dopita, M., Sutherland, R., Heisler, C., \& Trevena, J. 2001, The
  Astrophysical Journal, 556, \dodoi{10.1086/321545}

\bibitem[{Kewley {et~al.}(2006)Kewley, Groves, Kauffmann, \&
  Heckman}]{Classify_AGN_Kewley_et_al_06}
Kewley, L.~J., Groves, B., Kauffmann, G., \& Heckman, T. 2006, Monthly Notices
  of the Royal Astronomical Society, 372, 961,
  \dodoi{10.1111/j.1365-2966.2006.10859.x}

\bibitem[{Kirkpatrick {et~al.}(2015)Kirkpatrick, Pope, Sajina, Roebuck, Yan,
  Armus, D{\'{\i}}az-Santos, \& Stierwalt}]{Kirkpatrick_2015}
Kirkpatrick, A., Pope, A., Sajina, A., {et~al.} 2015, The Astrophysical
  Journal, 814, 9, \dodoi{10.1088/0004-637x/814/1/9}

\bibitem[{{Kormendy} \& {Ho}(2013)}]{Kormendy:2013ve}
{Kormendy}, J., \& {Ho}, L.~C. 2013, \araa, 51, 511,
  \dodoi{10.1146/annurev-astro-082708-101811}

\bibitem[{{Lacy} {et~al.}(2004){Lacy}, {Storrie-Lombardi}, {Sajina},
  {Appleton}, {Armus}, {Chapman}, {Choi}, {Fadda}, {Fang}, {Frayer},
  {Heinrichsen}, {Helou}, {Im}, {Marleau}, {Masci}, {Shupe}, {Soifer},
  {Surace}, {Teplitz}, {Wilson}, \& {Yan}}]{Lacy_2004ApJS..154..166L}
{Lacy}, M., {Storrie-Lombardi}, L.~J., {Sajina}, A., {et~al.} 2004, \apjs, 154,
  166, \dodoi{10.1086/422816}

\bibitem[{{Lambrides} {et~al.}(2020){Lambrides}, {Chiaberge}, {Heckman},
  {Gilli}, {Vito}, \& {Norman}}]{Lambrides_2020}
{Lambrides}, E.~L., {Chiaberge}, M., {Heckman}, T., {et~al.} 2020, \apj, 897,
  160, \dodoi{10.3847/1538-4357/ab919c}

\bibitem[{{Lang}(2014)}]{unWISE_Lang2014AJ....147..108L}
{Lang}, D. 2014, \aj, 147, 108, \dodoi{10.1088/0004-6256/147/5/108}

\bibitem[{{Law} {et~al.}(2009){Law}, {Kulkarni}, {Dekany}, {Ofek}, {Quimby},
  {Nugent}, {Surace}, {Grillmair}, {Bloom}, {Kasliwal}, {Bildsten}, {Brown},
  {Cenko}, {Ciardi}, {Croner}, {Djorgovski}, {van Eyken}, {Filippenko}, {Fox},
  {Gal-Yam}, {Hale}, {Hamam}, {Helou}, {Henning}, {Howell}, {Jacobsen},
  {Laher}, {Mattingly}, {McKenna}, {Pickles}, {Poznanski}, {Rahmer}, {Rau},
  {Rosing}, {Shara}, {Smith}, {Starr}, {Sullivan}, {Velur}, {Walters}, \&
  {Zolkower}}]{Law2009PASP..121.1395L}
{Law}, N.~M., {Kulkarni}, S.~R., {Dekany}, R.~G., {et~al.} 2009, \pasp, 121,
  1395, \dodoi{10.1086/648598}

\bibitem[{{Lawrence} {et~al.}(2007){Lawrence}, {Warren}, {Almaini}, {Edge},
  {Hambly}, {Jameson}, {Lucas}, {Casali}, {Adamson}, {Dye}, {Emerson},
  {Foucaud}, {Hewett}, {Hirst}, {Hodgkin}, {Irwin}, {Lodieu}, {McMahon},
  {Simpson}, {Smail}, {Mortlock}, \&
  {Folger}}]{UKIRT_Lawrence2007MNRAS.379.1599L}
{Lawrence}, A., {Warren}, S.~J., {Almaini}, O., {et~al.} 2007, \mnras, 379,
  1599, \dodoi{10.1111/j.1365-2966.2007.12040.x}

\bibitem[{{Liu} {et~al.}(2019){Liu}, {Liu}, {Dong}, {Zhou}, {Wang}, {Lu}, \&
  {Yuan}}]{Liu2019ApJS..243...21L}
{Liu}, H.-Y., {Liu}, W.-J., {Dong}, X.-B., {et~al.} 2019, \apjs, 243, 21,
  \dodoi{10.3847/1538-4365/ab298b}

\bibitem[{Luo {et~al.}(2015)Luo, Zhao, Zhao, Deng, Liu, Jing, Wang, Zhang, Shi,
  Cui, Chu, Li, Bai, Wu, Cai, Cao, Cao, Carlin, Chen, Chen, Chen, Chen, Chen,
  Chen, Chen, Christlieb, Chu, Cui, Dong, Du, Fan, Feng, Fu, Gao, Gong, Gu,
  Guo, Han, He, Hou, Hou, Hou, Hu, Hu, Hu, Huo, Jia, Jiang, Jiang, Jiang, Jin,
  Kong, Kong, Lei, Li, Li, Li, Li, Li, Li, Li, Li, Li, Li, Li, Li, Liang, Lin,
  Liu, Liu, Liu, Liu, Lu, Luo, Mao, Newberg, Ni, Qi, Qi, Shen, Shi, Song, Song,
  Su, Su, Tang, Tao, Tian, Wang, Wang, Wang, Wang, Wang, Wang, Wang, Wang,
  Wang, Wang, Wang, Wang, Wang, Wang, Wang, Wang, Wang, Wang, Wang, Wang, Wei,
  Wei, Wu, Wu, Wu, Wu, Xing, Xu, Xu, Xu, Yan, Yang, Yang, Yang, Yang, Yao, Yu,
  Yuan, Yuan, Yuan, Yuan, Zhai, Zhang, Zhang, Zhang, Zhang, Zhang, Zhang,
  Zhang, Zhang, Zhao, Zhou, Zhou, Zhu, Zhu, Zou, \& Zuo}]{Luo_2015}
Luo, A.-L., Zhao, Y.-H., Zhao, G., {et~al.} 2015, Research in Astronomy and
  Astrophysics, 15, 1095, \dodoi{10.1088/1674-4527/15/8/002}

\bibitem[{{Lyke} {et~al.}(2020){Lyke}, {Higley}, {McLane}, {Schurhammer},
  {Myers}, {Ross}, {Dawson}, {Chabanier}, {Martini}, {Busca}, {Mas des
  Bourboux}, {Salvato}, {Streblyanska}, {Zarrouk}, {Burtin}, {Anderson},
  {Bautista}, {Bizyaev}, {Brandt}, {Brinkmann}, {Brownstein}, {Comparat},
  {Green}, {de la Macorra}, {Mu{\~n}oz Guti{\'e}rrez}, {Hou}, {Newman},
  {Palanque-Delabrouille}, {P{\^a}ris}, {Percival}, {Petitjean}, {Rich},
  {Rossi}, {Schneider}, {Smith}, {Vivek}, \&
  {Weaver}}]{SDSS_Quasar_Catalog_DR16__2020ApJS..250....8L}
{Lyke}, B.~W., {Higley}, A.~N., {McLane}, J.~N., {et~al.} 2020, \apjs, 250, 8,
  \dodoi{10.3847/1538-4365/aba623}

\bibitem[{{MacLeod} {et~al.}(2011){MacLeod}, {Brooks}, {Ivezi{\'c}},
  {Kochanek}, {Gibson}, {Meisner}, {Koz{\l}owski}, {Sesar}, {Becker}, \& {de
  Vries}}]{Macleod2011ApJ...728...26M}
{MacLeod}, C.~L., {Brooks}, K., {Ivezi{\'c}}, {\v Z}., {et~al.} 2011, \apj,
  728, 26, \dodoi{10.1088/0004-637X/728/1/26}

\bibitem[{{Magorrian} {et~al.}(1998){Magorrian}, {Tremaine}, {Richstone},
  {Bender}, {Bower}, {Dressler}, {Faber}, {Gebhardt}, {Green}, {Grillmair},
  {Kormendy}, \& {Lauer}}]{1998AJ....115.2285M}
{Magorrian}, J., {Tremaine}, S., {Richstone}, D., {et~al.} 1998, \aj, 115,
  2285, \dodoi{10.1086/300353}

\bibitem[{{Maller} {et~al.}(2009){Maller}, {Berlind}, {Blanton}, \&
  {Hogg}}]{2009ApJ...691..394M}
{Maller}, A.~H., {Berlind}, A.~A., {Blanton}, M.~R., \& {Hogg}, D.~W. 2009,
  \apj, 691, 394, \dodoi{10.1088/0004-637X/691/1/394}

\bibitem[{Martin {et~al.}(2005)Martin, Fanson, Schiminovich, Morrissey,
  Friedman, Barlow, Conrow, Grange, Jelinsky, Milliard, \&
  et~al.}]{inital_GALEX_paper}
Martin, D.~C., Fanson, J., Schiminovich, D., {et~al.} 2005, The Astrophysical
  Journal, 619, L1–L6, \dodoi{10.1086/426387}

\bibitem[{{Mezcua} {et~al.}(2018){Mezcua}, {Civano}, {Marchesi}, {Suh},
  {Fabbiano}, \& {Volonteri}}]{Mezcua2018MNRAS.478.2576M}
{Mezcua}, M., {Civano}, F., {Marchesi}, S., {et~al.} 2018, \mnras, 478, 2576,
  \dodoi{10.1093/mnras/sty1163}

\bibitem[{{Molina} {et~al.}(2021){Molina}, {Reines}, {Latimer}, {Baldassare},
  \& {Salehirad}}]{Molina2021ApJ...922..155M}
{Molina}, M., {Reines}, A.~E., {Latimer}, C.~J., {Baldassare}, V., \&
  {Salehirad}, S. 2021, \apj, 922, 155, \dodoi{10.3847/1538-4357/ac1ffa}

\bibitem[{{Moran} {et~al.}(2014){Moran}, {Shahinyan}, {Sugarman}, {V{\'e}lez},
  \& {Eracleous}}]{Moran2014AJ....148..136M}
{Moran}, E.~C., {Shahinyan}, K., {Sugarman}, H.~R., {V{\'e}lez}, D.~O., \&
  {Eracleous}, M. 2014, \aj, 148, 136, \dodoi{10.1088/0004-6256/148/6/136}

\bibitem[{{Natarajan}(2014)}]{natarajan2014GReGr..46.1702N}
{Natarajan}, P. 2014, General Relativity and Gravitation, 46, 1702,
  \dodoi{10.1007/s10714-014-1702-6}

\bibitem[{{Pardo} {et~al.}(2016){Pardo}, {Goulding}, {Greene}, {Somerville},
  {Gallo}, {Hickox}, {Miller}, {Reines}, \&
  {Silverman}}]{Pardo2016ApJ...831..203P}
{Pardo}, K., {Goulding}, A.~D., {Greene}, J.~E., {et~al.} 2016, \apj, 831, 203,
  \dodoi{10.3847/0004-637X/831/2/203}

\bibitem[{{Reines} {et~al.}(2020){Reines}, {Condon}, {Darling}, \&
  {Greene}}]{Reines2020ApJ...888...36R}
{Reines}, A.~E., {Condon}, J.~J., {Darling}, J., \& {Greene}, J.~E. 2020, \apj,
  888, 36, \dodoi{10.3847/1538-4357/ab4999}

\bibitem[{Reines {et~al.}(2013)Reines, Greene, \& Geha}]{Reines:2013pia}
Reines, A.~E., Greene, J.~E., \& Geha, M. 2013, Astrophys. J., 775, 116,
  \dodoi{10.1088/0004-637X/775/2/116}

\bibitem[{Rodriguez-Pascual {et~al.}(1997)Rodriguez-Pascual, Alloin, Clavel,
  Crenshaw, Horne, Kriss, Krolik, Malkan, Netzer, O{\textquotesingle}Brien,
  Peterson, Reichert, Wamsteker, Alexander, Barr, Blandford, Bregman, Carone,
  Clements, Courvoisier, Robertis, Dietrich, Dottori, Edelson, Filippenko,
  Gaskell, Huchra, Hutchings, Kollatschny, Koratkar, Korista, Laor, MacAlpine,
  Martin, Maoz, McCollum, Morris, Perola, Pogge, Ptak, Recondo-Gonzalez,
  Rodriguez-Espinoza, Rokaki, Santos-Lleo, Sekiguchi, Shull, Snijders, Sparke,
  Stirpe, Stoner, Sun, Wagner, Wanders, Wilkes, Winge, \&
  Zheng}]{Rodriguez_Pascual_1997}
Rodriguez-Pascual, P.~M., Alloin, D., Clavel, J., {et~al.} 1997, The
  Astrophysical Journal Supplement Series, 110, 9, \dodoi{10.1086/312996}

\bibitem[{{Schlegel} {et~al.}(1998){Schlegel}, {Finkbeiner}, \&
  {Davis}}]{1998ApJ...500..525S}
{Schlegel}, D.~J., {Finkbeiner}, D.~P., \& {Davis}, M. 1998, \apj, 500, 525,
  \dodoi{10.1086/305772}

\bibitem[{{Skrutskie} {et~al.}(2006){Skrutskie}, {Cutri}, {Stiening},
  {Weinberg}, {Schneider}, {Carpenter}, {Beichman}, {Capps}, {Chester},
  {Elias}, {Huchra}, {Liebert}, {Lonsdale}, {Monet}, {Price}, {Seitzer},
  {Jarrett}, {Kirkpatrick}, {Gizis}, {Howard}, {Evans}, {Fowler}, {Fullmer},
  {Hurt}, {Light}, {Kopan}, {Marsh}, {McCallon}, {Tam}, {Van Dyk}, \&
  {Wheelock}}]{2MASS_2006AJ....131.1163S}
{Skrutskie}, M.~F., {Cutri}, R.~M., {Stiening}, R., {et~al.} 2006, \aj, 131,
  1163, \dodoi{10.1086/498708}

\bibitem[{{Smee} {et~al.}(2013){Smee}, {Gunn}, {Uomoto}, {Roe}, {Schlegel},
  {Rockosi}, {Carr}, {Leger}, {Dawson}, {Olmstead}, {Brinkmann}, {Owen},
  {Barkhouser}, {Honscheid}, {Harding}, {Long}, {Lupton}, {Loomis}, {Anderson},
  {Annis}, {Bernardi}, {Bhardwaj}, {Bizyaev}, {Bolton}, {Brewington}, {Briggs},
  {Burles}, {Burns}, {Castander}, {Connolly}, {Davenport}, {Ebelke}, {Epps},
  {Feldman}, {Friedman}, {Frieman}, {Heckman}, {Hull}, {Knapp}, {Lawrence},
  {Loveday}, {Mannery}, {Malanushenko}, {Malanushenko}, {Merrelli}, {Muna},
  {Newman}, {Nichol}, {Oravetz}, {Pan}, {Pope}, {Ricketts}, {Shelden},
  {Sandford}, {Siegmund}, {Simmons}, {Smith}, {Snedden}, {Schneider},
  {SubbaRao}, {Tremonti}, {Waddell}, \&
  {York}}]{SDSS_spect_2013AJ....146...32S}
{Smee}, S.~A., {Gunn}, J.~E., {Uomoto}, A., {et~al.} 2013, \aj, 146, 32,
  \dodoi{10.1088/0004-6256/146/2/32}

\bibitem[{{Spergel} {et~al.}(2007){Spergel}, {Bean}, {Dor{\'e}}, {Nolta},
  {Bennett}, {Dunkley}, {Hinshaw}, {Jarosik}, {Komatsu}, {Page}, {Peiris},
  {Verde}, {Halpern}, {Hill}, {Kogut}, {Limon}, {Meyer}, {Odegard}, {Tucker},
  {Weiland}, {Wollack}, \& {Wright}}]{spergel2007}
{Spergel}, D.~N., {Bean}, R., {Dor{\'e}}, O., {et~al.} 2007, \apjs, 170, 377,
  \dodoi{10.1086/513700}

\bibitem[{Stern {et~al.}(2012)Stern, Assef, Benford, Blain, Cutri, Dey,
  Eisenhardt, Griffith, Jarrett, Lake, Masci, Petty, Stanford, Tsai, Wright,
  Yan, Harrison, \& Madsen}]{Stern_2012}
Stern, D., Assef, R.~J., Benford, D.~J., {et~al.} 2012, The Astrophysical
  Journal, 753, 30, \dodoi{10.1088/0004-637x/753/1/30}

\bibitem[{{The Astropy Collaboration} {et~al.}(2018){The Astropy
  Collaboration}, {Price-Whelan}, {Sip{\\H o}cz}, {G{\\"u}nther}, {Lim},
  {Crawford}, {Conseil}, {Shupe}, {Craig}, {Dencheva}, {Ginsburg},
  {VanderPlas}, {Bradley}, {P{\\\'e}rez-Su{\\\'a}rez}, {de Val-Borro}, {Paper
  Contributors}, {Aldcroft}, {Cruz}, {Robitaille}, {Tollerud}, {Coordination
  Committee}, {Ardelean}, {Babej}, {Bach}, {Bachetti}, {Bakanov}, {Bamford},
  {Barentsen}, {Barmby}, {Baumbach}, {Berry}, {Biscani}, {Boquien}, {Bostroem},
  {Bouma}, {Brammer}, {Bray}, {Breytenbach}, {Buddelmeijer}, {Burke},
  {Calderone}, {Cano Rodr{\\\'{\\i}}guez}, {Cara}, {Cardoso}, {Cheedella},
  {Copin}, {Corrales}, {Crichton}, {D{\\rsquo}Avella}, {Deil}, {Depagne},
  {Dietrich}, {Donath}, {Droettboom}, {Earl}, {Erben}, {Fabbro}, {Ferreira},
  {Finethy}, {Fox}, {Garrison}, {Gibbons}, {Goldstein}, {Gommers}, {Greco},
  {Greenfield}, {Groener}, {Grollier}, {Hagen}, {Hirst}, {Homeier}, {Horton},
  {Hosseinzadeh}, {Hu}, {Hunkeler}, {Ivezi{\\\'c}}, {Jain}, {Jenness},
  {Kanarek}, {Kendrew}, {Kern}, {Kerzendorf}, {Khvalko}, {King}, {Kirkby},
  {Kulkarni}, {Kumar}, {Lee}, {Lenz}, {Littlefair}, {Ma}, {Macleod},
  {Mastropietro}, {McCully}, {Montagnac}, {Morris}, {Mueller}, {Mumford},
  {Muna}, {Murphy}, {Nelson}, {Nguyen}, {Ninan}, {N{\\"o}the}, {Ogaz}, {Oh},
  {Parejko}, {Parley}, {Pascual}, {Patil}, {Patil}, {Plunkett}, {Prochaska},
  {Rastogi}, {Reddy Janga}, {Sabater}, {Sakurikar}, {Seifert}, {Sherbert},
  {Sherwood-Taylor}, {Shih}, {Sick}, {Silbiger}, {Singanamalla}, {Singer},
  {Sladen}, {Sooley}, {Sornarajah}, {Streicher}, {Teuben}, {Thomas},
  {Tremblay}, {Turner}, {Terr{\\\'o}n}, {van Kerkwijk}, {de la Vega},
  {Watkins}, {Weaver}, {Whitmore}, {Woillez}, {Zabalza}, \&
  {Contributors}}]{astropy_2018AJ....156..123T}
{The Astropy Collaboration}, {Price-Whelan}, A.~M., {Sip{\\H o}cz}, B.~M.,
  {et~al.} 2018, \\aj, 156, 123, \dodoi{10.3847/1538-3881/aabc4f}

\bibitem[{{Trammell} {et~al.}(2007){Trammell}, {Vanden Berk}, {Schneider},
  {Richards}, {Hall}, {Anderson}, \& {Brinkmann}}]{2007AJ....133.1780T}
{Trammell}, G.~B., {Vanden Berk}, D.~E., {Schneider}, D.~P., {et~al.} 2007,
  \aj, 133, 1780, \dodoi{10.1086/511817}

\bibitem[{Trump {et~al.}(2015)Trump, Sun, Zeimann, Luck, Bridge, Grier, Hagen,
  Juneau, Montero-Dorta, Rosario, \& et~al.}]{Trump_2015}
Trump, J.~R., Sun, M., Zeimann, G.~R., {et~al.} 2015, The Astrophysical
  Journal, 811, 26, \dodoi{10.1088/0004-637x/811/1/26}

\bibitem[{Ulrich {et~al.}(1997)Ulrich, Maraschi, \& Urry}]{Ulrich_et_al_1997}
Ulrich, M.-H., Maraschi, L., \& Urry, C.~M. 1997, Annual Review of Astronomy
  and Astrophysics, 35, 445, \dodoi{10.1146/annurev.astro.35.1.445}

\bibitem[{Vaughan {et~al.}(2003)Vaughan, Edelson, Warwick, \&
  Uttley}]{Characterize_Variable_Vaughan_et_al_03}
Vaughan, S., Edelson, R., Warwick, R.~S., \& Uttley, P. 2003, Monthly Notices
  of the Royal Astronomical Society, 345, 1271,
  \dodoi{10.1046/j.1365-2966.2003.07042.x}

\bibitem[{Volonteri(2010)}]{VolonteriARAA}
Volonteri, M. 2010, A\&A Rev., 18

\bibitem[{{Volonteri} {et~al.}(2008){Volonteri}, {Lodato}, \&
  {Natarajan}}]{Volonteri2008MNRAS.383.1079V}
{Volonteri}, M., {Lodato}, G., \& {Natarajan}, P. 2008, \mnras, 383, 1079,
  \dodoi{10.1111/j.1365-2966.2007.12589.x}

\bibitem[{{Wright} {et~al.}(2010){Wright}, {Eisenhardt}, {Mainzer}, {Ressler},
  {Cutri}, {Jarrett}, {Kirkpatrick}, {Padgett}, {McMillan}, {Skrutskie},
  {Stanford}, {Cohen}, {Walker}, {Mather}, {Leisawitz}, {Gautier}, {McLean},
  {Benford}, {Lonsdale}, {Blain}, {Mendez}, {Irace}, {Duval}, {Liu}, {Royer},
  {Heinrichsen}, {Howard}, {Shannon}, {Kendall}, {Walsh}, {Larsen}, {Cardon},
  {Schick}, {Schwalm}, {Abid}, {Fabinsky}, {Naes}, \&
  {Tsai}}]{WISE_Wright2010AJ....140.1868W}
{Wright}, E.~L., {Eisenhardt}, P. R.~M., {Mainzer}, A.~K., {et~al.} 2010, \aj,
  140, 1868, \dodoi{10.1088/0004-6256/140/6/1868}

\bibitem[{{York} {et~al.}(2000){York}, {Adelman}, {Anderson}, {Anderson},
  {Annis}, {Bahcall}, {Bakken}, {Barkhouser}, {Bastian}, {Berman}, {Boroski},
  {Bracker}, {Briegel}, {Briggs}, {Brinkmann}, {Brunner}, {Burles}, {Carey},
  {Carr}, {Castander}, {Chen}, {Colestock}, {Connolly}, {Crocker}, {Csabai},
  {Czarapata}, {Davis}, {Doi}, {Dombeck}, {Eisenstein}, {Ellman}, {Elms},
  {Evans}, {Fan}, {Federwitz}, {Fiscelli}, {Friedman}, {Frieman}, {Fukugita},
  {Gillespie}, {Gunn}, {Gurbani}, {de Haas}, {Haldeman}, {Harris}, {Hayes},
  {Heckman}, {Hennessy}, {Hindsley}, {Holm}, {Holmgren}, {Huang}, {Hull},
  {Husby}, {Ichikawa}, {Ichikawa}, {Ivezi{\'c}}, {Kent}, {Kim}, {Kinney},
  {Klaene}, {Kleinman}, {Kleinman}, {Knapp}, {Korienek}, {Kron}, {Kunszt},
  {Lamb}, {Lee}, {Leger}, {Limmongkol}, {Lindenmeyer}, {Long}, {Loomis},
  {Loveday}, {Lucinio}, {Lupton}, {MacKinnon}, {Mannery}, {Mantsch}, {Margon},
  {McGehee}, {McKay}, {Meiksin}, {Merelli}, {Monet}, {Munn}, {Narayanan},
  {Nash}, {Neilsen}, {Neswold}, {Newberg}, {Nichol}, {Nicinski}, {Nonino},
  {Okada}, {Okamura}, {Ostriker}, {Owen}, {Pauls}, {Peoples}, {Peterson},
  {Petravick}, {Pier}, {Pope}, {Pordes}, {Prosapio}, {Rechenmacher}, {Quinn},
  {Richards}, {Richmond}, {Rivetta}, {Rockosi}, {Ruthmansdorfer}, {Sandford},
  {Schlegel}, {Schneider}, {Sekiguchi}, {Sergey}, {Shimasaku}, {Siegmund},
  {Smee}, {Smith}, {Snedden}, {Stone}, {Stoughton}, {Strauss}, {Stubbs},
  {SubbaRao}, {Szalay}, {Szapudi}, {Szokoly}, {Thakar}, {Tremonti}, {Tucker},
  {Uomoto}, {Vanden Berk}, {Vogeley}, {Waddell}, {Wang}, {Watanabe},
  {Weinberg}, {Yanny}, {Yasuda}, \& {SDSS Collaboration}}]{york2000}
{York}, D.~G., {Adelman}, J., {Anderson}, John~E., J., {et~al.} 2000, \aj, 120,
  1579, \dodoi{10.1086/301513}

\bibitem[{{Zhu} {et~al.}(2010){Zhu}, {Blanton}, \&
  {Moustakas}}]{2010ApJ...722..491Z}
{Zhu}, G., {Blanton}, M.~R., \& {Moustakas}, J. 2010, \apj, 722, 491,
  \dodoi{10.1088/0004-637X/722/1/491}

\end{thebibliography}
